\RequirePackage{lineno} 
\documentclass[reprint,superscriptaddress,twocolumn,showpacs,preprintnumbers,amsmath,amssymb]{revtex4}
\usepackage{graphicx}% Include figure files
\usepackage{dcolumn}% Align table columns on decimal point
\usepackage{bm}% bold math
\usepackage{color}% bold math
\usepackage{natbib}

\def \nobreakseq {\nobreak \hskip 0pt \hbox}

\definecolor{gris25}{gray}{0.75}
\definecolor{violet}{rgb}{0.5,0,0.5}
\definecolor{marron}{rgb}{0.88,0.41,0.}
\begin{document}
%\linenumbers

%\preprint{Release Candidate 6 - 21/10/13}

\title{Result of the search for neutrinoless double-$\beta$ decay in $^{100}$Mo \\with the NEMO-3 experiment}

\author{R.~Arnold}
\affiliation{IPHC, ULP, CNRS/IN2P3\nobreakseq{,} F-67037 Strasbourg, France}
\author{C.~Augier} 
\affiliation{LAL, Univ Paris-Sud\nobreakseq{,} CNRS/IN2P3\nobreakseq{,}
F-91405 Orsay\nobreakseq{,} France}
\author{J.D.~Baker$^{\footnote[2]{Deceased}}$}
\affiliation{Idaho National Laboratory\nobreakseq{,} Idaho Falls, ID 83415, U.S.A.}
\author{A.S.~Barabash}
\affiliation{ITEP, 117218 Moscow, Russia}
\author{A.~Basharina-Freshville} 
\affiliation{UCL, London WC1E 6BT\nobreakseq{,} United Kingdom}
\author{S.~Blondel} 
\affiliation{LAL, Univ Paris-Sud\nobreakseq{,} CNRS/IN2P3\nobreakseq{,}
F-91405 Orsay\nobreakseq{,} France}
\author{S.~Blot}
\affiliation{University of Manchester\nobreakseq{,} Manchester M13
  9PL\nobreakseq{,}~United Kingdom}
\author{M.~Bongrand} 
\affiliation{LAL, Univ Paris-Sud\nobreakseq{,} CNRS/IN2P3\nobreakseq{,}
F-91405 Orsay\nobreakseq{,} France}
\author{V.~Brudanin} 
\affiliation{JINR, 141980 Dubna, Russia}
\affiliation{National Research Nuclear University MEPhI, 115409 Moscow, Russia}
\author{J.~Busto} 
\affiliation{CPPM, Universit\'e de Marseille\nobreakseq{,} CNRS/IN2P3\nobreakseq{,} F-13288 Marseille\nobreakseq{,} France}
\author{A.J.~Caffrey}
\affiliation{Idaho National Laboratory\nobreakseq{,} Idaho Falls, ID 83415, U.S.A.}
\author{S. Calvez}
\affiliation{LAL, Univ Paris-Sud\nobreakseq{,} CNRS/IN2P3\nobreakseq{,}
F-91405 Orsay\nobreakseq{,} France}
\author{C.~Cerna} 
\affiliation{CENBG\nobreakseq{,} Universit\'e de Bordeaux\nobreakseq{,} CNRS/IN2P3\nobreakseq{,} F-33175 Gradignan\nobreakseq{,} France}
\author{J.~P.~Cesar}
\affiliation{University of Texas at Austin\nobreakseq{,}
  Austin\nobreakseq{,} TX 78712\nobreakseq{,}~U.S.A.}
\author{A.~Chapon} 
\affiliation{LPC Caen\nobreakseq{,} ENSICAEN\nobreakseq{,} Universit\'e de
Caen\nobreakseq{,} CNRS/IN2P3\nobreakseq{,} F-14050 Caen\nobreakseq{,} France}
\author{E.~Chauveau} 
\affiliation{University of Manchester\nobreakseq{,} Manchester M13
  9PL\nobreakseq{,}~United Kingdom}
\author{D.~Duchesneau} 
\affiliation{LAPP, Universit\'e de Savoie\nobreakseq{,}
CNRS/IN2P3\nobreakseq{,} F-74941 Annecy-le-Vieux\nobreakseq{,} France}
\author{D.~Durand} 
\affiliation{LPC Caen\nobreakseq{,} ENSICAEN\nobreakseq{,} Universit\'e de
Caen\nobreakseq{,} CNRS/IN2P3\nobreakseq{,} F-14050 Caen\nobreakseq{,} France}
\author{V.~Egorov}
\affiliation{JINR, 141980 Dubna, Russia}
\author{G.~Eurin} 
\affiliation{LAL, Univ Paris-Sud\nobreakseq{,} CNRS/IN2P3\nobreakseq{,}
F-91405 Orsay\nobreakseq{,} France}
\affiliation{UCL, London WC1E 6BT\nobreakseq{,} United Kingdom}
\author{J.J.~Evans} 
\affiliation{University of Manchester\nobreakseq{,} Manchester M13
  9PL\nobreakseq{,}~United Kingdom}
\author{L.~Fajt} 
\affiliation{Institute of Experimental and Applied Physics\nobreakseq{,} Czech Technical University in Prague\nobreakseq{,} CZ-12800 Prague\nobreakseq{,} Czech Republic}
\author{D.~Filosofov} 
\affiliation{JINR, 141980 Dubna, Russia}
\author{R.~Flack} 
\affiliation{UCL, London WC1E 6BT\nobreakseq{,} United Kingdom}
\author{X.~Garrido} 
\affiliation{LAL, Univ Paris-Sud\nobreakseq{,} CNRS/IN2P3\nobreakseq{,}
F-91405 Orsay\nobreakseq{,} France}
\author{H.~G\'omez} 
\affiliation{LAL, Univ Paris-Sud\nobreakseq{,} CNRS/IN2P3\nobreakseq{,}
F-91405 Orsay\nobreakseq{,} France}
\author{B.~Guillon} 
\affiliation{LPC Caen\nobreakseq{,} ENSICAEN\nobreakseq{,} Universit\'e de
Caen\nobreakseq{,} CNRS/IN2P3\nobreakseq{,} F-14050 Caen\nobreakseq{,} France}
\author{P.~Guzowski} 
\affiliation{University of Manchester\nobreakseq{,} Manchester M13
  9PL\nobreakseq{,}~United Kingdom}
\author{R.~Hod\'{a}k} 
\affiliation{Institute of Experimental and Applied Physics\nobreakseq{,} Czech
Technical University in Prague\nobreakseq{,} CZ-12800
Prague\nobreakseq{,} Czech Republic}
\author{A.~Huber} 
\affiliation{CENBG\nobreakseq{,} Universit\'e de Bordeaux\nobreakseq{,} CNRS/IN2P3\nobreakseq{,} F-33175 Gradignan\nobreakseq{,} France}
\author{P.~Hubert} 
\affiliation{CENBG\nobreakseq{,} Universit\'e de Bordeaux\nobreakseq{,} CNRS/IN2P3\nobreakseq{,} F-33175 Gradignan\nobreakseq{,} France}
\author{C.~Hugon}
\affiliation{CENBG\nobreakseq{,} Universit\'e de Bordeaux\nobreakseq{,} CNRS/IN2P3\nobreakseq{,} F-33175 Gradignan\nobreakseq{,} France}
\author{S.~Jullian} 
\affiliation{LAL, Univ Paris-Sud\nobreakseq{,} CNRS/IN2P3\nobreakseq{,}
F-91405 Orsay\nobreakseq{,} France}
\author{A.~Klimenko} 
\affiliation{JINR, 141980 Dubna, Russia}
\author{O.~Kochetov} 
\affiliation{JINR, 141980 Dubna, Russia}
\author{S.I.~Konovalov} 
\affiliation{ITEP, 117218 Moscow, Russia}
\author{V.~Kovalenko}
\affiliation{JINR, 141980 Dubna, Russia}
\author{D.~Lalanne} 
\affiliation{LAL, Univ Paris-Sud\nobreakseq{,} CNRS/IN2P3\nobreakseq{,}
F-91405 Orsay\nobreakseq{,} France}
\author{K.~Lang} 
\affiliation{University of Texas at Austin\nobreakseq{,}
  Austin\nobreakseq{,} TX 78712\nobreakseq{,}~U.S.A.}
\author{Y.~Lemi\`ere} 
\affiliation{LPC Caen\nobreakseq{,} ENSICAEN\nobreakseq{,} Universit\'e de
Caen\nobreakseq{,} CNRS/IN2P3\nobreakseq{,} F-14050 Caen\nobreakseq{,} France}
\author{T.~Le~Noblet} 
\affiliation{LAPP, Universit\'e de Savoie\nobreakseq{,} CNRS/IN2P3\nobreakseq{,} F-74941 Annecy-le-Vieux\nobreakseq{,} France}
\author{Z.~Liptak} 
\affiliation{University of Texas at Austin\nobreakseq{,}
  Austin\nobreakseq{,} TX 78712\nobreakseq{,}~U.S.A.}
\author{P.~Loaiza} 
\affiliation{LAL, Univ Paris-Sud\nobreakseq{,} CNRS/IN2P3\nobreakseq{,}
F-91405 Orsay\nobreakseq{,} France}
\author{G.~Lutter} 
\affiliation{CENBG\nobreakseq{,} Universit\'e de Bordeaux\nobreakseq{,} CNRS/IN2P3\nobreakseq{,} F-33175 Gradignan\nobreakseq{,} France}
\author{F.~Mamedov}
\affiliation{Institute of Experimental and Applied Physics\nobreakseq{,} Czech
Technical University in Prague\nobreakseq{,} CZ-12800
Prague\nobreakseq{,} Czech Republic}
\author{C.~Marquet} 
\affiliation{CENBG\nobreakseq{,} Universit\'e de Bordeaux\nobreakseq{,} CNRS/IN2P3\nobreakseq{,} F-33175 Gradignan\nobreakseq{,} France}
\author{F.~Mauger} 
\affiliation{LPC Caen\nobreakseq{,} ENSICAEN\nobreakseq{,} Universit\'e de
Caen\nobreakseq{,} CNRS/IN2P3\nobreakseq{,} F-14050 Caen\nobreakseq{,} France}
\author{B.~Morgan} 
\affiliation{University of Warwick\nobreakseq{,} Coventry CV4
7AL\nobreakseq{,} United Kingdom}
\author{J.~Mott} 
\affiliation{UCL, London WC1E 6BT\nobreakseq{,} United Kingdom}
\author{I.~Nemchenok} 
\affiliation{JINR, 141980 Dubna, Russia}
\author{M.~Nomachi} 
\affiliation{Osaka University\nobreakseq{,} 1-1 Machikaney arna
Toyonaka\nobreakseq{,} Osaka 560-0043\nobreakseq{,} Japan}
\author{F.~Nova} 
\affiliation{University of Texas at Austin\nobreakseq{,}
  Austin\nobreakseq{,} TX 78712\nobreakseq{,}~U.S.A.}
\author{F.~Nowacki} 
\affiliation{IPHC, ULP, CNRS/IN2P3\nobreakseq{,} F-67037 Strasbourg, France}
\author{H.~Ohsumi} 
\affiliation{Saga University\nobreakseq{,} Saga 840-8502\nobreakseq{,}
  Japan}
\author{R.B.~Pahlka}
\affiliation{University of Texas at Austin\nobreakseq{,}
  Austin\nobreakseq{,} TX 78712\nobreakseq{,}~U.S.A.}
\author{F.~Perrot} 
\affiliation{CENBG\nobreakseq{,} Universit\'e de Bordeaux\nobreakseq{,} CNRS/IN2P3\nobreakseq{,} F-33175 Gradignan\nobreakseq{,} France}
\author{F.~Piquemal} 
\affiliation{CENBG\nobreakseq{,} Universit\'e de Bordeaux\nobreakseq{,} CNRS/IN2P3\nobreakseq{,} F-33175 Gradignan\nobreakseq{,} France}
\affiliation{Laboratoire Souterrain de Modane\nobreakseq{,} F-73500
Modane\nobreakseq{,} France}
\author{P.~Povinec}
\affiliation{FMFI,~Comenius~Univ.\nobreakseq{,}~SK-842~48~Bratislava\nobreakseq{,}~Slovakia}
\author{P.~P\v{r}idal} 
\affiliation{Institute of Experimental and Applied Physics\nobreakseq{,} Czech Technical University in Prague\nobreakseq{,} CZ-12800 Prague\nobreakseq{,} Czech Republic}
\author{Y.A.~Ramachers} 
\affiliation{University of Warwick\nobreakseq{,} Coventry CV4
7AL\nobreakseq{,} United Kingdom}
\author{A.~Remoto}
\affiliation{LAPP, Universit\'e de Savoie\nobreakseq{,}
CNRS/IN2P3\nobreakseq{,} F-74941 Annecy-le-Vieux\nobreakseq{,} France}
\author{J.L.~Reyss} 
\affiliation{LSCE\nobreakseq{,} CNRS\nobreakseq{,} F-91190
  Gif-sur-Yvette\nobreakseq{,} France}
\author{B.~Richards} 
\affiliation{UCL, London WC1E 6BT\nobreakseq{,} United Kingdom}
\author{C.L.~Riddle} 
\affiliation{Idaho National Laboratory\nobreakseq{,} Idaho Falls, ID 83415, U.S.A.}
\author{E.~Rukhadze} 
\affiliation{Institute of Experimental and Applied Physics\nobreakseq{,} Czech
Technical University in Prague\nobreakseq{,} CZ-12800
Prague\nobreakseq{,} Czech Republic}
\author{R.~Saakyan} 
\affiliation{UCL, London WC1E 6BT\nobreakseq{,} United Kingdom}
\author{X.~Sarazin} 
\affiliation{LAL, Univ Paris-Sud\nobreakseq{,} CNRS/IN2P3\nobreakseq{,}
F-91405 Orsay\nobreakseq{,} France}
\author{Yu.~Shitov} 
\affiliation{JINR, 141980 Dubna, Russia}
\affiliation{Imperial College London\nobreakseq{,} London SW7
2AZ\nobreakseq{,} United Kingdom}
\author{L.~Simard} 
\affiliation{LAL, Univ Paris-Sud\nobreakseq{,} CNRS/IN2P3\nobreakseq{,}
F-91405 Orsay\nobreakseq{,} France}
\affiliation{Institut Universitaire de France\nobreakseq{,} F-75005 Paris\nobreakseq{,} France}
\author{F.~\v{S}imkovic} 
\affiliation{FMFI,~Comenius~Univ.\nobreakseq{,}~SK-842~48~Bratislava\nobreakseq{,}~Slovakia}
\author{A.~Smetana}
\affiliation{Institute of Experimental and Applied Physics\nobreakseq{,} Czech
Technical University in Prague\nobreakseq{,} CZ-12800
Prague\nobreakseq{,} Czech Republic}
\author{K.~Smolek} 
\affiliation{Institute of Experimental and Applied Physics\nobreakseq{,} Czech
Technical University in Prague\nobreakseq{,} CZ-12800
Prague\nobreakseq{,} Czech Republic}
\author{A.~Smolnikov} 
\affiliation{JINR, 141980 Dubna, Russia}
\author{S.~S\"oldner-Rembold}
\affiliation{University of Manchester\nobreakseq{,} Manchester M13
  9PL\nobreakseq{,}~United Kingdom}
\author{B.~Soul\'e}
\affiliation{CENBG\nobreakseq{,} Universit\'e de Bordeaux\nobreakseq{,} CNRS/IN2P3\nobreakseq{,} F-33175 Gradignan\nobreakseq{,} France}
\author{I.~\v{S}tekl} 
\affiliation{Institute of Experimental and Applied Physics\nobreakseq{,} Czech Technical University in Prague\nobreakseq{,} CZ-12800 Prague\nobreakseq{,} Czech Republic}
\author{J.~Suhonen} 
\affiliation{Jyv\"askyl\"a University\nobreakseq{,} FIN-40351 Jyv\"askyl\"a\nobreakseq{,} Finland}
\author{C.S.~Sutton} 
\affiliation{MHC\nobreakseq{,} South Hadley\nobreakseq{,} Massachusetts 01075\nobreakseq{,} U.S.A.}
\author{G.~Szklarz}
\affiliation{LAL, Univ Paris-Sud\nobreakseq{,} CNRS/IN2P3\nobreakseq{,}
F-91405 Orsay\nobreakseq{,} France}
\author{J.~Thomas} 
\affiliation{UCL, London WC1E 6BT\nobreakseq{,} United Kingdom}
\author{V.~Timkin} 
\affiliation{JINR, 141980 Dubna, Russia}
\author{S.~Torre} 
\affiliation{UCL, London WC1E 6BT\nobreakseq{,} United Kingdom}
\author{Vl.I.~Tretyak} 
\affiliation{Institute for Nuclear Research\nobreakseq{,} MSP 03680\nobreakseq{,} Kyiv\nobreakseq{,} Ukraine}
\author{V.I.~Tretyak}
\affiliation{JINR, 141980 Dubna, Russia}
\author{V.I.~Umatov} 
\affiliation{ITEP, 117218 Moscow, Russia}
\author{I.~Vanushin} 
\affiliation{ITEP, 117218 Moscow, Russia}
\author{C.~Vilela} 
\affiliation{UCL, London WC1E 6BT\nobreakseq{,} United Kingdom}
\author{V.~Vorobel} 
\affiliation{Charles University in Prague\nobreakseq{,} Faculty of Mathematics
and Physics\nobreakseq{,} CZ-12116 Prague\nobreakseq{,} Czech Republic}
\author{D.~Waters} 
\affiliation{UCL, London WC1E 6BT\nobreakseq{,} United Kingdom}
\author{A.~\v{Z}ukauskas}
\affiliation{Charles University in Prague\nobreakseq{,} Faculty of Mathematics
and Physics\nobreakseq{,} CZ-12116 Prague\nobreakseq{,} Czech Republic}
\collaboration{NEMO-3 Collaboration}
\noaffiliation

\date{\today}% It is always \today, today,

\begin{abstract}

The NEMO-3 detector, which had been operating in the Modane Underground Laboratory from 2003 to 2010, was designed to search for neutrinoless double $\beta$ ($0\nu\beta\beta$) decay. 
We report final results of a search for $0\nu\beta\beta$ decays with $6.914$~kg of $^{100}$Mo using the entire NEMO-3 data set with a detector live time of $4.96$~yr, which corresponds to an exposure of 34.3~kg$\cdot$yr. We perform a detailed study of the expected background in the $0\nu\beta\beta$ signal region and  find no evidence of $0\nu\beta\beta$ decays in the data. The level of observed background in the $0\nu\beta\beta$ signal region $[2.8-3.2]$~MeV is $0.44 \pm 0.13$~counts/yr/kg, and no events are observed in the interval $[3.2-10]$~MeV. 
We therefore derive a lower limit on the half-life of $0\nu\beta\beta$ decays in $^{100}$Mo of $T_{1/2}(0\nu\beta\beta)> 1.1 \times 10^{24}$~yr at the $90\%$ Confidence Level, under the hypothesis of decay kinematics similar to that for light Majorana neutrino exchange.
Depending on the model used for calculating nuclear matrix elements, the limit for the effective Majorana neutrino mass lies in the range $\langle m_{\nu} \rangle < 0.33$--$0.62$~eV. We also report constraints on other lepton-number violating mechanisms for $0\nu\beta\beta$ decays. 
\end{abstract}


\maketitle

\newpage

%\tableofcontents

%\linenumbers

\section{Introduction}

Since neutrinos are the only fermions that carry no electric charge, they can be represented by a Majorana field for which the distinction between matter and antimatter vanishes. The Majorana nature of neutrinos could play a fundamental role in many extensions of the Standard Model. For instance, the see-saw mechanism~\cite{Mohapatra1980}, which requires the existence of a Majorana neutrino, naturally explains the origin of small neutrino masses. A Majorana neutrino would provide a framework for lepton number violation, and in particular for the Leptogenesis process~\cite{Leptogenesis},  which could explain the observed matter-antimatter asymmetry in the Universe. 

The observation of neutrinoless double $\beta$ ($0\nu\beta\beta$) decay would prove that neutrinos are Majorana particles~\cite{valle} and that lepton number is not conserved.  
The most commonly studied mechanism of $0\nu\beta\beta$ decay is the exchange of a Majorana neutrino. However, other mechanisms such as the existence of right-handed currents in the electroweak interaction, the exchange of supersymmetric particles with $R$-parity violating couplings, or the additional emission of a Majoron particle, are possible. 
Except for the case of Majoron emission, the experimental signature of $0\nu\beta\beta$ decays is the emission of two electrons with a total energy $E_{\rm tot}$ that is equal to the transition energy $Q_{\beta\beta}$ of the decay. 

For a given mechanism and isotope, the $0\nu\beta\beta$ decay half-life depends on the phase space factors and on the nuclear matrix element (NME). 
The decay half-lives of different isotopes can differ by a few orders of magnitude with large theoretical uncertainties of the NME calculations. It is therefore essential to search for $0\nu\beta\beta$ decays in several isotopes. 

The NEMO-3 detector~\cite{nemo3-tdr-2005} was operated from 2003 until 2010 in the Modane Underground Laboratory (LSM) to measure two-neutrino double $\beta$ ($2\nu\beta\beta$) decays of seven isotopes in the form of thin foils and to search for $0\nu\beta\beta$ decays.
The full topology of double $\beta$ decays is reconstructed
by combining information from a calorimeter and a tracking detector that are both distinct from the source foils. We measure the contributions from different background processes separately by exploiting specific event topologies. The NEMO-3 design and its capacity to identify electrons, positrons, $\gamma$ rays, and $\alpha$ particles are unique in enabling us to reject background processes very efficiently. 

The isotope $^{100}$Mo  represents the largest source sample in NEMO-3 with a mass of $6.914$~kg and $Q_{\beta\beta}=3034.40 \pm 0.17$~keV~\cite{Rahaman2007}.
A result based on a subset of the data had previously been published in~\cite{article0nuphase1}. 
We reported as rapid communications~\cite{nemo3-mo100-pr-shortcomm} the results of a search for $0\nu\beta\beta$ decays for the entire data set, corresponding to a live time of $4.96$~yr and an exposure of 34.3~kg$\cdot$yr of $^{100}$Mo. 
In this Article, we describe this analysis in more detail.

The NEMO-3 detector is introduced in Section~\ref{sec:detector}, and the energy and timing calibration of the detector are described in Section~\ref{sec:calib}. Selection criteria for $0\nu\beta\beta$ candidates are given in Section~\ref{sec:bb-selection}. The methodology and the results of the measurement of the different background components are presented in Section~\ref{sec:bkg}. Results of the search for $0\nu\beta\beta$ decays are summarised in Section~\ref{sec:bb0nu}.

\section{The NEMO-3 detector}
\label{sec:detector}

%In a double $\beta$ decay, two electrons are simultaneously emitted from a common vertex in one of the source foils.
The distinctive feature of the NEMO-3 detection method is a full reconstruction of the double $\beta$ decay topology using tracking in three dimensions as well as calorimetric and timing information.  
It provides not only the total energy $E_{\rm tot}$ of the two simultaneously emitted electrons, but also the single energy of each electron and their angular distribution at the emission point from the foil. 
A detailed description of the NEMO-3 detector can be found in \cite{nemo3-tdr-2005}.

\begin{figure}[htbp]
\begin{center}
\includegraphics[scale=0.33]{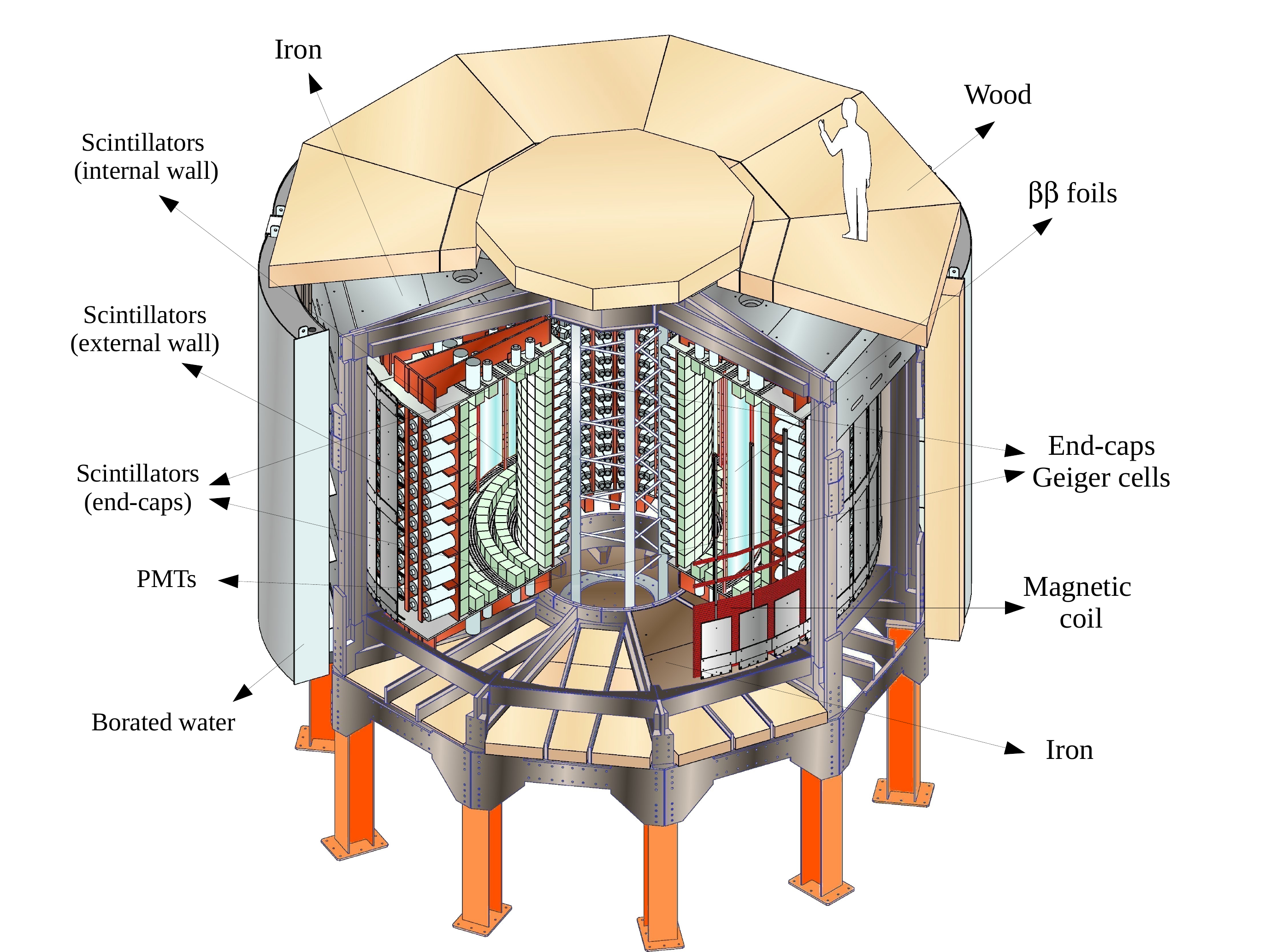}
\caption{A schematic view of the NEMO-3 detector, showing the double $\beta$ source foils, the tracking chamber, the calorimeter composed of scintillator blocks and PMTs, the magnetic coil and the shield.}
\label{fig:nemo3-layout}
\end{center}
\end{figure}

The thin source foils with a density of $40$--$60$~mg/cm$^2$ containing the active double $\beta$ decay isotope are surrounded by a tracking detector comprising open drift cells and a calorimeter composed of plastic scintillators. The source
foils are distributed over a cylindrical surface of about 20~m$^2$, which is segmented into
20 sectors of equal size, as shown in Figure~\ref{fig:nemo3-layout}. 

Several double $\beta$ decay sources are installed in the detector.
The main isotope used to search for $0\nu\beta\beta$ decays is $^{100}$Mo with a total mass of $6.914$~kg.
Smaller amounts of other isotopes are mainly used to measure $2\nu\beta\beta$ decays, comprising
 $^{82}$Se (0.932~kg, 2 sectors), $^{116}$Cd (0.405~kg, 1 sector), $^{130}$Te (0.454~kg, 2 sectors), $^{150}$Nd (36.55~g), $^{96}$Zr (9.4~g), and $^{48}$Ca (7~g).
In addition, 1.5 sectors of natural tellurium, corresponding to $0.614$~kg of TeO$_2$, and 1~sector equipped with pure copper ($0.621$~kg) are used to perform measurements of backgrounds from processes other than double $\beta$ decay. 
There are seven foil strips in each sector.
The mean length of the strips is 2480 mm with a width of 65~mm for the five central strips and 63~mm for the two edge strips.

There are two types of $^{100}$Mo foils, metallic and composite. 
The metallic foils were produced in vacuum by heating and rolling $^{100}$Mo mono-crystals in the form of foils. 
To produce the composite foils, thin and chemically purified $^{100}$Mo powder was mixed with polyvinyl alcohol (PVA) glue and then deposited between mylar foils with a thickness of 19~$\mu$m. 
The metallic Mo foils were placed in Sectors 02, 03 and 04. 
There are also five additional strips in Sector 1 and two strips in Sector 5.
The total surface of metallic foils is $43924$~cm$^2$.
The total mass of $^{100}$Mo in metallic foils is 2479~g and the average percentage of $^{100}$Mo enrichment is $97.7\%$. 
It corresponds to an average surface density of the metallic foils of 57.9~mg/cm$^2$. 
The composite Mo foils were placed in Sectors 10, 11, 12, 13, 14, 15, and 16. 
There are also two additional strips in Sector 01 and three strips in Sector 05.
The total surface of composite Mo foils is $84410$~cm$^2$.
The total mass of $^{100}$Mo in composite foils is 4435~g and the average percentage of $^{100}$Mo enrichment is $96.5\%$. 
The total mass of components (Mo, PVA and Mylar) is 5569~g. 
It corresponds to an average surface density of the composite foils of 66.0~mg/cm$^2$. 

On both sides of the source foils, a gaseous tracking detector comprising $6180$ open drift cells operating in Geiger mode provides three-dimensional track information. 
We use a cylindrical coordinate system with the $z$-axis pointing upwards.
The drift cells are oriented parallel to the $z$-axis and provide measurements of the transverse and longitudinal coordinates of the track.
To minimize multiple scattering, the gas is a mixture of $94.9\%$ helium, $4\%$ ethyl alcohol, $1\%$ argon, and $0.1\%$ water vapor for a total volume of about 28.5~m$^3$. 
 
%----------------------------------
% added for referee
%{\color{red} 
The basic cell consists of a central anode wire surrounded by eight ground wires. 
All the wires are 50~$\mu$m in diameter and 2.7~m long.
On each end of the cell is the cathode ring. 
When a charged particle crosses a cell the ionized gas yields around six electrons per centimeter. 
These secondary electrons drift towards the anode wire at a speed of around 1 to 2 cm/$\mu$s depending on the distance of the electrons to the anode. 
Measurements of these drift anodic times are used to reconstruct the transverse position of the particle in the cells. 
In the Geiger regime, the avalanche near the anode wire develops into a Geiger plasma which propagates along the wire at a speed of about 6.5~cm/$\mu$s. 
The arrival of the plasma at the two ends of the wire is detected with the cathode rings mentioned above. 
The two propagation times of the plasma are used to determine the longitudinal position of the particle as it passes through the cell.

To readout the drift cells, the analog Geiger signals from the anode wires and the two cathode rings signals are first amplified and then compared to anode and cathode thresholds. 
For signals exceeding the thresholds, the anode signal starts four Time-to-Digital Converter (TDC) scalers. 
The first three are for the anode and the two cathode contents which are measured with a 12-bit TDC and give times between 0 and 82~$\mu$s.  The last TDC scaler is 17-bits, which can provide time measurements between 0 and 2.6~ms. It is used for delayed $\alpha$ particle tagging. The cathode TDCs are stopped by the cathode signals while the anode TDCs are stopped by a signal sent by the general trigger.
%}
%----------------------------------

%----------------------------------
% added for referee
%{\color{red} 
The methode for the track reconstruction and its calibration, and the tracking performances are presented in \cite{nemo3-tdr-2005}.
%}
%----------------------------------
The average transverse and longitudinal resolutions of the Geiger cells are $0.5$~mm and 
$0.8$~cm, respectively.  If the two electron tracks from a double $\beta$ decay
are constrained to originate from the same vertex in the foil, the transverse and longitudinal vertex resolutions, defined as the r.m.s. of the distance between the intersection points of
the two individual tracks with the foil, are $0.6$~cm and $1.0$~cm, respectively.
These resolutions are sufficient to discriminate between decays from different source foils and isotopes.

The energy and time-of-flight of particles are measured by polystyrene scintillators surrounding the tracking detector. We use the time-of-flight
to discriminate between signal events emitted from the foil and 
background events where particles crossed the foil.
To further increase acceptance,
 the end caps (the top and bottom parts of the detector, named petals) are also equipped with scintillators in the spaces between the drift cell layers. 
The calorimeter is composed of $1940$ optical modules, which consist of large scintillator blocks, with a typical size of 
$20 \times 20 \times 10$~cm$^3$, coupled to low radioactive 
photomultipliers (PMTs).

%------------------------------
% added for referee
%{\color{red} 
The analog PMTs signals are sent to both a low and a high threshold leading edge discriminator.
If the PMT signal exceeds the lower level threshold it starts a TDC measurement and opens a charge
integration gate for 80 ns. The high threshold discriminator works as a one shot that delivers a calorimeter event signal to the trigger logic which reflects the number of channels that have exceeded the upper threshold. 
This level is used to trigger the system (first level trigger) if the desired multiplicity of active PMTs is achieved. 
The trigger logic then produces a signal called STOP-PMT, which is sent to all the calorimeter electronic channels, to save their data.
So the TDCs are stopped and the integrated charge is stored. Then digital conversions begin.
At the same time, a signal is sent to the calorimeter acquisition processor, which permits the read out of the digitized times and charges for the active channels.
The analog-to-digital conversions of the charge and the timing signal are made with two 12-bit ADCs. 
The energy resolution is 0.36~pC/channel (about 3~keV/channel) and the time resolution is 53~ps/channel. 
If any PMT signal exceeds the high level threshold then the TDC measurement and charge integration are aborted and the system resets after 200 ns.
%}
%------------------------------

The external wall of the calorimeter is equipped with $5$-inch PMTs and the internal wall 
with $3$-inch PMTs, and the end caps with both PMT types. 
The average energy resolution of the calorimeter is $\sigma_E/E = 5.8 \%/\sqrt{E(\mathrm{MeV})}$ for the scintillators equipped with $5$-inch PMTs, and $\sigma_E/E = 7.2 \%/\sqrt{E(\mathrm{MeV})}$ for the scintillators equipped with $3$-inch PMTs.

Photons are identified as hits in the calorimeter where no electron
track points at the scintillator block. The scintillator blocks with a thickness of $10$~cm yield a high photon detection 
efficiency of
$51\%$ $(33\%)$ for a photon of energy $1$~MeV ($3$~MeV) at a normal angle of incident.

A solenoidal magnet surrounding the detector provides a magnetic field of $25$~G used to discriminate between electrons and positrons with an efficiency of about $95 \%$ 
at an electron/positron energy of $1$~MeV.
An external shield with a thickness of $19$~cm constructed of low-radioactive iron, a borated water shield, and a wood shield  
surround the detector to reduce background from external $\gamma$ rays and neutrons. 
Calibrations are performed by inserting calibrated radioactive sources into the detector through dedicated tubes installed between each sector in the plane of the foils.

During the first data taking period, labeled Phase~I, from February 2003 until October 2004, the dominant background to the $0\nu\beta\beta$ signal was contamination from radon ($^{222}$Rn) in the tracking chamber.
Radon contamination in the tracking chamber is measured by
detecting electrons from $\beta$ decay of $^{214}\mathrm{Bi}$,
accompanied by a delayed $\alpha$ particle from $^{214}$Po decay.  
To detect delayed $\alpha$ particles,
every hit inside the wire chamber arriving with a delay 
of up to $700$~$\mu$s is read out with dedicated electronics.
The $^{222}$Rn activity of about $30$~mBq/m$^3$ inside the tracking chamber during Phase~I is caused by 
a low rate of diffusion of $^{222}$Rn from the laboratory hall, with an activity of around $15$~Bq/m$^3$, into the detector.
This contamination was significantly reduced, by a factor of about 6, by the installation of a radon-tight tent enclosing the detector and a radon trapping facility in December 2004.
The second data taking period between December 2004 until the end of running in December 2010 
(Phase~II) therefore has a reduced radon gas contamination of around 5~mBq/m$^3$. Data from both Phases are presented in this Article.

The trigger conditions used for recording double $\beta$ candidate events require at least one PMT signal with an amplitude greater than 50~mV, corresponding to an energy of $>150$~keV deposited in the associated scintillator, in coincidence with at least three hits in
the tracking detector within a time window of $6$~$\mu$s recorded
in the same half-sector of the detector as the scintillator hit.
Additional PMT signals with an amplitude of $>10$~mV, corresponding to an energy 
deposit of $>30$~keV, are also recorded if they coincide within a time window of $80$~ns.
The trigger rates of the data acquisition are about 7~Hz for Phase~I and about 5~Hz for Phase-II.
The dead time of the data acquisition is measured to be $1\%$ and is treated as an inefficiency.

Monte Carlo (MC) simulations are performed with a {\sc geant3}-based~\cite{geant3} 
detector simulation using the {\sc \mbox{decay0}}~\cite{decay-generator} event generator.
The time-dependent status and conditions of the detector and its performance are
taken into account in the detector simulation.

In this Article, we present a search for $0\nu\beta\beta$ decays using 
data recorded between February 2003 and October 2010, 
with a live time of $1.02$~yr in Phase~I and $3.94$~yr in Phase~II, and a total mass of $6914$~g of $^{100}$Mo in the form of metallic and composite foils.
 This corresponds to a total exposure  of 34.3~kg$\cdot$yr. 

\section{Calibration}
\label{sec:calib}

\subsection{Energy scale calibration and resolution}
\label{sec:energy-calib}

Absolute energy calibrations of the calorimeter optical modules were carried out every month using $^{207}$Bi sources which provide internal conversion electrons with energies of $482$~keV and $976$~keV from the $K$ lines, 
with branching ratios of $1.5\%$ and $7.1\%$, respectively.  
Each calibration run has a length of about 24~hours. In addition,
a dedicated long run was performed using a $^{90}$Sr source since 
the end point of the $\beta$ spectrum of $^{90}$Y, a daughter nucleus of $^{90}$Sr, provides an additional high-energy point at an energy of $2279$~keV. 

\begin{figure}[htbp] 
\begin{center}
\includegraphics[scale=0.6]{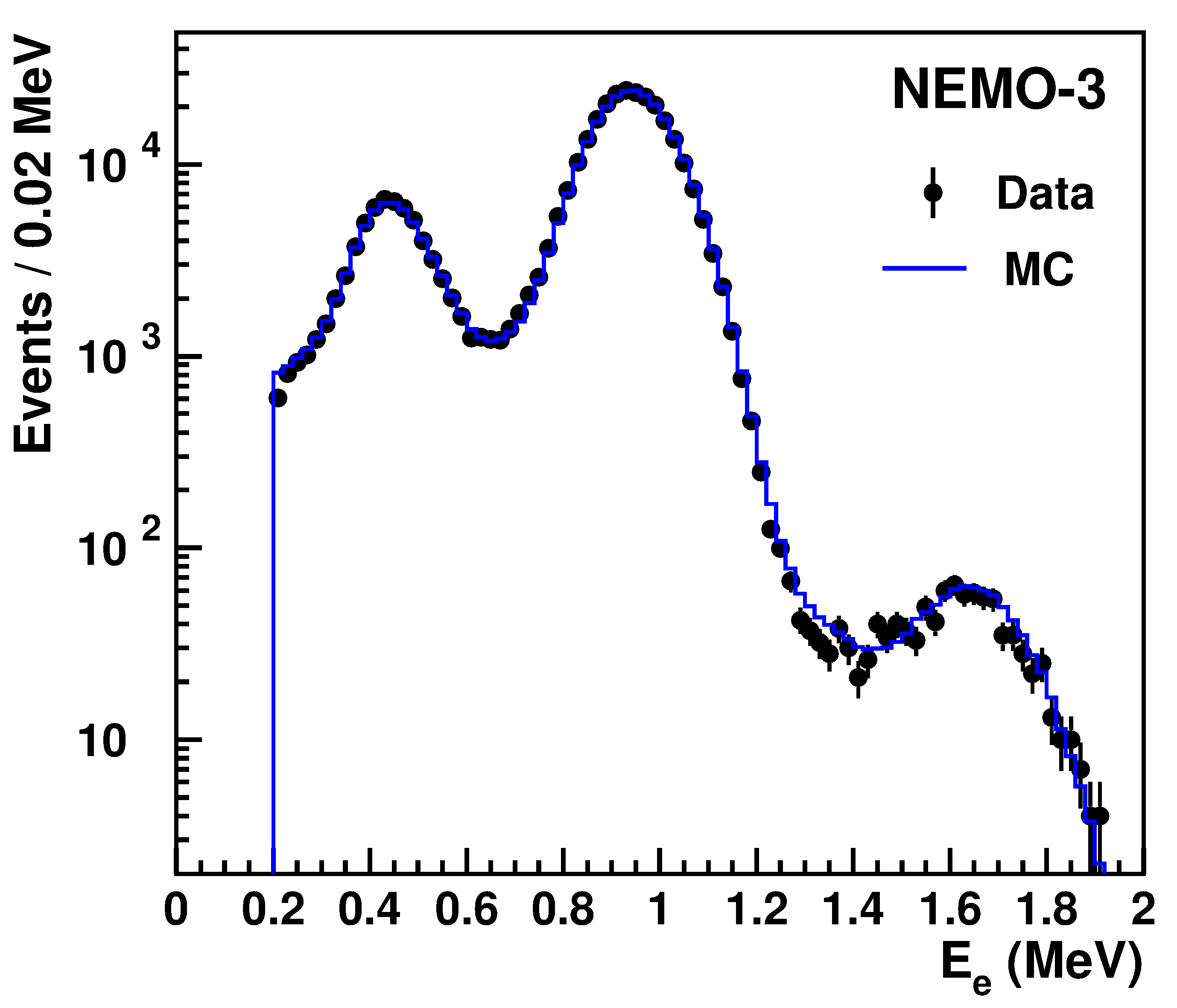}
\caption{Energy spectrum of a typical scintillator block, measured with the $^{207}$Bi calibration sources and summed over all the calibration runs. The data points are compared to a histogram of the energy spectrum calculated by the MC simulation. The peaks correspond to the energies of electrons from the main $482$~keV, $976$~keV and $1682$~keV internal conversion $K$ lines of $^{207}$Bi.}
\label{fig:bi207-1682keV}
\end{center}
\end{figure}

The response of each scintillator block to electrons with an energy of $976$~keV is measured as a function of the impact position of the electron track on the entrance surface of the scintillator using $^{207}$Bi calibration runs. A dependence on impact position was previously observed with data obtained with the electron spectrometer during the NEMO-3 calorimeter assembly. 
The impact position is sampled by dividing the entrance surface of the scintillator blocks in $3 \times 3$ equal squares for the  blocks equipped with $3$-inch PMTs, and $5 \times 5$ equal squares for the  blocks equipped with $5$-inch PMTs, corresponding to $3 \times 3$ and $5 \times 5$ corrections points respectively. 
The impact corrections are small for the scintillator blocks equipped with $3$-inch PMTs, typically $1\%-2\%$, but they can increase up to $10\%$ for $5$-inch PMT scintillator blocks. This effect is corrected offline by applying different impact correction factors for each scintillator block type.

%-------------------------------
% Referee
%{\color{red} 
The linearity of the PMTs has been verified with a dedicated light injection test during the construction phase. Upper limits on the non-linearity of the PMTs were found to be $<1\%$ for energies $<4$~MeV, corresponding to the energy range of interest for double $\beta$ decay measurements. 
It is shown by Monte-Carlo simulations that a non-linearity lower than 1\% has no effect in the final $\beta\beta 0 \nu$ analysis.
%}
%-------------------------------

The linear fit combining the energy calibration obtained with the two $^{207}$Bi energy peaks and the end point of the $^{90}$Y $\beta$ spectrum does not intersect with the origin, because the scintillator response for electrons at low energies (below the energy threshold of 200~keV) is non-linear. The extrapolated energy offset at a charge of $Q_{\rm ADC}=0$ is on average $33 \pm 3$~keV. It is determined after subtracting the electronic pedestal of the Analog-to-Digital Converters (ADCs) used
to read out the PMTs and accounting for an impact point correction. 
This offset is taken into account in the energy calculation. 
%-------------------------------
% Referee
%{\color{red} 
It is shown by Monte-Carlo simulations that the uncertainty on the energy offset measurement is negligible for the final $\beta\beta 0 \nu$ analysis.
%}
%-------------------------------
An example of a linear fit for one counter can be found in~Ref~\cite{nemo3-tdr-2005}.

\begin{figure}[htbp]  
\begin{center}
\includegraphics[width=0.48\textwidth]{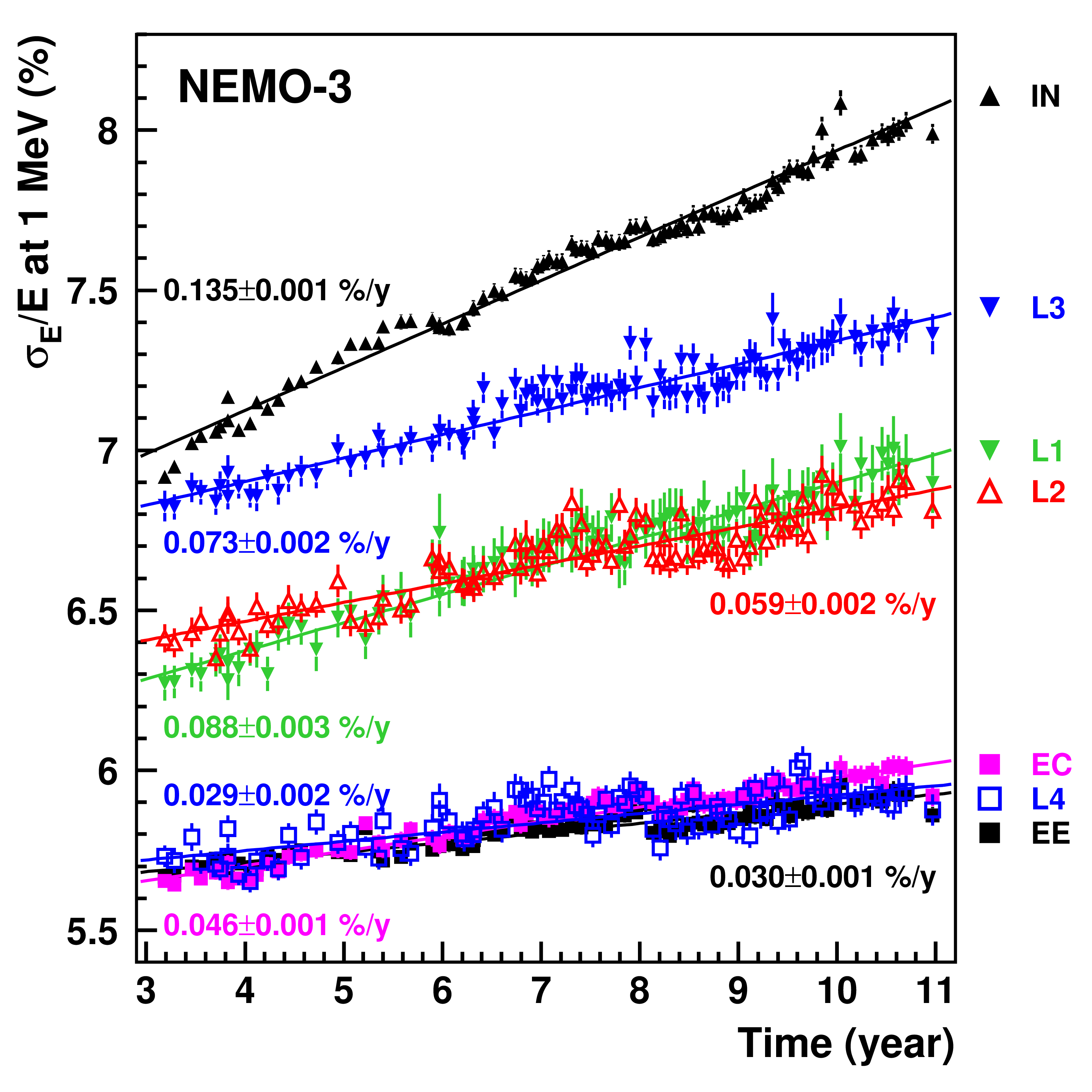}
\caption{Average energy resolution $\sigma_E / E$ measured at $E=1$~MeV  for the different types of scintillator blocks and sizes of PMTs as a function of NEMO-3 running time. Here, IN refers to the calorimeter blocks with 3-inch PMTs located in the central tower (inner wall of the calorimeter), {L1}, {L2} and {L3} refer to 3-inch PMTs located on the upper and lower end caps, {EC} and {EE} refer to 5-inch PMTs located on the external wall, and {L4} to 5-inch PMTs on the upper and lower 
end caps of the calorimeter (see Figure~3 in \protect\cite{nemo3-tdr-2005} for the exact location of the different types of scintillator).}
\label{fig:energy-resol}
\end{center}
\end{figure}

The rare internal conversion electron $K$ line of $^{207}$Bi 
with an energy of $1682$~keV has a small branching ratio of $0.02\%$. 
It is used to determine the systematic uncertainty on the energy scale from
the difference between the reconstructed peak position in data and MC simulation,
which is $<0.2\%$ for $99\%$ of the optical modules. 
It is shown by Monte-Carlo simulations that a $0.2\%$ uncertainty on the energy scale is negligible for the final  $\beta\beta 0 \nu$ analysis.
The remaining optical modules of the calorimeter with incorrect reconstruction of the energy peak are rejected in the analysis. 
A typical energy spectrum measured with a single optical module is shown in Figure~\ref{fig:bi207-1682keV}. 

Figure~\ref{fig:energy-resol} shows the average energy resolution as a function of 
running time for the different types of scintillator blocks and PMTs. The resolution at
 an electron energy of $1$~MeV ranges from $\sigma_E/E = 5.7\%$ to $8.0\%$, depending on the type of block and the data taking period. A deterioration of the energy resolution of $0.03\%$--$0.05\%$ and $0.06\%$--$0.14\%$ per year is observed for the blocks equipped with $5$  and $3$-inch PMTs, respectively. 
This drift might be caused by the residual helium concentration in the air surrounding the detector that leads to after-pulsing of the PMTs. The helium concentration in the central tower of the NEMO-3 detector, where most of the $3$-inch PMTs are located, is higher than in other regions of the detector, which could explain the larger drift in this region. 

The expected full width at half maximum (FWHM) of the spectrum of two electrons energy sum in $0\nu\beta\beta$ decays is $350$~keV. It is a convolution of the energy resolution of the calorimeter and of the non-Gaussian fluctuations in the electron energy loss, which occur mainly in the source foil and to a lesser extent in the tracking detector. In the absence of energy loss fluctuations in the foil, the expected FWHM would be about 250~keV.
%Should we give the contributions here?

\begin{figure}[htbp]  
\begin{center}
\includegraphics[width=0.47\textwidth]{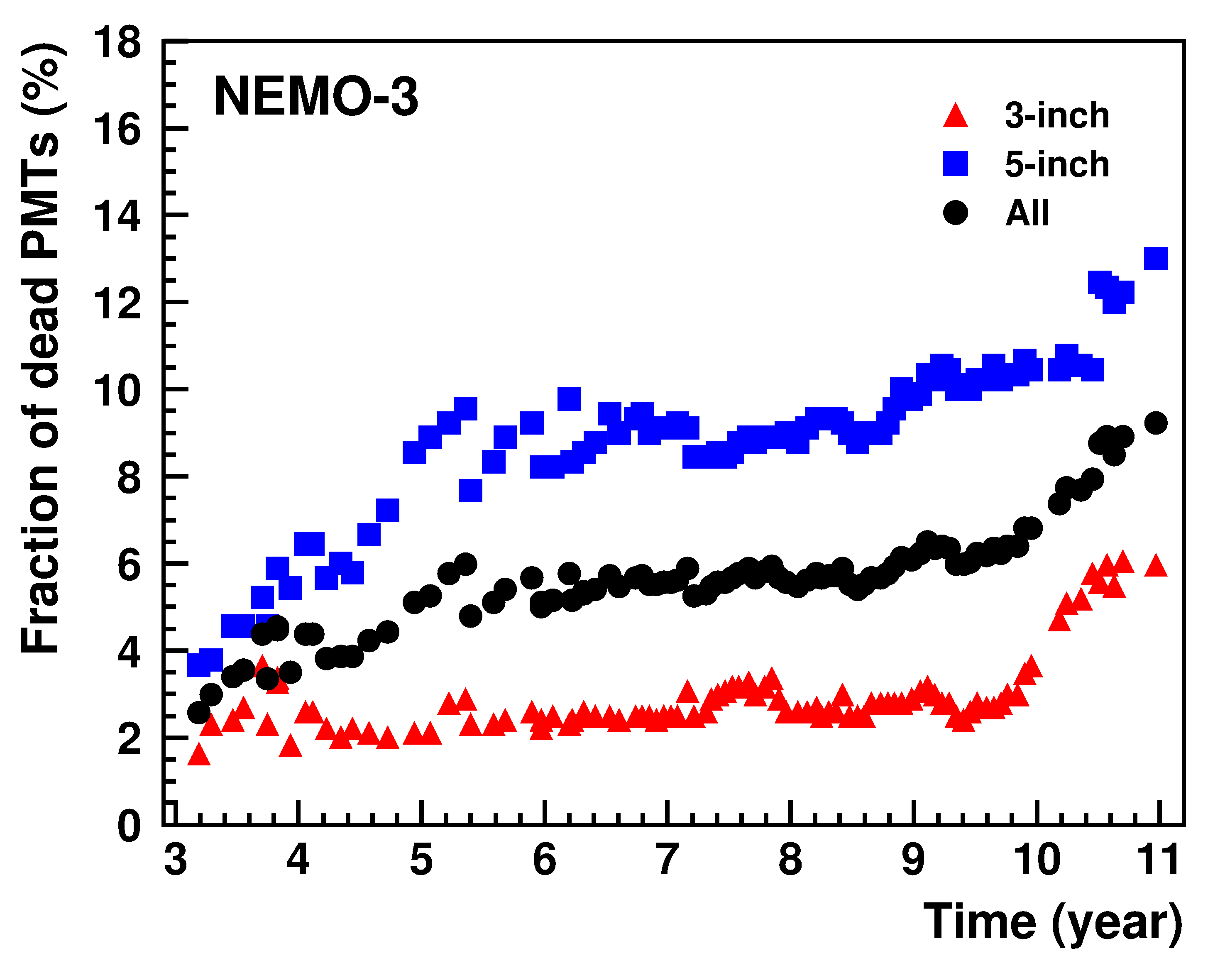}
\caption{Fraction of dead PMTs (in $\%$) as a function of NEMO-3 running time for 3 and 5-inch PMTs, and for all PMTs.}
\label{fig:dead-pm}
\end{center}
\end{figure}

After close to eight years of stable operation of the experiment, fewer than $10\%$ of PMTs had to be turned off because they displayed unstable gain or noisy signals. The fraction of dead PMTs as a function of the NEMO-3 running time is presented in Figure~\ref{fig:dead-pm}. 
The fraction of PMTs with noisy signals in the recorded data is estimated by measuring the random coincidence rate of scintillator hits with a constant timing distribution. The same fraction of PMTs is randomly rejected in the MC simulation, leading to a reduction of the $0\nu\beta\beta$ detection efficiency of $0.9\%$ in Phase~I and $2.4\%$ in Phase~II.

\subsection{Laser survey}

The stability of the PMT gains between two consecutive absolute $^{207}$Bi calibration runs is maintained using dedicated laser runs which were performed twice daily. 
The laser beam is split and transmitted to two different devices to calibrate the $3$ and $5$-inch PMTs separately. 
A description of the laser system is given in~\cite{nemo3-tdr-2005}.
Data taking is divided into successive laser survey periods that
are separated by major incidents such as a general shutdown of the high voltage crates or any other event that could cause a discontinuity in the operating conditions of the PMTs.

The laser survey measures the time dependence of the average response of all PMTs of the same size to monitor the variation of gains. 
The mean energies $\langle e_{3}(t)\rangle$ and $\langle e_{5}(t)\rangle$ of the two different sets of $3$ and $5$-inch PMTs calculated for each laser run at a recording time $t$ are given by
\begin{equation}
\langle e_{3,5}(t) \rangle =\frac{\sum_{k}g^{\rm calib}_k(t) Q_{\rm ADC}(k,t)}{N_{3,5}(t)},
\end{equation}
where the sum extends over the $3$ or $5$-inch PMTs.
Here, $Q_{\rm ADC}(k,t)$ and $g^{\rm calib}_k(t)$ are the recorded charge after pedestal subtraction and the laser calibration constant for the PMT labeled $k$ and for the laser run recorded at time $t$. 
The numbers of $3$ and $5$-inch PMTs recorded during a laser run are $N_3(t)$ and $N_5(t)$, respectively. The parameters $\eta(k,t)$ are calculated for each PMT
\begin{equation}
\eta(k,t)=\frac{g^{\rm calib}_k(t)Q_{\rm ADC}(k,t)}{\langle e_{3,5}(t) \rangle}
\end{equation}
depending on its type. The parameters
 $\eta(k,t)$ are divided by $\langle \eta_{0}(k)\rangle$, which is the mean value of $\eta(k,t)$ during the associated absolute energy $^{207}$Bi calibration run, to calculate the final laser correction factor of
\begin{equation}
C(k,t)=\eta(k,t)/\langle \eta_{0}(k)\rangle.
\end{equation}

\begin{figure}[htbp]  
\begin{center}
\includegraphics[width=0.4\textwidth]{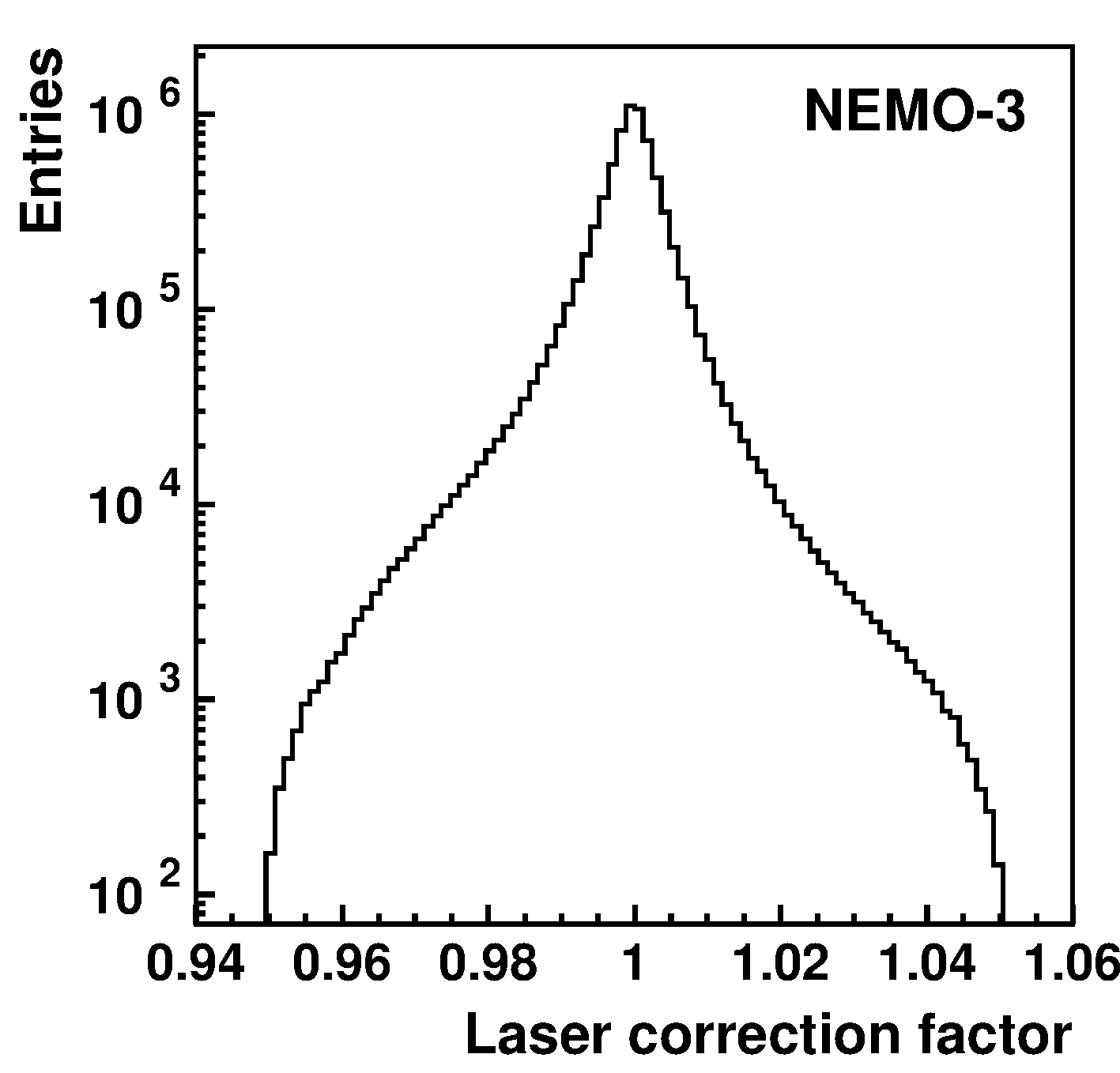}
\caption{Distribution of the laser correction factors calculated for all laser runs and for all  stable PMTs used in the analysis.}
\label{fig:laser-cor}
\end{center}
\end{figure}

The time dependence of the correction factors is analysed to characterize the level of stability of the PMT gains during data taking. 
A large change of the correction factor or discontinuities during a period between two absolute energy calibrations are interpreted as an instability, and the corresponding PMT and associated events are rejected for that period. 
For each PMT, we estimate its stability during that period, by determining the number of laser runs for which the correction factor deviates $5\%$.
During the entire data taking period, $82\%$ of PMTs are considered to be stable.
Taking into account that more than $90\%$ of data are recorded with a reliable laser survey, the efficiency to select a double $\beta$ event is reduced by $25.2\%$ in Phase~II when the laser survey is applied, as reported in Table~\ref{table:effbb0nu}. The efficiency reduction is $38.6\%$ in Phase~I because of a less stable laser during this first phase of data. 
The distribution of the laser correction factors for all laser runs is shown in Figure~\ref{fig:laser-cor} for the stable PMTs.

%\begin{figure}[htbp]  
%\begin{center}
% \includegraphics[width=0.45\textwidth]{desin_bi214.png}  \\
%(a)\\
% \includegraphics[width=0.45\textwidth]{desin_tl208.png} \\
% (b)
%\end{center}
%\caption{Simplified decay diagrams for (a) $^{214}$Bi and (b) $^{208}$Tl ~\cite{Firestone}. In the {\sc \mbox{decay0}} event generator used in the simulations, the full schemes of the decays are used.}
%\label{fig:decay-tl-bi}
%\end{figure}

The reliability of the laser survey procedure is validated by analysing a pure sample
of electrons
with an energy close to the end point
of $Q_{\beta} = 3.27$~MeV in the $\beta$ energy spectrum of $^{214}$Bi decays 
occurring in the tracking chamber.
Any excess in data over the MC expectation around $Q_\beta$ would be a sign of 
unstable PMT gains. The events are selected by requiring electrons in coincidence with
a delayed $\alpha$ track from the $^{214}$Bi-$^{214}$Po cascade (``BiPo events").

The entire data set is used in this analysis. The selection of BiPo events is similar to the one used for the radon background measurement, described in Section~\ref{sec:radon}. Here, only BiPo events with a vertex inside the tracking chamber are selected, and
the electron track length is restricted to $>45$~cm.
Electrons crossing the source foils are rejected, since they could have lost energy in the foils.

\begin{figure*}
\begin{center}
\includegraphics[width=1.01\textwidth]{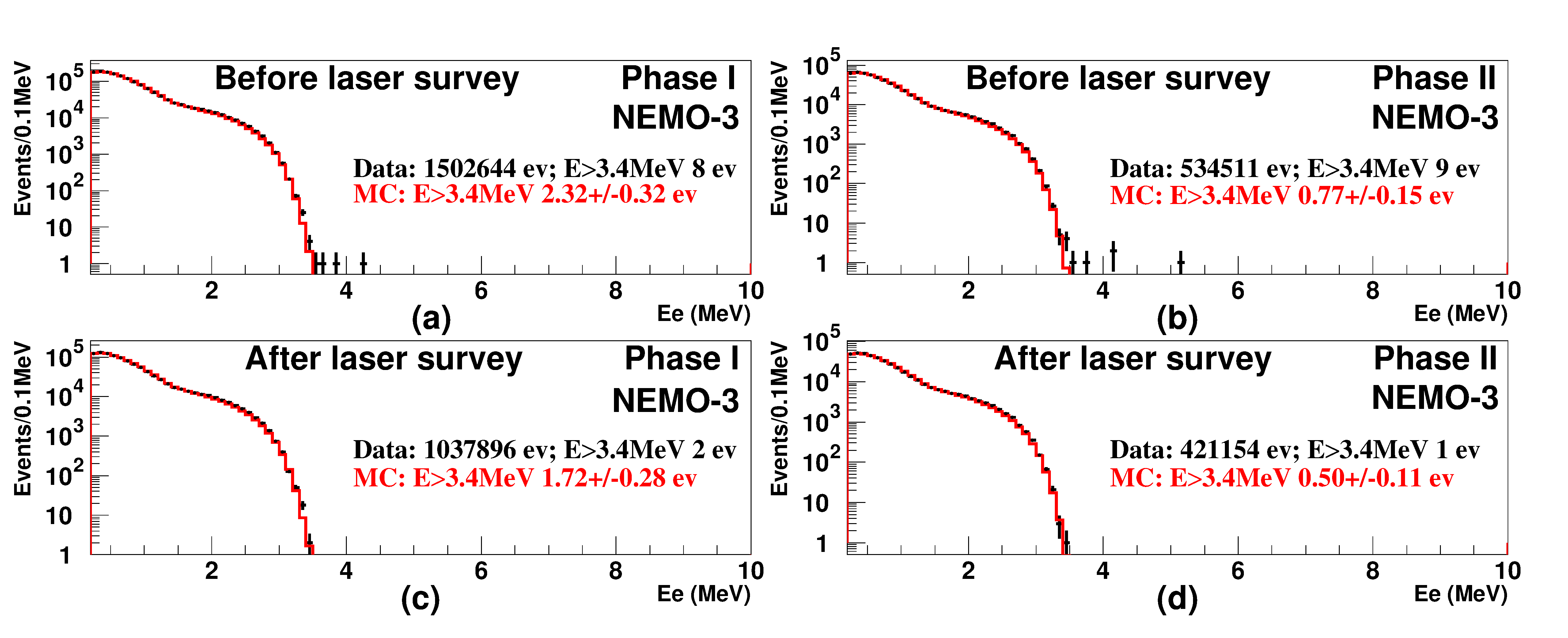}
\caption{Energy spectrum of electrons from $\beta$ decay  of $^{214}$Bi
measured using BiPo events inside the tracking detector, without laser survey (a,b) and after laser survey (c,d) for Phase~I (a,c) and Phase~II (b,d). The data are compared to a MC simulation. The excess of
electrons observed at $E_e>3.4$~MeV in data are caused by PMTs with unstable gains. They are rejected by the laser correction
 (see Table~\protect\ref{table:bipoevents}).}
\label{fig:ealpha-laser}
\end{center}
\end{figure*}

To minimize the proportion of re-firing Geiger cells, we require the $\alpha$ delay time to be
$>140$~$\mu$s for events with only one delayed hit, and  $>70$~$\mu$s for 
 $>1$ hits. The delay time distributions are analysed separately for each Geiger plane, on each side of the source foils, and for 1, 2, 3, and $>3$ delayed hits. 
Regions within the tracking chamber with a significantly increased fraction of random coincidences or re-firing cells are excluded. These effects are therefore negligible in
the selected data set.

\begin{table}[hbt]
\begin{center}
\begin{tabular}{l|cc|cc}
\hline\hline
  &   \multicolumn{2}{c|}{Phase 1}  & \multicolumn{2}{c}{Phase 2} \\
 & Data & MC & Data & MC\\
 \cline{2-5} 
No laser correction &  8 & $2.32 \pm 0.32$ & 9 & $0.77 \pm 0.15$ \\
With laser correction &  2 &  $1.72 \pm 0.28$ & 1 & $0.50 \pm 0.11$ \\
\hline\hline
\end{tabular}
\caption{Numbers of BiPo events with $E_e>3.4$~MeV}
\label{table:bipoevents}
\end{center}
\end{table} 

The electron energy spectra obtained from this analysis are shown in Figure~\ref{fig:ealpha-laser}, for Phases~I and~II separately, before and after applying the laser corrections. They are compared to the expected background from the MC simulation, assuming the $^{214}$Bi activity on the surfaces of wires and foils, and the $^{214}$Bi activity inside the foils described in Section~\ref{sec:bkg}. 
The number of data and MC events  with $E_e>3.4$~MeV, without laser correction and after applying the laser correction is given in Table~\ref{table:bipoevents}. 
%-------------------------
% Referee
%{\color{red} 
Spurious events observed beyond the end-point are well rejected after applying the laser survey. It demonstrates the reliability of the laser survey to reject false high energy events with a wrong recorded energy.
%}
%-------------------------

\subsection{Timing calibration and time-of-flight}

Time-of-flight measurements are used to discriminate between two-electron events 
from double $\beta$ decays emitted from the source foil and events where
an external electron crosses the detector and foil. The crossing electron in these events could be reconstructed as two separate
tracks with a common vertex. 

%------------------------------------------------------------------------------------------
% Referee
%{\color{red} 
The time calibration of the optical modules takes into account both the individual absolute time shift of each optical module  and a time-vs-charge dependence induced by the effect of leading edge discriminators. 
The calibrated  time, $t(i)$, used for a time-of-flight calculation for counter number $i$ is:
\begin{equation}
t(i)= tdc(i) - t_s(i) - f(Q(i))
\end{equation}
where $tdc(i)$ is the TDC measurement, $t_s(i,t)$ is the  time shift, and $f(Q(i))$ is the time-charge correction function, which correct the measurement of the TDC as a function of the charge $Q$ (with formula given in [4]).

The absolute time shifts are measured individually for each optical module, using ($e^-,\gamma$) events selected from the absolute energy calibration runs carried out with $^{207}$Bi sources using the relation
%The 130.5~ps half-life of the metastable state $^{207}$Pb in the $^{207}$Bi decay is taken into account.
\begin{equation}
\label{eq:time-shift}
t_s(i) = \frac{\sum_{N_i}(\Delta t_{tof}^{j,i} + \Delta tdc^{j,i} + f(Q(j)) - f(Q(i)) + t_s(j))}{N_i}
\end{equation}
where $\Delta t_{tof}^{j,i} = t_{tof}(j)-t_{tof}(i)$ is the difference between the calculated time-of-flights $t_{tof}$ of the electron and the $\gamma$, $\Delta tdc^{j,i} = tdc(j)-tdc(i)$, and $N_i$ is the number of selected ($e^-,\gamma$) events with the optical module $i$ hitted by the electron.

The time-vs-charge correction functions $f(Q)$ are measured for each of the seven scintillator block types by using crossing-electron events from a dedicated  run with an external Am-Be fast neutron source. 
Fast neutrons are thermalized mostly in the scintillators. Then $\gamma$ are created by the capture of the thermalized neutrons in the
copper walls. If a $\gamma$ produce an electron by Compton effect in the scintillator, this electron can escape and crosses the tracking chamber, producing a crossing-electron event. The correction functions are measured using the relation
\begin{equation}
\label{eq:time-charge}
f(Q(i)) =  \Delta tdc^{i,j}  - (t_s(i) - t_s(j)) + f(Q(j)) 
\end{equation}
where $\Delta tdc^{i,j} = t_{tof}(i)+t_{tof}(j)$ is the calculated time-of-flight for the electron to cross the tracking detector from the optical module $j$ to the optical module $i$. 
The values of $f(Q(i))$ are groupped according to the seven scintillator block types, and then used to produce the time-vs-charge $f(Q)$ distribution that is then fitted with a formula using four parameters $p_k$.
\begin{equation}
f(Q) =  p_1 - \frac{p_2}{p_3 \times \sqrt{Q} + p_4}
\end{equation}

Since the absolute time shifts and the time-vs-charge correction functions are both used in the two calibration relations~\ref{eq:time-shift} and \ref{eq:time-charge}, an iterative procedure is required to determine them.
First the absolute time shifts are calculated according to Equation~\ref{eq:time-shift} with initial values of the time-vs-charge correction functions obtained with laser runs and initial values of the time shifts set to zero. Then the time-vs-charge correction functions are calculated using Equation~\ref{eq:time-charge}. These new corrections functions are then used to calculate the absolute time shifts, and so one. Successive iterations are performed until a convergence is obtained. 

The daily laser surveys are used to identify and correct any variation of the TDC response. 
This laser timing correction is calculated separately for each optical module and laser survey run.

%}
%------------------------------------------------------------------------------------------

The average timing resolution of a scintillator hit is about 250~ps for a 1~MeV electron.

\begin{figure}[htbp]  
\begin{center}
\includegraphics[width=0.48\textwidth]{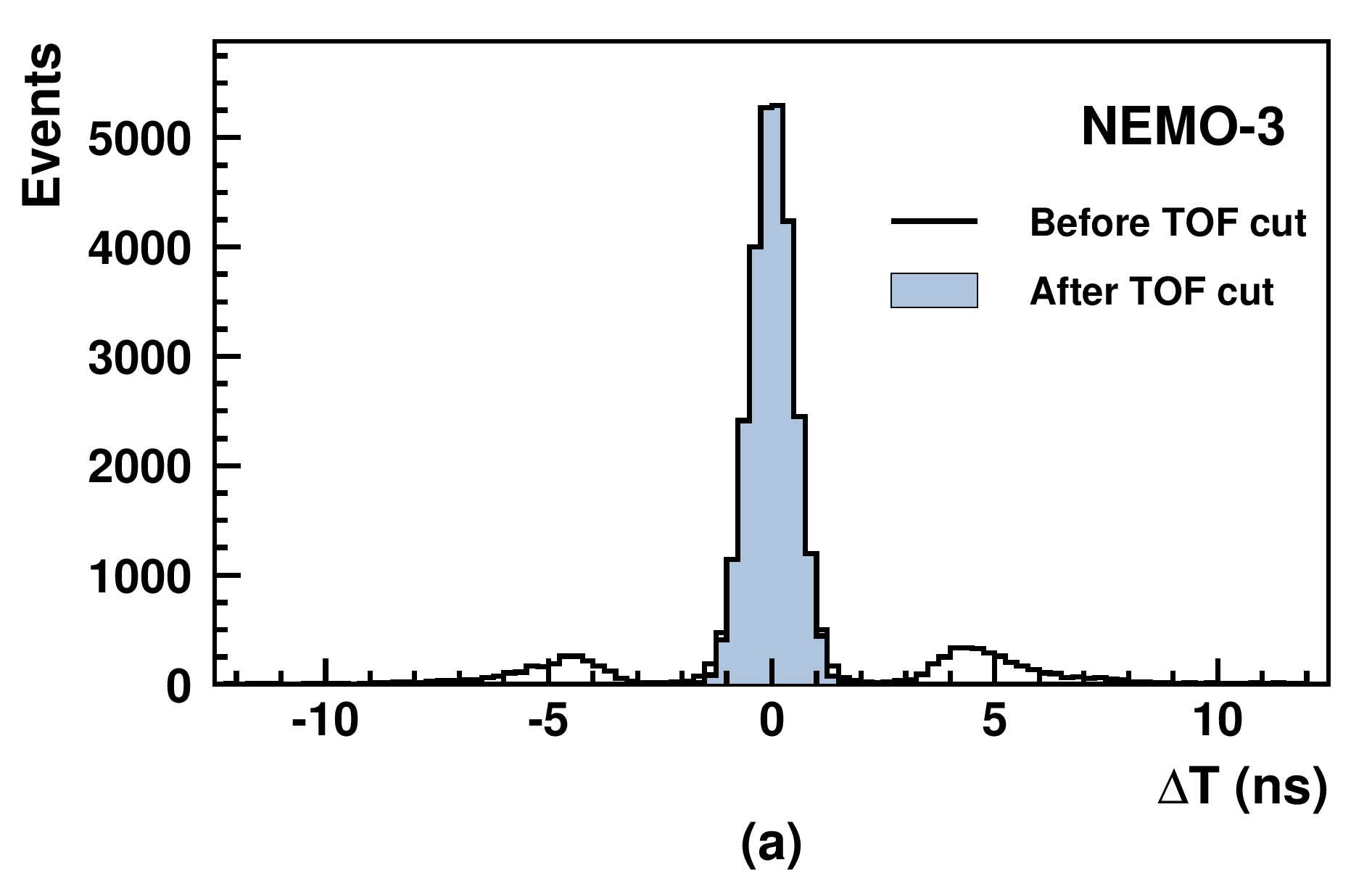}  \\ 
\includegraphics[width=0.48\textwidth]{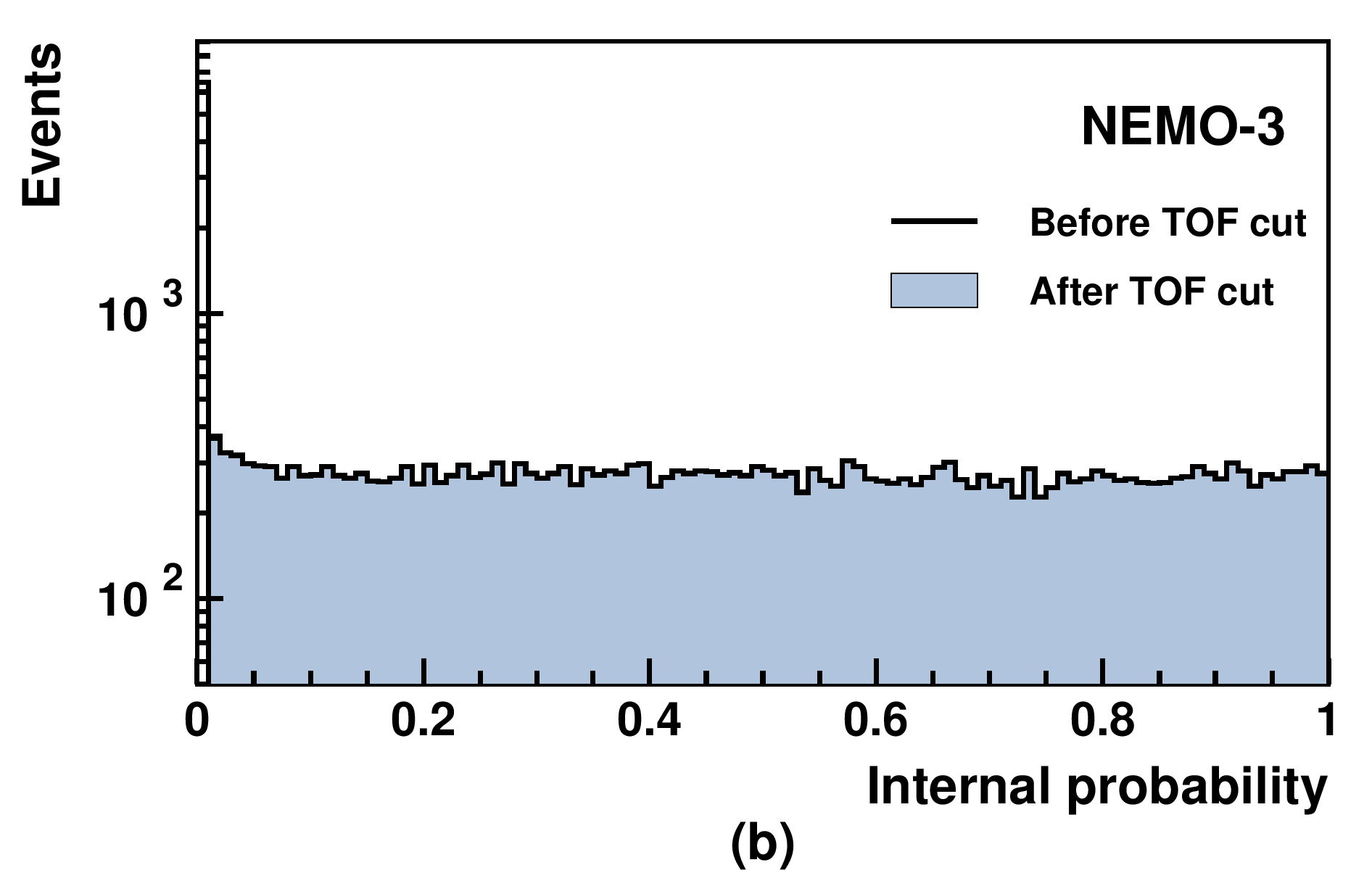} 
\caption{Distributions of the difference $\Delta T$ between the measured and expected time differences of scintillator hits for the internal hypothesis (a) and the internal probability $P_{\rm int}$ (b) for two-electron events.
The superimposed shaded histogram shows events with $P_{\rm int}>1\%$.} 
\label{fig:tof-betabeta}
\end{center}
\end{figure}

The time-of-flight analysis is based on a comparison between the measured and 
expected time differences of the two scintillator hits. The expected time-of-flight is calculated assuming two different hypotheses: the external hypothesis corresponding to a crossing electron and the internal hypothesis corresponding to two electrons being emitted simultaneously from the same vertex on the foil in a double $\beta$ decay. The time-of-flight calculation also accounts for the length of the tracks and the energy loss
in the tracking detector. 
To correctly take into account uncertainties on the timing measurement, we calculate separate probabilities for internal two-electron events ($P_{\rm int}$) and external crossing-electron events ($P_{\rm ext}$). 
%using the timing resolution $\sigma_i(t)$ of calorimeter block $i$ given by
%\begin{equation}
%\sigma_i^2(t) = \sigma_0^2 + \tau_{sc}^2 \times \frac{\sigma_i(E)^2}{E_i^2}, 
%\end{equation}
%where $E$ is the energy of the electron (in MeV) measured in calorimeter block $i$, $\sigma_i(E)$ the energy resolution of the block at an energy of $1$~MeV, and $\tau_{sc}=4.326$~ns and $\sigma_0=0.04$~ns.
The distributions of the difference $\Delta T$ between
the measured and theoretical time differences of the two scintillator hits, calculated assuming the internal 
hypothesis, is shown in Figure~\ref{fig:tof-betabeta}a for the full sample of two-electrons events selected using all criteria described in Section~\ref{sec:bb-selection}, except the requirement on the time-of-flight. 
The $P_{\rm int}$ distribution shown in Figure~\ref{fig:tof-betabeta}b is constant above $P_{\rm int}=1\%$, as expected for double $\beta$ decays, 
while the peak at $P_{\rm int}<1\%$ corresponds to crossing-electron events.
Internal double $\beta$ events emitted from the source are centred around $\Delta T=0$~ns, while crossing-electron events from external background sources have $|\Delta T|>3$~ns. The r.m.s. of the $\Delta T$ 
distribution for $P_{\rm int}>1\%$ is $490$~ps.

\section{Selection of double $\beta$ decay events and efficiency}
\label{sec:bb-selection}

\begin{figure*}  
\begin{center}
\includegraphics[width=0.8\textwidth]{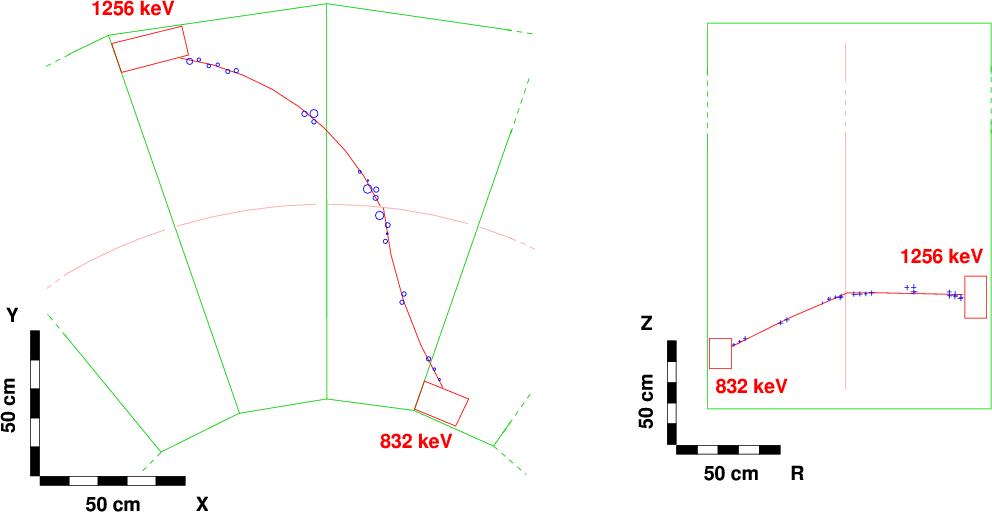}
\caption{Transverse and longitudinal view of a
 reconstructed double $\beta$ data event. Tracks are reconstructed from a single vertex in the source foil, with an electron-like curvature in the magnetic field, and are each associated to an energy deposit in a calorimeter block.}
\label{fig:nemo3-event}
\end{center}
\end{figure*}

Candidate double $\beta$ decay events are selected by requiring exactly two electron tracks. Events with more than two tracks are rejected.
\begin{itemize}
\item Each track must be associated with a scintillator hit, and the extrapolated track must hit the front face of the scintillator block and not the lateral side of petal blocks.
The associated scintillator hits must be isolated, i.e.,~no hits are found in neighboring scintillator blocks, and only a single track can be associated with the scintillator block.
Events with a $\gamma$ candidate, defined by a scintillator hit that is not associated to a track, are
rejected.
\item
The two electron tracks must originate from a common vertex in the $^{100}$Mo source foil. 
We therefore require that the transverse and longitudinal components of the distance between their intersection points with the foil are
less than $4$~cm and $8$~cm, respectively.  
\item To reject background from $^{214}$Bi decays near the foil, the number of unassociated 
hits in the tracking detector close to the vertex should not exceed one.
When the two tracks are on the same side of the foil, there must be no unassociated hit on the opposite side of the foil close to the vertex.
\item
The energy of each electron as measured in the calorimeter must be $>200$~keV.
\item 
The curvature of both tracks must be negative to reject positrons.
\item The time-of-flight must correspond to the two electrons being emitted from the same vertex
in the source foil, requiring $P_{\rm int}>1\%$ and $P_{\rm ext}<1\%$.
To ensure a reliable time-of-flight measurement, the track length of each track must exceed $50$~cm. 
Events with hits in scintillator blocks from the innermost circle of petals are rejected. 
\item Events with delayed tracker hits close to the electron tracks are rejected to reduce $^{214}$Bi and radon background (see Section~\ref{sec:radon}). 
The delay time of these hits is required to be greater than $100$~$\mu$s for events with only one delayed hit, and 40~$\mu$s, 20~$\mu$s, and 4~$\mu$s for events with 2, 3, or $>3$ delayed hits, respectively. These criteria reduce the sensitivity to spurious hits in cells close to the electron track.  
\item Events are rejected if a scintillator hit is linked to a PMT that has been flagged by the laser survey
as having unstable gain. 
\end{itemize}

A typical double $\beta$ event is shown in Figure~\ref{fig:nemo3-event}.
Only events with an energy sum $E_{\rm tot}>2$~MeV for the two electrons are considered in the $0\nu\beta\beta$ search. 
The efficiencies to select $0\nu\beta\beta$ events are calculated using the MC simulation, and are given in Table~\ref{table:effbb0nu} after each successive application of the selection criteria.
The $0\nu\beta\beta$ signal selection efficiency is $11.3\%$ for Phase~I and II combined
and $E_{\rm tot}>2$~MeV. It reduces to $4.7\%$ in the energy window $E_{\rm tot}=[2.8-3.2]$~MeV. 
%-------------------------------------
% Added for Referee
%{\color{red} 
This reduction is due to the fact that the $E_{\rm tot}$ energy spectrum of the $\beta\beta0\nu$ signal peaks around 2.8~MeV, i.e. 200~keV below the theoretical $Q_{\beta\beta}$ value, because of the energy losses of the electrons in the foil and in the wire chamber.
%}
%-------------------------------------
If the inefficiency due to noisy Geiger cells and unstable or dead PMTs is removed,  
these efficiencies increase to $20.3\%$ for $E_{\rm tot}>2$~MeV, and $8.5\%$ in the energy window $E_{\rm tot}=[2.8-3.2]$~MeV. 

The uncertainty on the signal efficiency is determined using dedicated runs with two calibrated $^{207}$Bi sources, with a low activity of around 180~Bq,  at four different locations inside the detector. The runs were taken in March 2004, June 2004 and April 2006. The two conversion electrons emitted simultaneously by the $^{207}$Bi sources are selected. 
%-------------------------------------
% Added for Referee
%{\color{red} 
The criteria to select the two electrons events are the same as the ones used to select the double $\beta$ events, except that the energy of the electrons must correspond to the expected energy of the conversion electrons and the common vertex of the two electron tracks must originate from the calibration sources. 
%}
%-------------------------------------
The reconstructed $^{207}$Bi activities are in agreement with the nominal values within $5\%$, which is consistent
within the expected systematic uncertainty.

\begin{table}[hbt]
\begin{center}
\begin{tabular}{l|c|c|c}
\hline \hline
 Selection Criteria & Ideal & Phase~I & Phase~II\\
\hline
 Trigger                               & 0.973 & 0.973 & 0.973 \\
 Two tracks reconstructed                      & 0.480 & 0.478 & 0.462 \\
 Track-scintillator association                & 0.352 & 0.348 & 0.331 \\
 Associated PMTs not dead                      & 0.352 & 0.321 & 0.288 \\
 No extra scintillator hit                               & 0.313 & 0.287 & 0.258 \\
 Scintillator correctly calibrated                 & 0.313 & 0.281 & 0.245 \\
 Common track vertex in foil         & 0.280 & 0.251 & 0.218 \\
 Tracks have hits near foil                 & 0.273 & 0.244 & 0.211 \\
 No extra prompt hits near vertex       & 0.271 & 0.242 & 0.209 \\
 Track length $> 50$ cm                        & 0.252 & 0.225 & 0.194 \\
 Scintillator energy $> 200$ keV                    & 0.245 & 0.219 & 0.189 \\
 Negative track curvature                      & 0.223 & 0.199 & 0.172 \\
 Isolated scintillator blocks                  & 0.219 & 0.195 & 0.169 \\
 No scintillator at petals near foil                & 0.209 & 0.186 & 0.161 \\
 Timing requirement                                    & 0.206 & 0.184 & 0.159 \\
 Reject $\alpha$ particles                                 & 0.206 & 0.184 & 0.159 \\
 Energy laser survey                           & 0.206 & 0.113 & 0.119 \\
\hline
 $E_{\rm tot}>2$~MeV                                & 0.204 & 0.111 & 0.117 \\
 $E_{\rm tot}>2.8$~MeV                              & 0.085 & 0.047 & 0.049 \\
\hline\hline
\end{tabular}
\caption{Evolution of the $0\nu\beta\beta$ efficiency as a function of the successive criteria of selection for Phase~I and II. ``Ideal" refers to the detector without any noisy Geiger cell neither unstable or noisy PMTs.}
\label{table:effbb0nu}
\end{center}
\end{table}

%----------------------------------------------------------------------

\section{Background measurements}
\label{sec:bkg}

\begin{figure*}  
\centering
\includegraphics[width=0.9\textwidth]{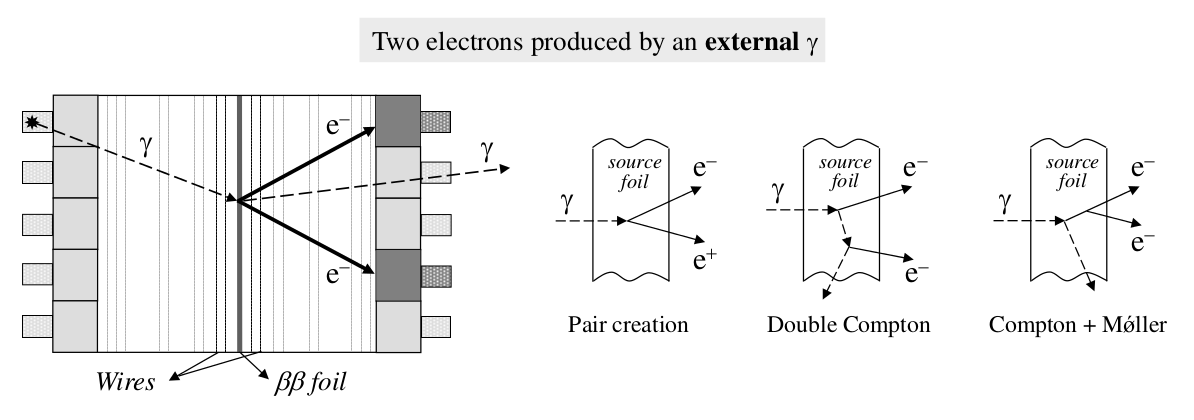}
\includegraphics[width=0.9\textwidth]{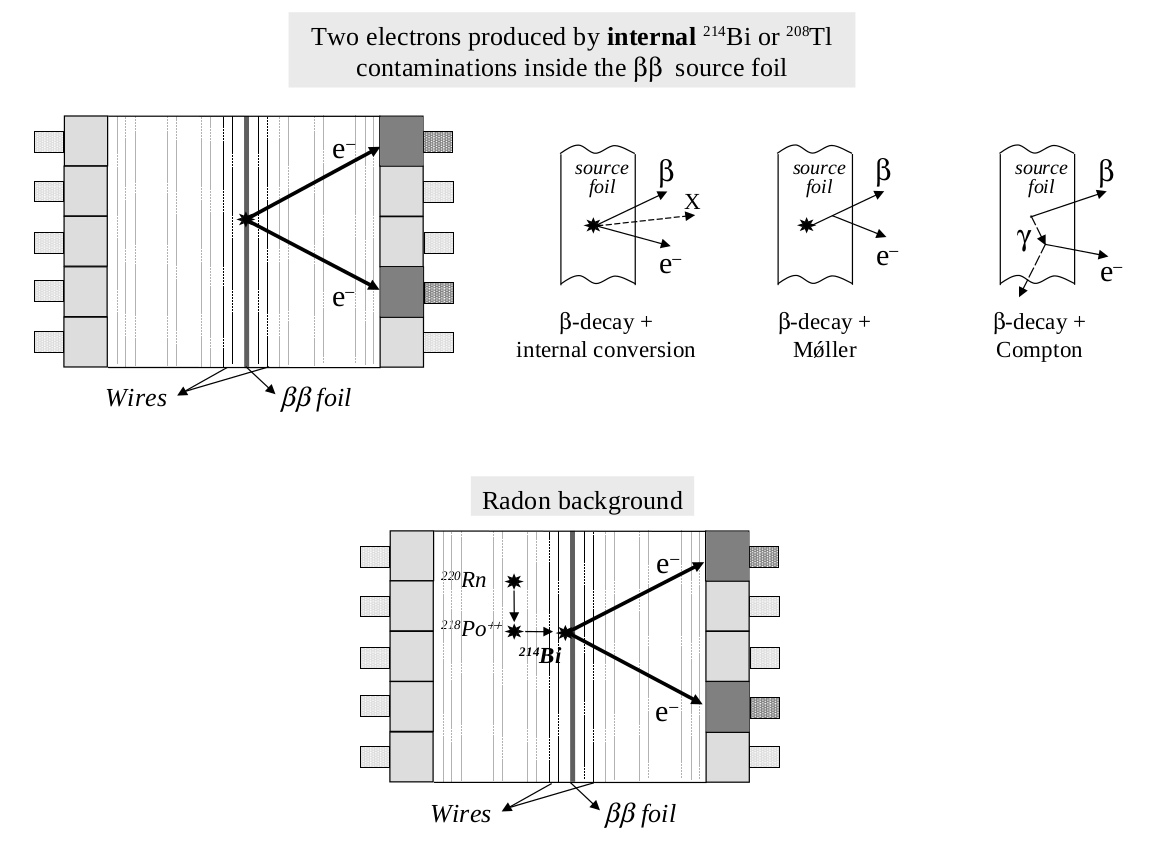}
\caption{Schematic view of the different components of the two-electron background: the external background produced by an external 
$\gamma$ ray, the internal background produced by internal $^{214}$Bi and $^{208}$Tl contaminations in the $^{100}$Mo source foil, and the radon contamination inside the tracking detector.}
\label{fig:bkg-schema-1}
\end{figure*}

The NEMO-3 detector is unique in its ability to identify electrons, positrons, $\gamma$ rays and delayed  $\alpha$ particles by combining information from the tracking detector, calorimeter, and the track curvature in the magnetic field. 
This allows the separation of different non double $\beta$ background processes by exploiting differences in their event topologies and final states. 
We distinguish three background components, as illustrated in Figure~\ref{fig:bkg-schema-1}, namely the external background, the internal background, and the background from radon.
We first measure the external background. Then, the radon and thoron backgrounds 
inside the tracking detector are measured, setting the external backgrounds to their measured values. Finally, the internal $^{208}$Tl and $^{214}$Bi contaminations inside the $\beta \beta$ source foils are determined, with all other backgrounds fixed.
A full description of the background analysis and preliminary background measurements with part of the NEMO-3 data set are given in Ref.~\cite{nemo3-bkg-2009}. Here, we 
report the results of the background measurements using the full data set.

\subsection{External background}

External background is produced by the interaction of external $\gamma$ rays originating from the natural radioactivity of the detector outside of the source, by external neutrons 
undergoing neutron capture that results in emission of $\gamma$ rays, or by cosmic rays. 
If an external $\gamma$ ray is not detected by a scintillator, it can reach the source foil without being tagged. It can then mimic a $\beta \beta$ event by creating an $e^+e^-$ pair,  
if the two photons from a subsequent positron annihilation remain undetected or the sign of the positron track curvature is incorrectly measured. Double or single Compton scattering followed by M\o ller scattering can also lead to a double $\beta$-like topology. The different mechanisms are illustrated in Figure~\ref{fig:bkg-schema-1}.

\begin{figure*}[htbp]  
\centering
\includegraphics[width=0.8\textwidth]{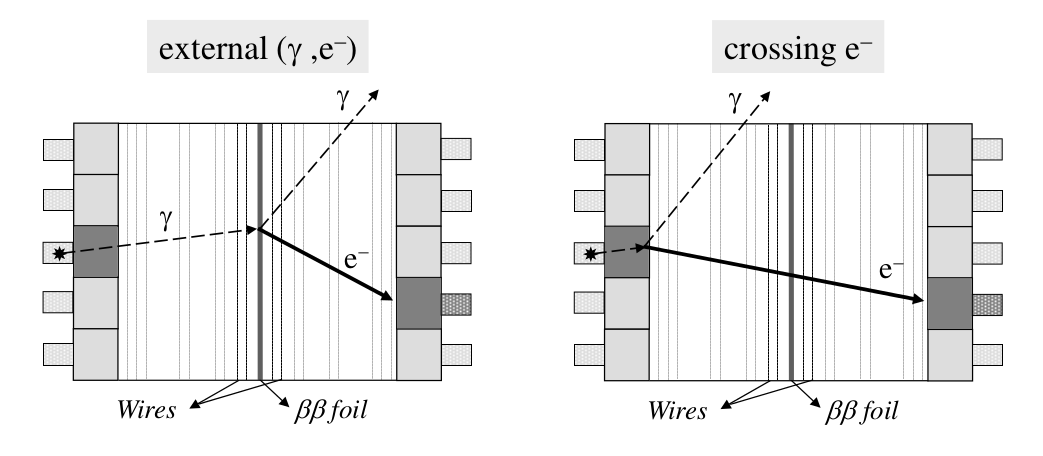}
\caption{The two events topologies used to measure the external background: external ($\gamma,e^-$) events and 
crossing-electrons events.}
\label{fig:bkg-schema-2}
\end{figure*}

We measure the external background using both
external ($\gamma,e^-$) and  crossing-electron events, as illustrated in Figure~\ref{fig:bkg-schema-2}. 
External ($\gamma,e^-$) events are selected requiring one isolated scintillator hit,
assumed to be from the $\gamma$ ray, and one electron track coming from the source foil and associated with a different scintillator. The time difference between the scintillator hits
must agree with the hypothesis that an external $\gamma$ ray has hit the first scintillator block before producing a Compton electron in the foil.

Crossing electrons leave a track that traverses the detector and is associated with a scintillator hit on either side with a time-of-flight and a curvature consistent with a crossing electron. In this topology, an external $\gamma$ hits the first scintillator block from outside
 and then creates an electron by Compton scattering in the last few millimeters of the scintillator closest to the tracking detector. This Compton electron crosses the detector including the foil before hitting the second scintillator, depositing its entire energy.

The external background is modelled by fitting the data in both channels
assuming contaminations of $^{214}$Bi from $^{238}$U and $^{208}$Tl from
$^{232}$Th decays, $^{40}$K inside the PMTs, scintillators, iron shield and iron structure,
cosmogenic $^{60}$Co inside the mechanical structure, and external $\gamma$ rays from the laboratory environment.

The $^{208}$Tl and $^{214}$Bi contaminations inside the PMTs are the dominant components of the external background in the range $E_{\rm tot}>2$~MeV. Their activities have been set to the values quoted in our previous background measurement with part of the NEMO-3 data set~\cite{nemo3-bkg-2009}. 
Activities of other components in the MC simulation are fitted to the data using a 
combined fit to the distributions of the electron energy $E_{e^-}$, the $\gamma$ energy $E_\gamma$, the sum of the energy $E_{e^-}+E_\gamma$, and the angle between the reconstructed $\gamma$ direction and electron track. 

\begin{figure*}[htbp] 
\begin{center}
\includegraphics[width=0.42\textwidth]{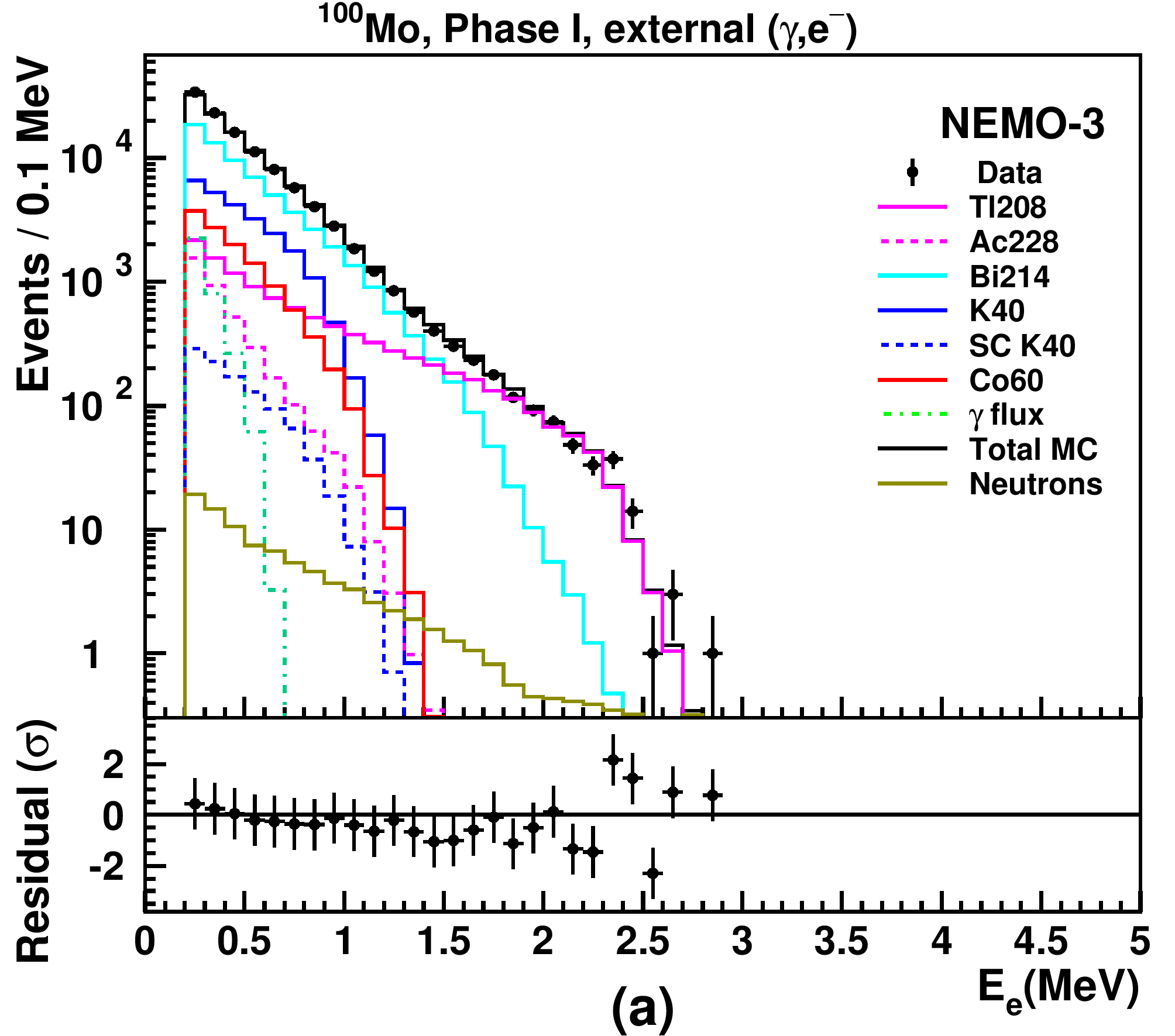}
\includegraphics[width=0.42\textwidth]{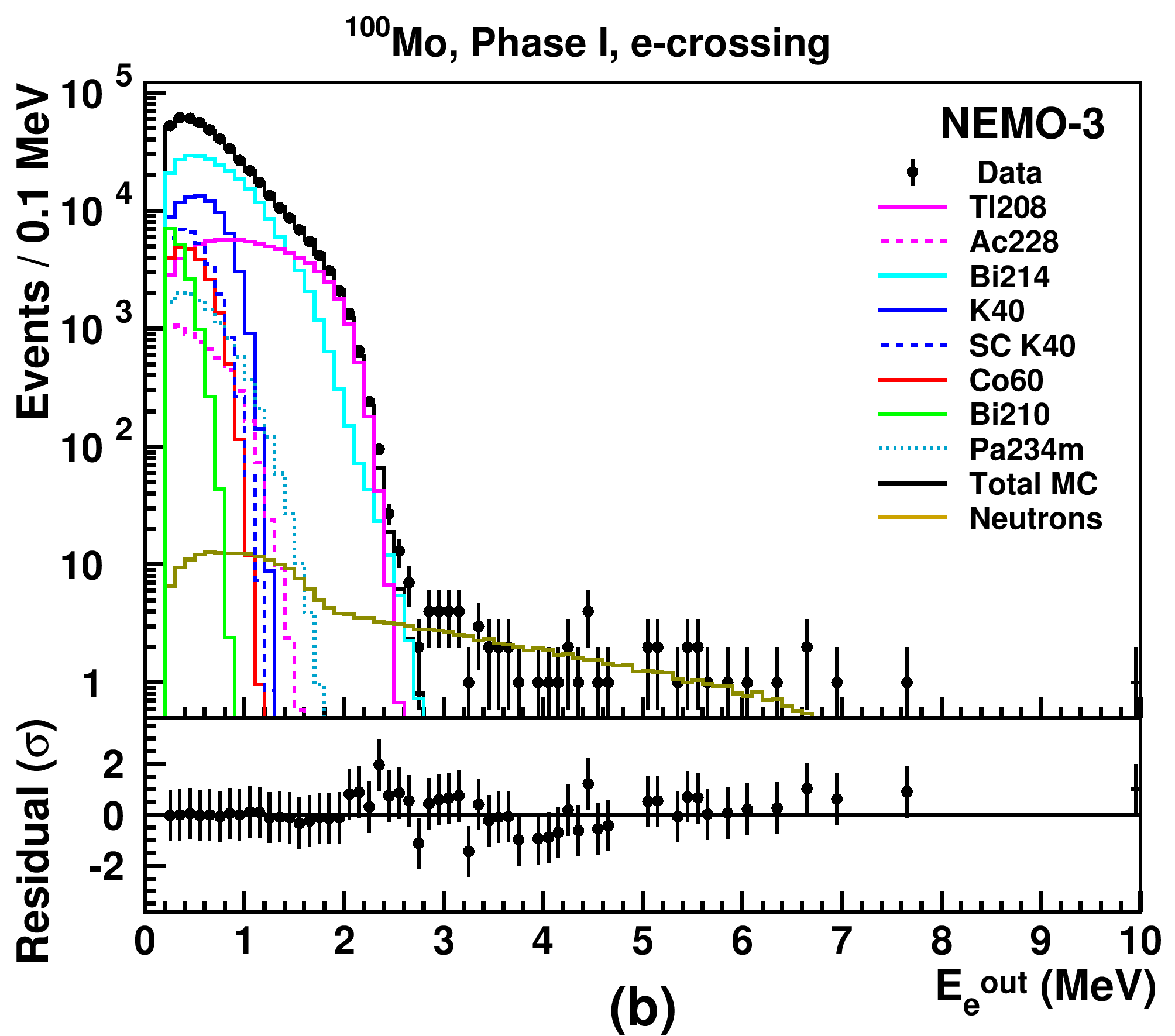}\\
\includegraphics[width=0.42\textwidth]{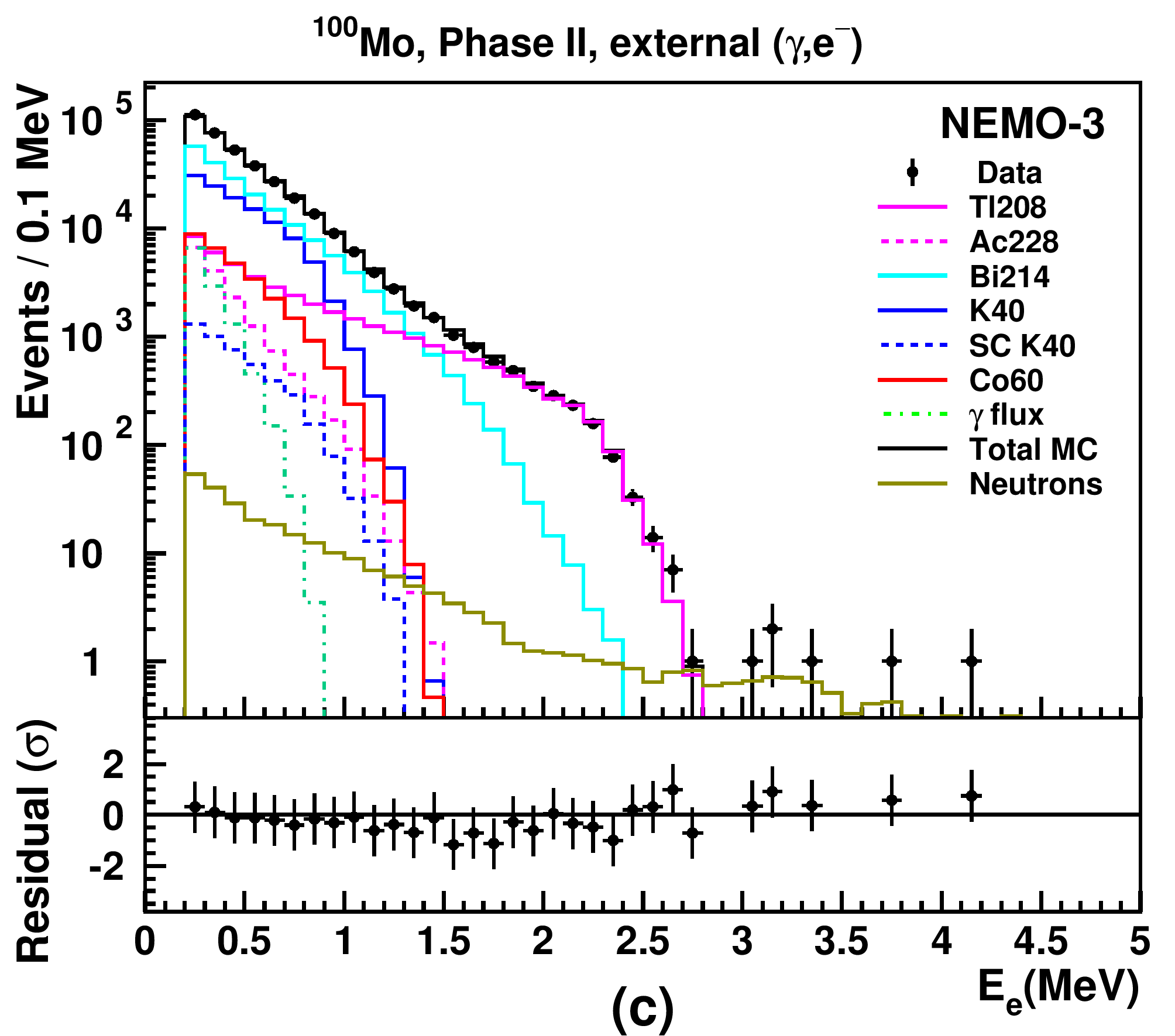}
\includegraphics[width=0.42\textwidth]{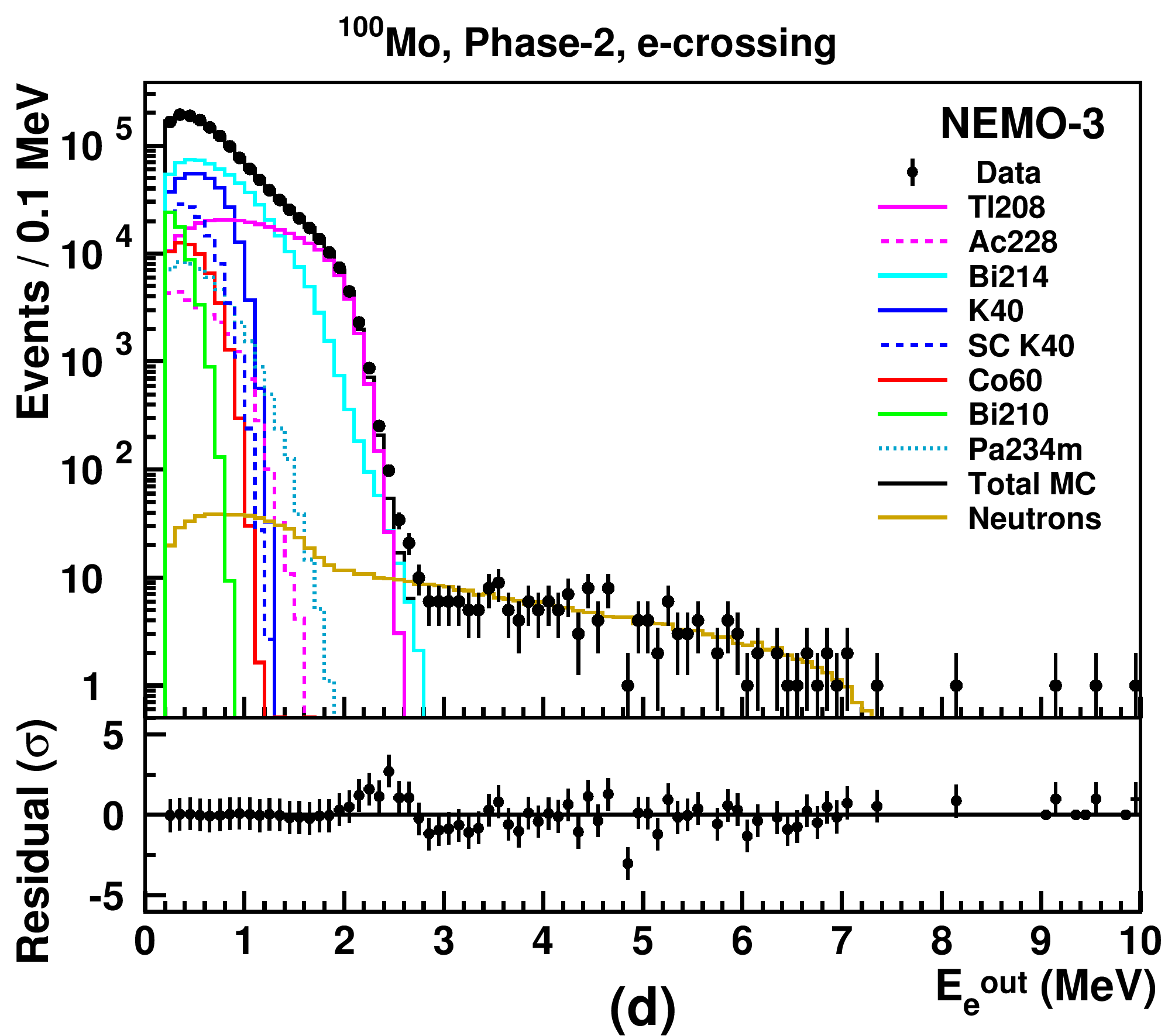}
\end{center}
\caption{Result of the fit of the external background
to data for the total $^{100}$Mo exposure of 34.3~kg$\cdot$yr, 
for the electron energy $E_e$ of
external ($\gamma,e^-$) events (a,c) and the
energy $E_{e}^{\rm out}$ measured in the last scintillator block hit 
in crossing-electron events (b,d). 
The distributions are shown separately for Phases~I (a,b) and II (c,d). SC K40 corresponds to $^{40}$K impurities inside the scintillators.
Lower panels show residuals between data and expected background, normalized to the Poisson error, ignoring bins with 0 events.}
\label{fig:nemo3-bkg-ext-result}
\end{figure*}

Figure~\ref{fig:nemo3-bkg-ext-result} shows the energy spectra of the Compton electrons for external ($\gamma,e^-$) events and the energy measured in the last scintillator block hit ($E_{e}^{\rm out}$) for crossing electrons. The fitted MC background model agrees with the data and lies within the $10\%$ systematic uncertainty of the previous results obtained with a smaller data set~\cite{nemo3-bkg-2009}. It is also consistent with the radioactivity measurements of the detector materials using high-purity germanium (HPGe) detectors before
installation~\cite{nemo3-bkg-2009}. 
 
The neutron contribution to the external background is measured with dedicated runs performed with an Am--Be neutron source located outside of the shield. The data provide the energy spectra of Compton electrons created by external neutrons in the  
($\gamma,e^-$) and crossing-electron channels. These spectra are then used in the fit of the external background model in Figure~\ref{fig:nemo3-bkg-ext-result}. The contribution of neutrons to the external background is negligible for $E_{\rm tot}<2.6$~MeV, which
corresponds to the energy of the $\gamma$ line of $^{208}$Tl, but neutrons dominate at higher energies. The good agreement between data and expected background from neutrons shows that the measurement performed with the Am--Be source correctly emulates the expected external background induced by neutrons
for $E_{\rm tot}>2.6$~MeV, and can be used to estimate the expected background in the $0\nu\beta\beta$ energy range. 
Only six double $\beta$-like events with vertices in the $^{100}$Mo foils and $2.8<E_{\rm tot}<3.2$~MeV are observed in the Am--Be neutron data. 
With the normalization factor obtained from the fit of the external background in Figure~\ref{fig:nemo3-bkg-ext-result}, we obtain a negligible expected background rate of $0.03 \pm 0.01$~events 
 for the combined Phase~I and II
data sets in the 
energy range $2.8<E_{\rm tot}<3.2$~MeV consistent with a $0\nu\beta\beta$ signal. The expected number of double $\beta$-like events for
$E_{\rm tot}>4$~MeV is $0.14 \pm 0.03$ and is also negligible.

\begin{figure}[htbp]  
\centering
\includegraphics[width=0.4\textwidth]{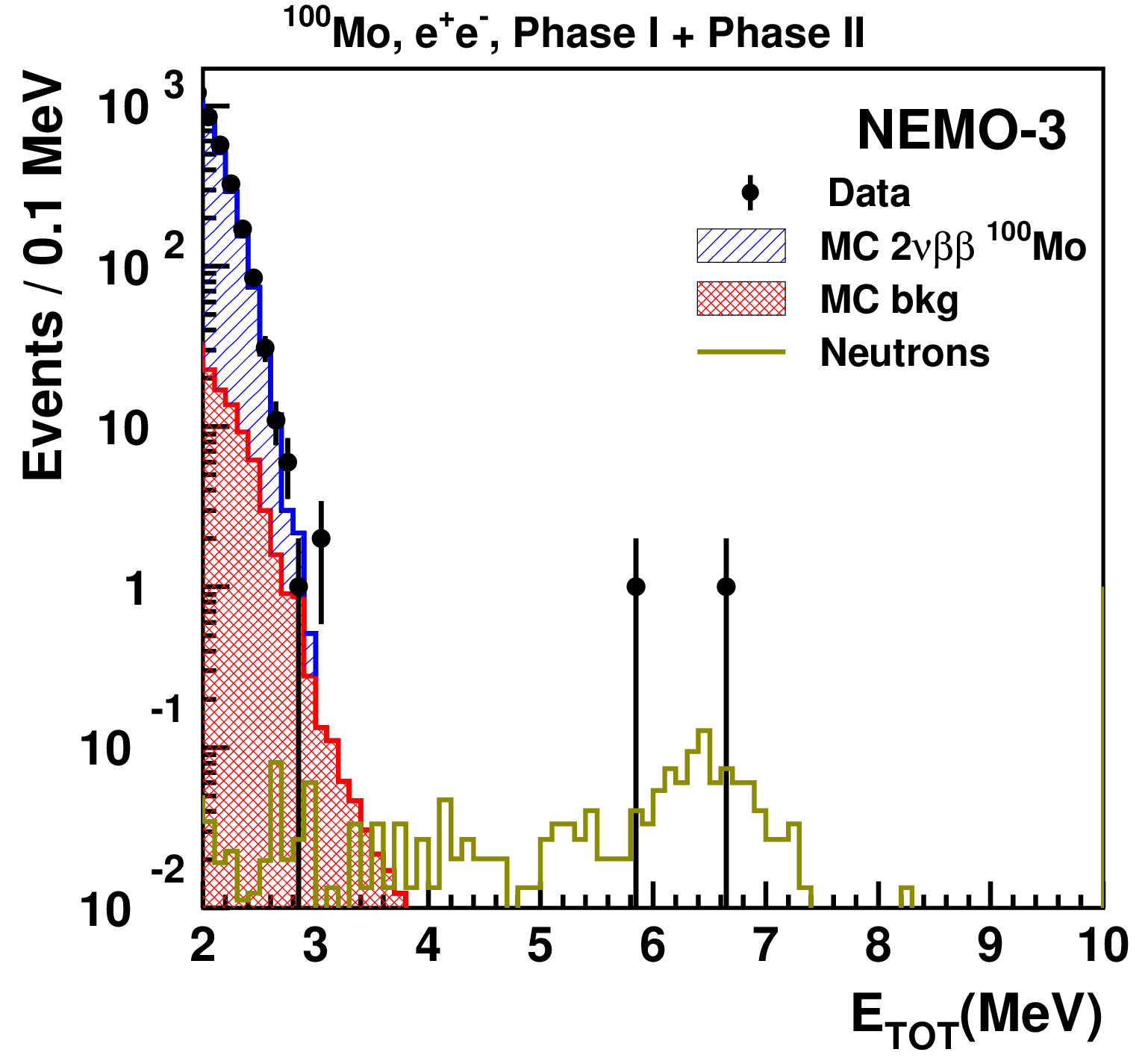}
\caption{Distribution of $E_{\rm tot}$ for $e^+e^-$ pair events 
consistent with being emitted from $^{100}$Mo foils for the entire
data set.  The data are compared to the sum of the
expected background from external neutrons, $2\nu\beta\beta$ events, and the other background components.}
\label{fig:neutron-e+e-}
\end{figure}

The neutron background model is further studied 
using events with $e^+e^-$ pairs, where
external neutrons are the only expected component of the background
for $E_{\rm tot}>4$~MeV. 
The criteria to select $e^+e^-$ events are the same as the ones used to select two-electrons events (see Section~\ref{sec:bb-selection}), except that the curvatures of the two tracks are required to be of opposite sign.
For  $E_{\rm tot}>4$~MeV, we observe 2 $e^+e^-$ events, in agreement with 
the expectation of $1.1 \pm 0.1$ neutron events.
The $E_{\rm tot}$ distribution for these events is shown in Figure~\ref{fig:neutron-e+e-}. 

\subsection{Radon and thoron contaminations}
\label{sec:radon}

Radon and thoron are both found inside the tracking detector. 
Radon ($^{222}$Rn) with a half-life of $T_{1/2}=3.824$~days and thoron ($^{220}$Rn) with $T_{1/2}=55.6$~s 
are $\alpha$-decay isotopes that have $^{214}$Bi and $^{208}$Tl as daughter isotopes in their respective decay chains. 
Radon and thoron emanate from the rock into the air, from where they diffuse into the detector and contaminate the interior of the tracking chamber.
They can also emanate directly from the detector materials inside the tracking chamber. 
Subsequent $\alpha$ decays of these rare gases produce $^{214}$Pb or $^{212}$Pb ions, which drift mainly to the cathode wires. 
If they are deposited on wires close to source foils, their decays can mimic a $\beta \beta$ decay, as illustrated in Figure~\ref{fig:bkg-schema-1}.
Contamination from thoron is much lower than from radon since the shorter half-life makes it less likely for thoron to emanate and diffuse into the detector. 

The radon contamination is measured by detecting BiPo events,
where the electron from $\beta$ decay of $^{214}$Bi, a daughter of $^{222}$Rn, is followed by a delayed $\alpha$ particle from the decay of $^{214}$Po, which has a short half-life of 164~$\mu$s. 
Additional photons may also be emitted and detected. 
A BiPo event in the NEMO-3 detector is identified by requiring an electron track inside the wire chamber associated with a scintillator hit, and at least one delayed hit in the tracking chamber close to the emission point of the electron, due to the delayed  $\alpha$ particle.
The delay time is required to be at least $100$~$\mu$s for events with only one delayed hit, and at least
$40$, $20$, and $4$~$\mu$s for events with $2$, $3$ and $>3$ delayed hits, respectively, to reject hits where electrons have caused neighboring Geiger cells to re-fire. 
Applying these criteria, the mean efficiency to select a BiPo event produced on the
surface of a wire is estimated by MC simulations to be $23 \%$. 

\begin{figure*}[htbp]
\centering
\includegraphics[width=0.8\textwidth]{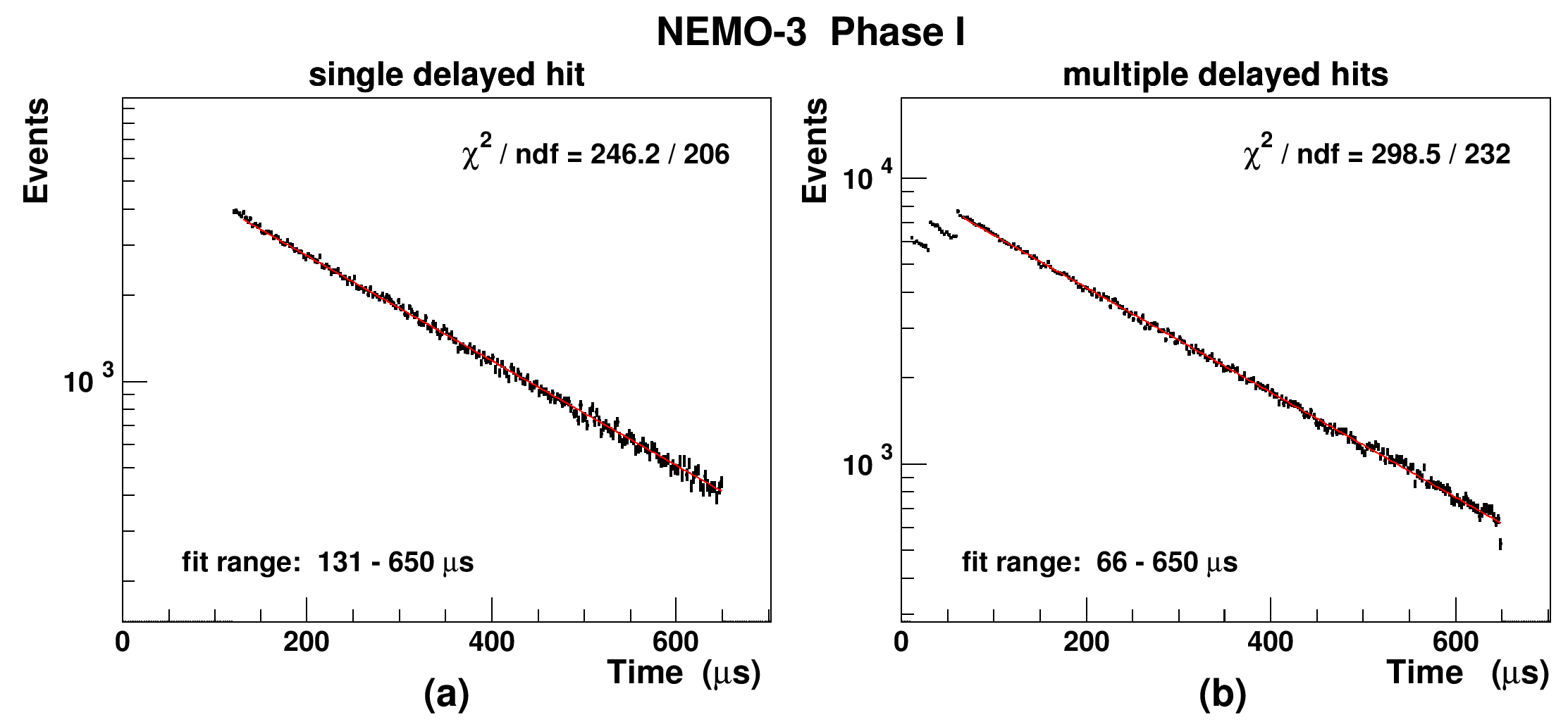}
\includegraphics[width=0.8\textwidth]{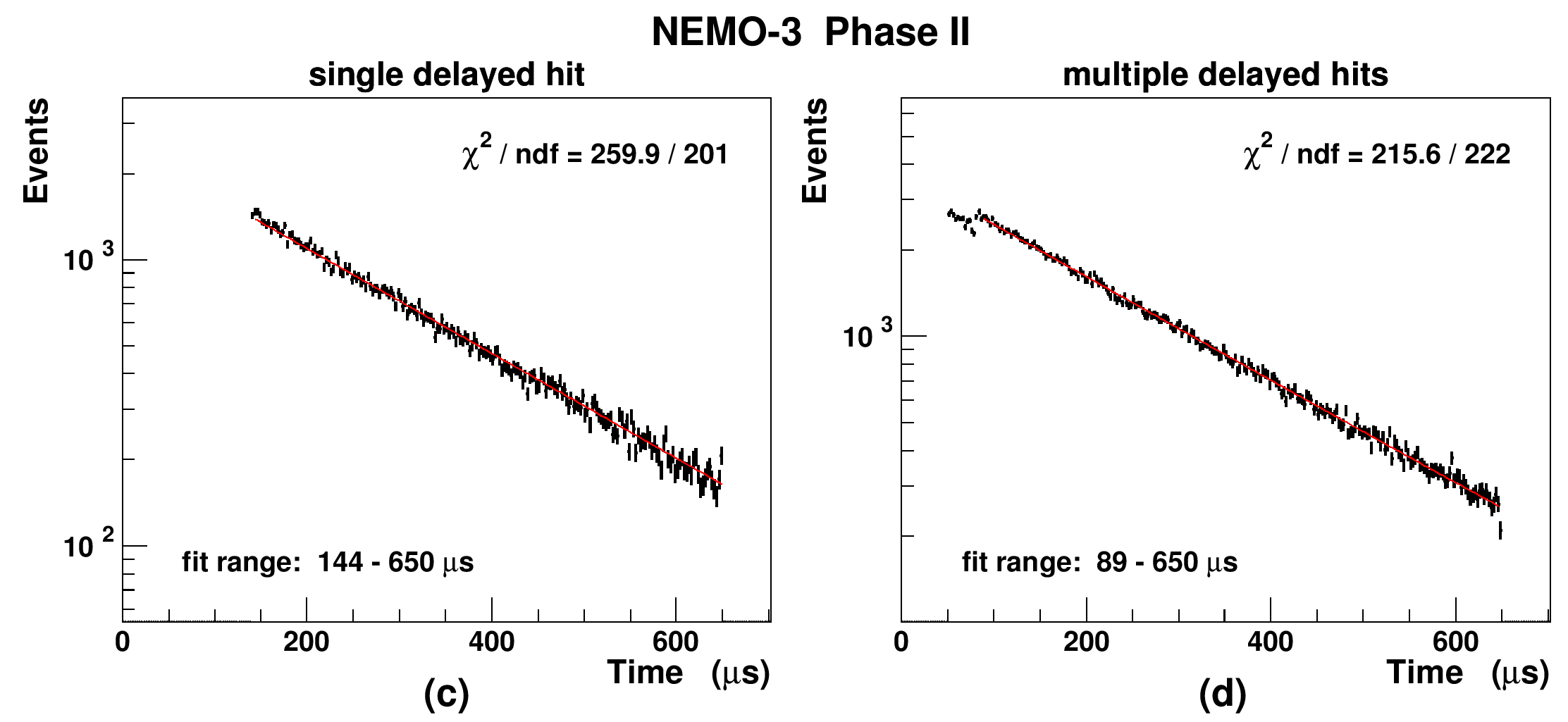}
\caption{Time distribution of delayed $\alpha$ tracks, measured for BiPo
decays emitted inside the tracking detector, for Phase~I ((a) and (b)) and Phase~II ((c) and (d)), and for single delayed Geiger hit ((a) and (c)) or multiple delayed Geiger hits ((b) and (d)). The distributions are fitted by the sum of an exponential function with $T_{1/2}$ set to the $^{214}$Po half-life of
$T_{1/2}=164$~$\mu$s and a constant term accounting for random coincidences.}
\label{fig:nemo3-delay-time}
\end{figure*}

The time distribution of delayed tracks, shown in Figure~\ref{fig:nemo3-delay-time}, is used to
demonstrate the purity of the event selection. 
We fit the sum of an exponential function and a constant term accounting for random coincidences to the data distributions, assuming a $^{214}$Po half-life of  $T_{1/2}=164$~$\mu$s. For Phase~II the fits are applied to delay times larger than 140~$\mu$s for events with only one delayed hit in the tracking detector, and 80~$\mu$s and 60~$\mu$s for events with 2 or $>2$ delayed hits, respectively. Slightly lower minimum delay times are used for Phase I.
 The very small excess of events over the extrapolated curve at low delay time provides the fraction of re-firing Geiger cells, and the constant term provides the fraction of random coincidences.
The contribution of random coincidences and Geiger re-firings, given in Table~\ref{table:rc-rf-radon}, depend on the number of delayed hits and the data taking period. In all cases, they are found to be negligible.

\begin{figure*}[htbp]  
\centering
\includegraphics[width=0.9\textwidth]{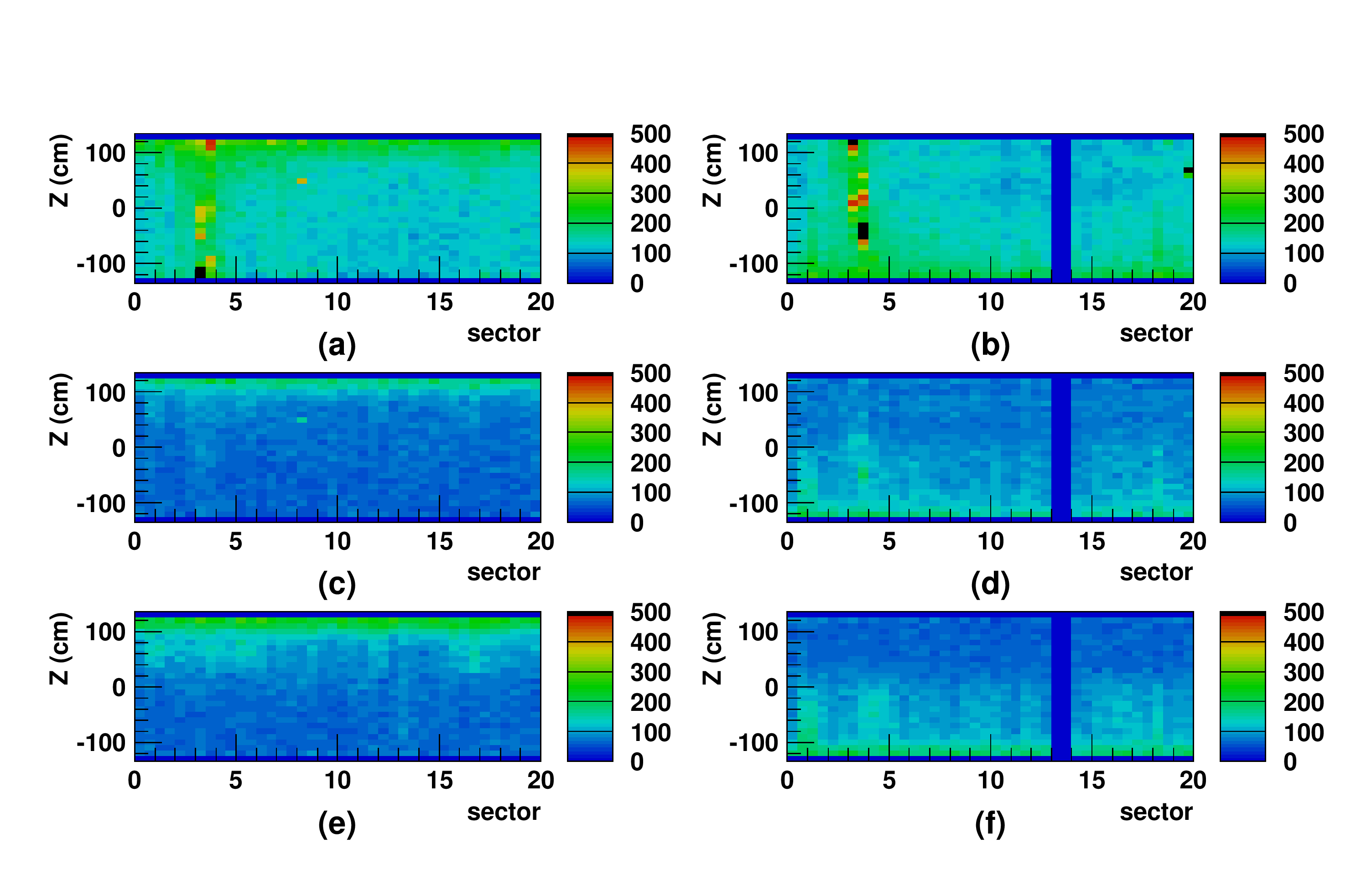}
\caption{The spatial distribution (vertical coordinate $z$ versus sector number) of the emission vertex of detected $^{214}$Bi-$^{214}$Po decay cascade events emitted inside the tracking detector close to the source foils for Phase~II. Left (right) correspond to events with an emission vertex on the internal (external) side of the source foil. (a) and (b) correspond to a vertex on the foil or on the wires of the first layer of Geiger cells close to the foil, (c) and (d) correspond to a vertex on the wires of the second layer of Geiger cells, and (e) and (f)  correspond to a vertex on the wires of the third layer of Geiger cells. The external side of Sector 13 is not represented because of noise observed for Geiger cells in this zone.}
\label{fig:radon-map}
\end{figure*}

This method allows a daily measurement of the radon activity inside the tracking detector. 
The average radon activity is about 30~mBq/m$^3$ in Phase~I and about 5~mBq/m$^3$ in Phase~II.
Figure~\ref{fig:radon-map} shows the spatial distribution of vertices for 
BiPo events that either originate on the foils or on one of the first two layers of Geiger cells inside the tracking chamber. The activity is larger in 
Sector 03, which hosts a  $^{100}$Mo source, than in other sectors.  
The radon model used for the background simulation includes the contributions of $^{214}$Bi deposited on the surface of wires and on the surface of foils. 

The systematic uncertainty on the $^{214}$Bi background contribution 
caused by radon contamination is dominated by the uncertainty on the efficiency of the tracking chamber to detect a delayed $\alpha$ decay of $^{214}$Bi. 
It is estimated by independently measuring the activities of the isotope $^{214}$Bi using ($e^-,\alpha$) and ($e^-,\gamma$) events. 
A large fraction of the $^{214}$Bi $\beta$ decays are accompanied by a high energy $\gamma$ ray 
emitted from the same point inside the tracking chamber. 
These  ($e^-,\gamma$) events are contaminated both by external $\gamma$ rays that Compton scatter on the wires of the Geiger cells, and by ($\beta, \gamma$) emitters in the wires. 
To suppress this background, only events with a $\gamma$  energy $>1$~MeV are selected.

The  $^{214}$Bi measurement using ($e^-,\gamma$) events suffers from larger background and has an approximately three times smaller detection efficiency compared to the method using delayed tracks.
It is sensitive to the systematic uncertainties on $\gamma$ detection,
but it is not affected by systematic uncertainties on the $\alpha$ 
detection efficiency.
The $^{214}$Bi and radon measurement using ($e^-,\gamma$) 
events agree within $10\%$ with the result using an electron and
a delayed $\alpha$ track~\cite{nemo3-bkg-2009}.

The $^{208}$Tl activity from thoron inside the tracking chamber is measured using ($e^-, \gamma \gamma$) and ($e^-, \gamma \gamma \gamma$) events (see next Section). The $^{208}$Tl activity is about $0.1$~mBq/m$^3$, both in Phase~I and in Phase~II. Taking into account the branching ratio of
$36\%$  for producing $^{208}$Tl in the $^{232}$Th decay chain yields a thoron activity of about $0.3$~mBq/m$^3$. 
The MC simulations predict that this thoron activity 
leads to a background for two-electron events 
with $E_{\rm tot}> 2$~MeV that 
is a factor of 50 smaller than the background originating from radon for Phase~I, and a factor of 8 for Phase~II. 
The $^{208}$Tl contribution
 is therefore negligible in the $0\nu\beta \beta$ energy region, and for 
decays with $E_{\rm tot}> 2.8$~MeV.

\begin{table}[hbt]
\begin{center}
\begin{tabular}{l|cc}
\hline\hline
Number of Delayed Hits               & $1$ &  $>1$   \\
\hline 
  & \multicolumn{2}{c}{Phase I}  \\
Random Coincidences & $< 0.03$\%   &  $< 2.7$\%  \\
Refiring            & $< 0.5$\%   &  $< 2.6$)\%     \\
 & \multicolumn{2}{c}{Phase II}\\
Random Coincidences & $< 0.05$\%   &  ($1.1 \pm 0.3$)\%   \\
Refiring            & $< 0.7 $\%   &  $< 0.7$\%       \\
\hline\hline
\end{tabular}
\caption{Contribution of random coincidences and Geiger refirings in the selection of BiPo events used for the Radon measurement, for the high radon period (Phase~I) and the low radon period (Phase~II), requiring either exactly one or several delayed Geiger hits. Upper limits are given at 90\%~C.L.}
\label{table:rc-rf-radon}
\end{center}
\end{table} 

\subsection{Internal backgrounds}

Internal backgrounds originating from radioactive contaminants inside the source foils
are mainly due to $\beta$ decay of $^{214}$Bi with $Q_{\beta}=3.27$~MeV and $^{208}$Tl  with $Q_{\beta}=4.99$~MeV. The two isotopes are products of the $^{238}$U and $^{232}$Th decay chains, respectively. 
As illustrated in Figure~\ref{fig:bkg-schema-1}, the presence of $^{214}$Bi and $^{208}$Tl can mimic $\beta \beta$ events by a $\beta$ decay accompanied by an internal conversion electron process. 
This is the dominant channel in the case of $^{208}$Tl with a conversion rate of $0.2\%$ for the 2615~keV $\gamma$ ray, which produces a conversion electron with an energy of $2527$~keV. 
Other processes are M\o ller scattering of the $\beta$-decay electrons in the source foil, or $\beta$ decay to an excited state followed by a $\gamma$ undergoing Compton scattering, which can be reconstructed as two-electron events if the $\gamma$ is not detected. 

\subsubsection{$^{208}$Tl contamination in the source foils}
\label{sec:tl_int}

The $\beta$ decay of $^{208}$Tl is usually accompanied by two or three $\gamma$ rays. The $^{208}$Tl contamination
inside the sources foils is therefore measured by selecting internal $(e^-, \gamma \gamma)$ and $(e^-,\gamma \gamma \gamma)$ events defined as one electron track originating from the source foil that is associated with a scintillator hit, and two or three isolated scintillator hits.
The time-of-flight must be consistent with
the hypothesis that all particles are emitted from the track intersection with the foil.

\begin{figure*}[htbp]
\centering
\includegraphics[width=1.\textwidth]{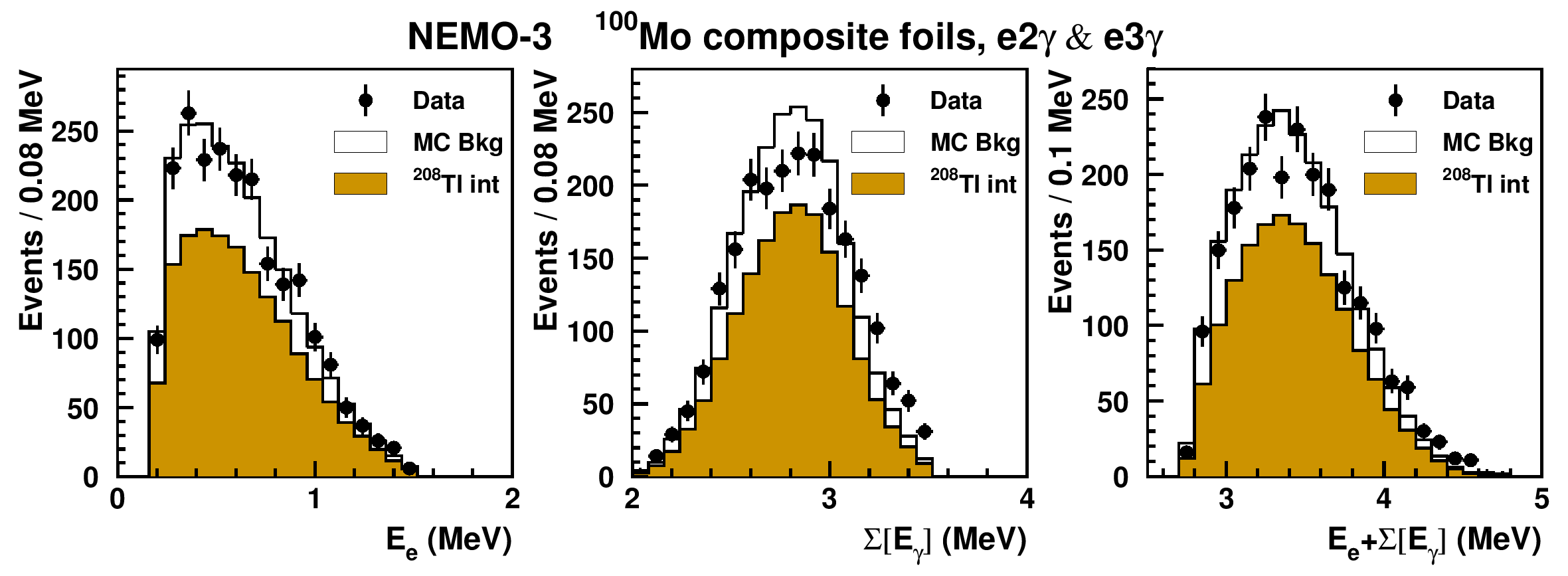}
\includegraphics[width=1.\textwidth]{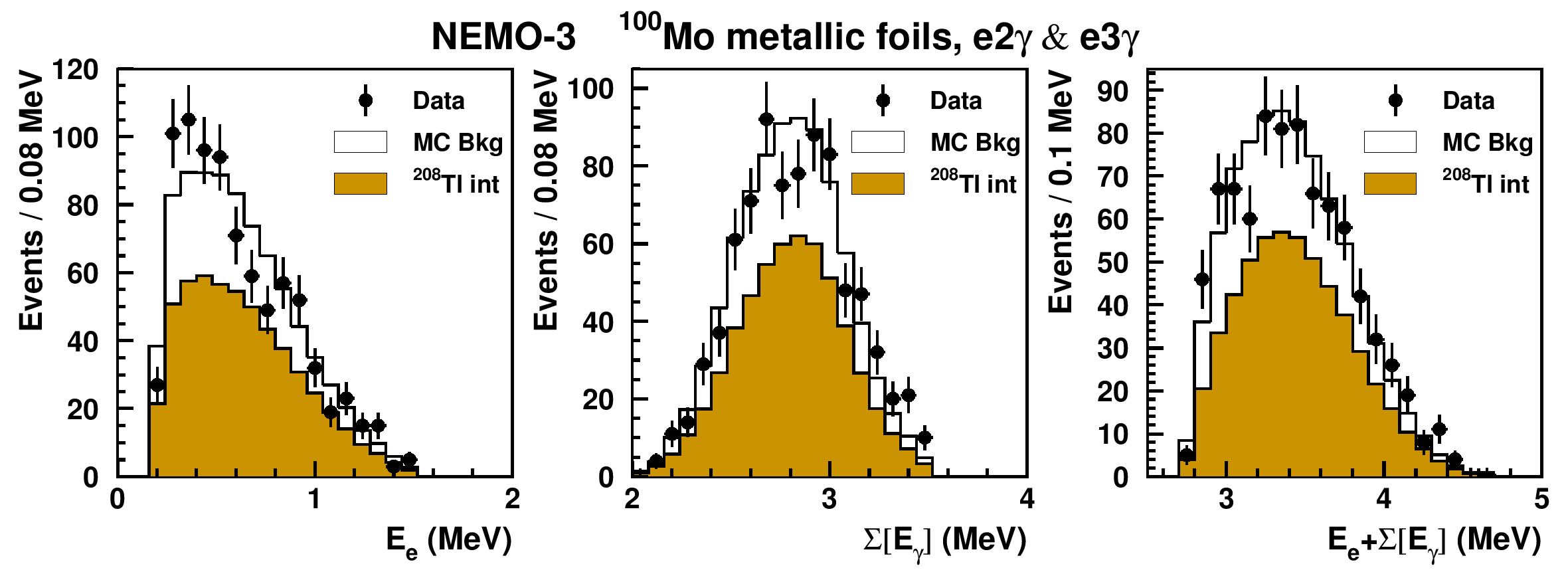}
\caption{Distributions of the energy of the electron, $E_{e^-}$, the energy sum $\sum E_{\gamma}$,  and $E_{e^-}+\sum E_{\gamma}$
using $(e^-,\gamma \gamma)$ and $(e^-,\gamma \gamma \gamma)$ events for the combined $^{100}$Mo data set. The top panels show the composite and the bottom panels the metallic foils. The data are compared to the sum of the expected background from
MC simulations and the fitted $^{208}$Tl activity inside the $^{100}$Mo foils.}
\label{fig:tl208-engamma-mo}
\end{figure*}

We require that the energy of the electron is in the range $0.2<E_{e^-}<1.5$~MeV, $E_{\gamma}>0.2$~MeV for all $\gamma$ energies, and that the sum 
$\sum E_{\gamma}<3.5$~MeV. The condition 
\begin{equation}
E_{e^-} \mathrm{(MeV)} > \left( 4\mathrm{(MeV)} - 1.5 \times \sum E_{\gamma}\mathrm{(MeV)} \right)
\end{equation}
rejects $^{214}$Bi background.
The highest energy photon must have $E_{\gamma}>1.7$~MeV to select the 2615~keV $\gamma$ line.
Finally, we require $P_{\rm int}>0.05$, $P_{\rm ext}<0.01$, and the $z$ coordinate of the emission vertex of the electron must satisfy $|z| < 120$~cm.
The distributions of $E_{e^-}$, $\sum E_{\gamma}$,  and the total energy
$E_{e^-}+ \sum E_{\gamma}$ are shown in Figure~\ref{fig:tl208-engamma-mo}.
The thoron and radon activities inside the tracking chamber are set to the values obtained from the prior measurements described in Section~\ref{sec:radon}. 

The measured $^{208}$Tl activities of the metallic and composite Mo source foils, and of the copper and tellurium foils are given in Table~\ref{table:tl208}, for both event topologies combined. 
The data are in agreement with the upper limits from the HPGe measurements of $^{208}$Tl activities, prior to the installation of the foils in the detector.
The two event topologies, $(e^-,\gamma \gamma)$ and $(e^-,\gamma \gamma \gamma)$, give consistent results when analysed separately. 
The $^{208}$Tl activities of the copper and tellurium foils are used in section~\ref{sec:validation-bkg-model} for the validation of the background model. 

\begin{table}[htbp]
\begin{center}
\begin{tabular}{l|cccccc}
\hline\hline
              & $N_{\rm obs}$ & $N_{B}$ & $S/B$ & $\epsilon$ & $A$          & $A$ (HPGe)\\
Source  Foil  &             &        &       & $(\%)$     & ($\mu$Bq/kg) & ($\mu$Bq/kg) \\
              &             &        &       &            &              & (90\% C.L.) \\
\hline 
$^{100}$Mo Metal. & 823   & 281  & 1.93   & 2.05  & $87 \pm 4$   & $<100$ \\
$^{100}$Mo Comp.  & 2241  & 617  & 2.63   & 2.15  & $128 \pm 3$   & $<170$                  \\
Copper               & 75    & 60   & 0.25  & 1.82  & $11 \pm 3$  &  $<33$             \\
$^{130}$Te        & 563  & 155   & 2.64  & 2.54  & $206 \pm 10$  &  $<500$             \\
Te-nat        & 741  & 121   & 5.14  & 2.18  & $301 \pm 12$  &  $<830$             \\
\hline\hline
\end{tabular}
\caption{Numbers of observed $(e^-,\gamma \gamma)$ and $(e^-,\gamma \gamma \gamma)$ events ($N_{\rm obs}$), expected number of background events ($N_B$), signal-to-background ratio, $^{208}$Tl signal efficiency ($\epsilon$), and measured  $^{208}$Tl activity of the $^{100}$Mo metallic (Metal.) and composite (Comp.) foils, the copper, $^{130}$Te and natural Te foils. The activities of the foils are compared to the HPGe measurements performed before their installation. Only statistical uncertainties are given.}
\label{table:tl208}
\end{center}
\end{table} 

The systematic uncertainty on the $^{208}$Tl activity is determined by using two $^{232}$U radioactive sources (the isotope $^{232}$U is a parent of $^{208}$Tl).
The $^{208}$Tl activities of the sources are first calibrated by gamma spectroscopy with a coaxial HPGe detector,  by measuring the intensity of the $\gamma$ line emitted in the decay of $^{212}$Pb to $^{212}$Bi with an energy of $238$~keV, while the two $\gamma$ lines emitted in the decay of $^{208}$Tl with energies of $583$~keV and $2615$~keV are used to check the results.
The HPGe detection efficiency is determined with a calibrated $^{232}$Th source that has an activity known to within $0.5\%$, and using a MC simulation of the setup. 
The sources are measured at four different distances between the source and the Ge crystal. 
%For each of the distances, experimental (what is that ?) and simulated detector efficiencies are obtained, yielding a difference of $\le 2\%$. 
The four activities obtained for each distance are combined to obtain a total statistical uncertainties of $0.7\%$ and a systematic uncertainty of $3\%$. 
The two calibrated $^{232}$U sources are then temporarily introduced into the NEMO-3 detector through the calibration tubes.
We select $(e^-,\gamma\gamma)$ and $(e^-,\gamma\gamma\gamma)$ events  and fit the activities of the two sources using a MC simulation  of $^{232}$U decays. The results are given in Table~\ref{table:U232}. 
The largest sources of systematic uncertainty are the knowledge
of the exact location of the sources ($3\%$) and the kinematic selection criteria (6\%). This systematic uncertainty is estimated by allowing a variation of the energy requirements, considering tracks that traverse only 
a single sector, tracks only on the inner or outer side of the foils, and by accepting or rejecting scintillator blocks with an energy $<150$~keV.

The results of the in-situ NEMO-3 and the HPGe measurements 
shown in Table~\ref{table:U232} are consistent within their systematic uncertainties. We assign a systematic uncertainty of $10\%$ to the $^{208}$Tl activity measurement, corresponding to the larger difference between the in-situ and the HPGe measurements obtained for the second $^{232}$U source. 

\begin{table}[htbp]
\begin{center}
\begin{tabular}{ l | c  c }
\hline\hline
  & \multicolumn{2}{c}{$^{232}$U Activity (Bq)} \\
  & Source (1) &  Source (2) \\
\hline 
NEMO-3 & $ 7.36 \pm 0.03 \pm 0.52$  & $ 14.56 \pm 0.05 \pm 1.02$    \\
HPGe   & $ 7.79 \pm 0.04 \pm 0.21$  & $ 15.91 \pm 0.09 \pm 0.43$    \\
\hline\hline
\end{tabular}
\caption{The $^{208}$Tl activities from  $^{232}$U sources obtained with the NEMO-3 
detector and with HPGe $\gamma$ spectrometers.}
\label{table:U232}
\end{center}
\end{table} 

\begin{figure}[htbp]
\centering
\includegraphics[width=0.45\textwidth]{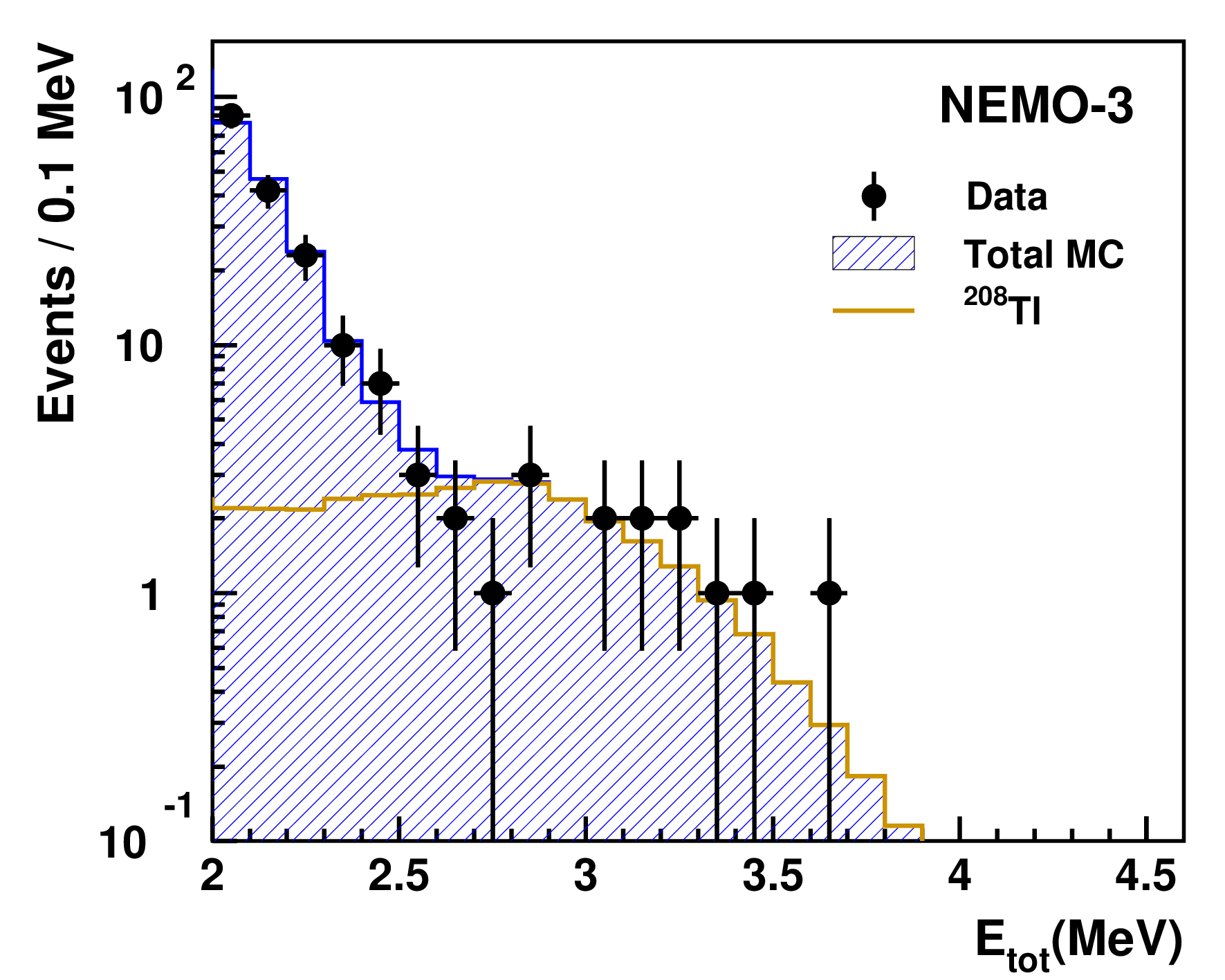}
\caption{Distribution of the total energy of two electrons  $E_{\rm tot}$ in the $(e^- e^-, N \gamma)$ channel for the $^{100}$Mo data set compared to the expected background from $^{208}$Tl contamination inside the foils and to the total expected background. The normalisations of the different background components are not fitted, but set to the measured values. No event is observed for $E_{\rm tot} > 3.7$~MeV.}
\label{fig:nemo3-bbng}
\end{figure}

The $^{208}$Tl background measurement is validated by using the two-electron channel with at least one associated $\gamma$ ray emitted in time from the source foil ($e^-e^-, N \gamma$). 
In the region where the sum of the two electrons energies satisfies $E_{\rm tot}>2.6$~MeV, $^{208}$Tl contamination inside the foil
dominates, whereas $2\nu\beta\beta$ decays are strongly suppressed by the selection criteria.
Figure~\ref{fig:nemo3-bbng} shows the total energy of two electrons $E_{\rm tot}$ for $(e^- e^-, N \gamma)$ events for the entire $^{100}$Mo data set.
The normalisations of the different background components are set to the previously measured values and are not fitted to this distribution.
The data are in good agreement with the expected background, which is dominated by $^{208}$Tl contamination inside the foils. 
We observe $7$~events in the $^{100}$Mo foils in the interval $[2.8-3.2]$~MeV whereas $8.8$~events are expected from the  simulation. This
independent check validates the estimation of the $^{208}$Tl activity
inside the foils within relatively large statistical uncertainties.

\subsubsection{$^{214}$Bi contamination in the source foil}
\label{sec:bi_int}

The $^{214}$Bi contamination inside the source foils is measured by analysing the distribution of the length of the delayed $\alpha$ tracks in BiPo events. It allows the discrimination of the $^{214}$Bi contamination inside the foils, and inside the mylar for composite foils, from the dominant radon background close to or on the surface of a foil.

The criteria for the selection of the BiPo events are similar to the 
selection used for the radon activity measurement, except that the common vertex of the electron track and the delayed $\alpha$ track must be in the foil or in the first layer of wires of the tracking chamber.
The $^{214}$Bi contamination inside the source foils is found by fitting the distribution of the delayed $\alpha$ track length, taking into account the other unknown activities as free parameters in the fit. These parameters are the $^{214}$Bi activities from radon deposition on the surface of the source foils and on the surface of the two closest layers of wires.
Only Phase~II data are used to reduce the radon background.

\begin{figure*}[htpb]
\begin{center}
\includegraphics[width=1.\textwidth]{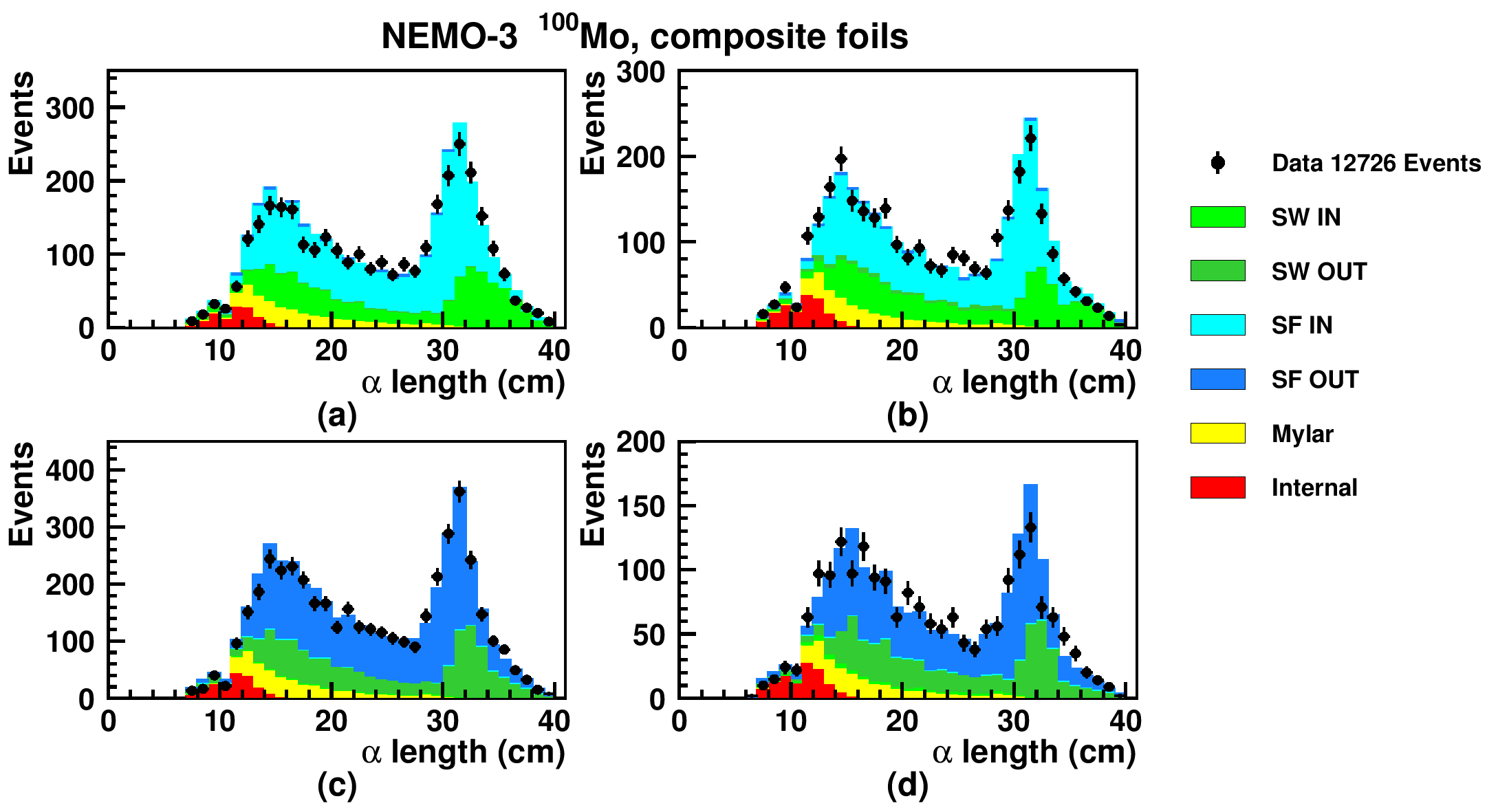}
\end{center}
\caption{Distribution of the lengths of delayed $\alpha$ tracks for composite $^{100}$Mo foils for Phase~II: (a,c) for electron and $\alpha$ tracks on the same side of the foils, (b,d) for electron and $\alpha$ tracks on opposite sides of the foils, (a,b) for $\alpha$ tracks on the inner side of the foils, (c,d) for $\alpha$ tracks on the outer side of the foils. 
The data are compared to the simulated background with
a normalisation determined by the fit of the different components of $^{214}$Bi background. 
``SW" corresponds to the $^{214}$Bi deposition on the surface of the wires, and ``SF" to the deposition on the surface on the foil, where
``IN" and ``OUT" corresponds to the components from the wires and
surfaces inside and outside relative to the position of the foil.
``Internal" $^{214}$Bi contamination originates inside the $^{100}$Mo foils, and the ``mylar" contamination from inside the mylar.}
\label{fig:bi214-foils-mo100-comp}
\end{figure*}

\begin{figure*}[htbp]
\begin{center}
\includegraphics[width=1.\textwidth]{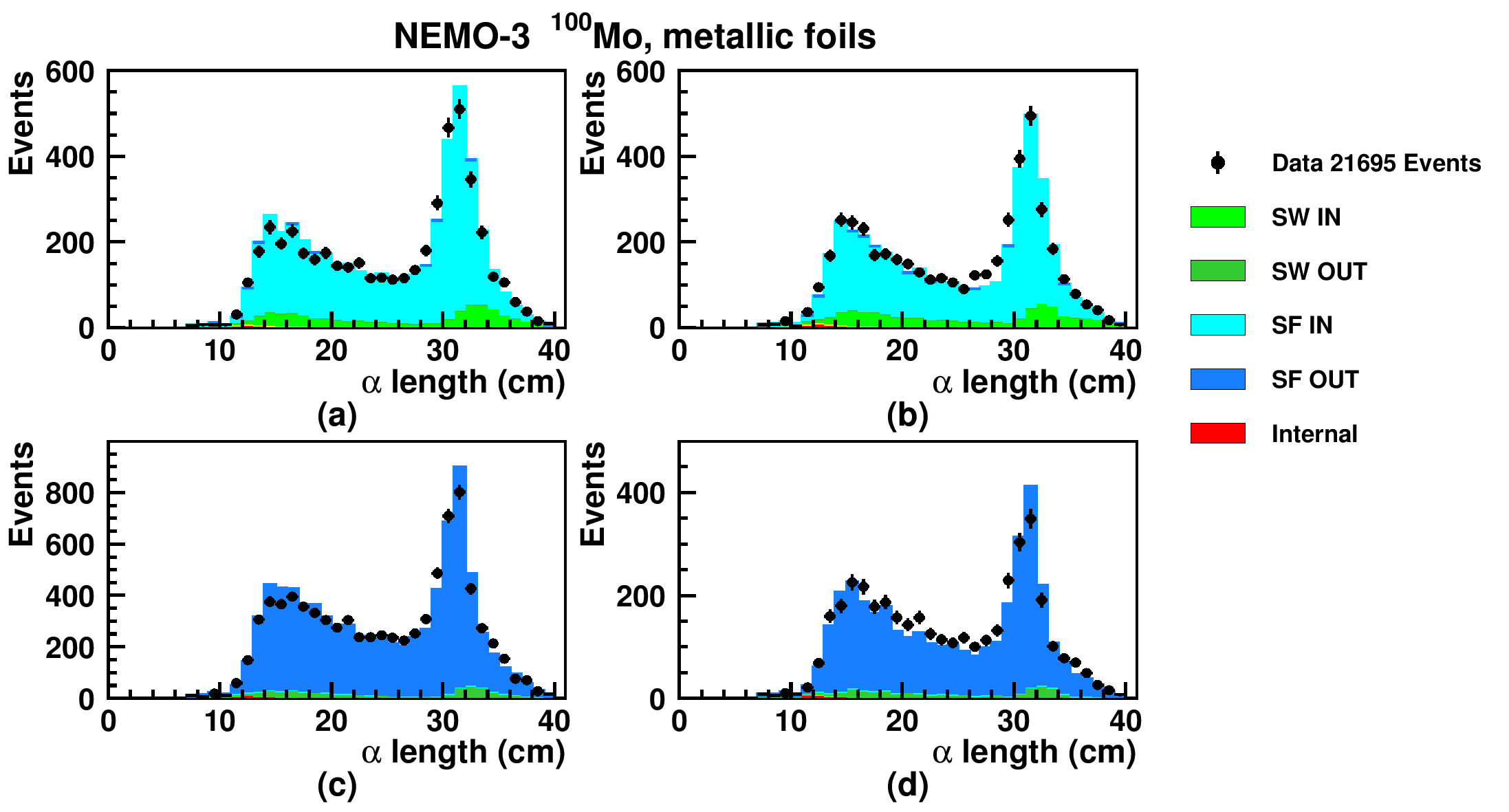}
\end{center}
\caption{Distribution of the delayed $\alpha$ track length for metallic $^{100}$Mo foils (see Figure~\ref{fig:bi214-foils-mo100-comp} caption for further details).}
\label{fig:bi214-foils-mo100-met}
\end{figure*}

The results of the fit are shown in Figures.~\ref{fig:bi214-foils-mo100-comp} and \ref{fig:bi214-foils-mo100-met} for the $^{100}$Mo composite and metallic foils, respectively.
The results of the $^{214}$Bi activity measurement are given in Table~\ref{table:bi214} for $^{100}$Mo foils, and also for copper, $^{130}$Te, and natural tellurium foils. They are in agreement with the upper limits obtained from HPGe measurements.

\begin{table}[htbp]
\begin{center}
\begin{tabular}{l|c|c|c|c}
\hline\hline
                 & Activity        & Activity        &  $A$ (HPGe)  &  $A$ (HPGe)   \\
                 &   Foil          & Mylar           &  Foil+Mylar  &  Mylar        \\
Source  Foil     &  (mBq/kg)       & (mBq/kg)        &  (mBq/kg)    &  (mBq/kg)     \\
\hline 
$^{100}$Mo Comp.  & $0.31 \pm 0.04$ & $1.05 \pm 0.06$ & $<0.34$      & $<0.67$        \\
$^{100}$Mo Metal. & $0.06 \pm 0.02$ &  No mylar       & $<0.39$      & No mylar       \\
Copper           & $0.16 \pm 0.04$ &  No mylar       & $<0.12$      & No mylar       \\
$^{130}$Te        & $0.41 \pm 0.06$ & $1.81 \pm 0.17$ & $<0.67$      & $3.3 \pm 0.5$  \\
Te-nat           & $0.37 \pm 0.05$ & $1.11 \pm 0.17$ & $<0.17$      & $1.7 \pm 0.5$  \\
\hline\hline
\end{tabular}
\caption{Measured $^{214}$Bi activity of the $^{100}$Mo metallic, $^{100}$Mo composite, copper, $^{130}$Te, and natural Te source foils, compared to the HPGe measurements performed before their installation. Only statistical uncertainties are given. 
The fraction of the mylar mass relative to the total mass of the foil is 
in the range $5\%$--$10\%$, depending on the foil.}
\label{table:bi214}
\end{center}
\end{table} 

The measured $^{214}$Bi contamination is checked by 
selecting two-electron events emitted from the $^{100}$Mo foils, where an associated delayed $\alpha$ track is emitted from the two-electron vertex ($e^-e^-,\alpha$). This channel is dominated by radon background close to the foil and by $^{214}$Bi contamination from inside the foil. The criteria to select the two electrons are the same as those used for the selection of double $\beta$ decay events (see Section~\ref{sec:bb-selection}). The criteria to select the delayed $\alpha$ track are identical to those used for the radon background measurement. 
Using all $^{100}$Mo foils, we observe six events with a 
($e^-e^-,\alpha$) topology in the energy range for
the two electrons of $E_{\rm tot}= [2.8-3.2]$~MeV in the combined
Phase~I and II data, while $9.4 \pm 0.4$ events are expected from simulations. Within large statistical uncertainties,
this result confirms the prediction for the $^{214}$Bi background contribution in the $0\nu\beta\beta$ signal region.

\subsection{Validation of background model with copper and tellurium foils}
\label{sec:validation-bkg-model}

\begin{figure*}[htbp]
\begin{center}
\includegraphics[width=0.38\textwidth]{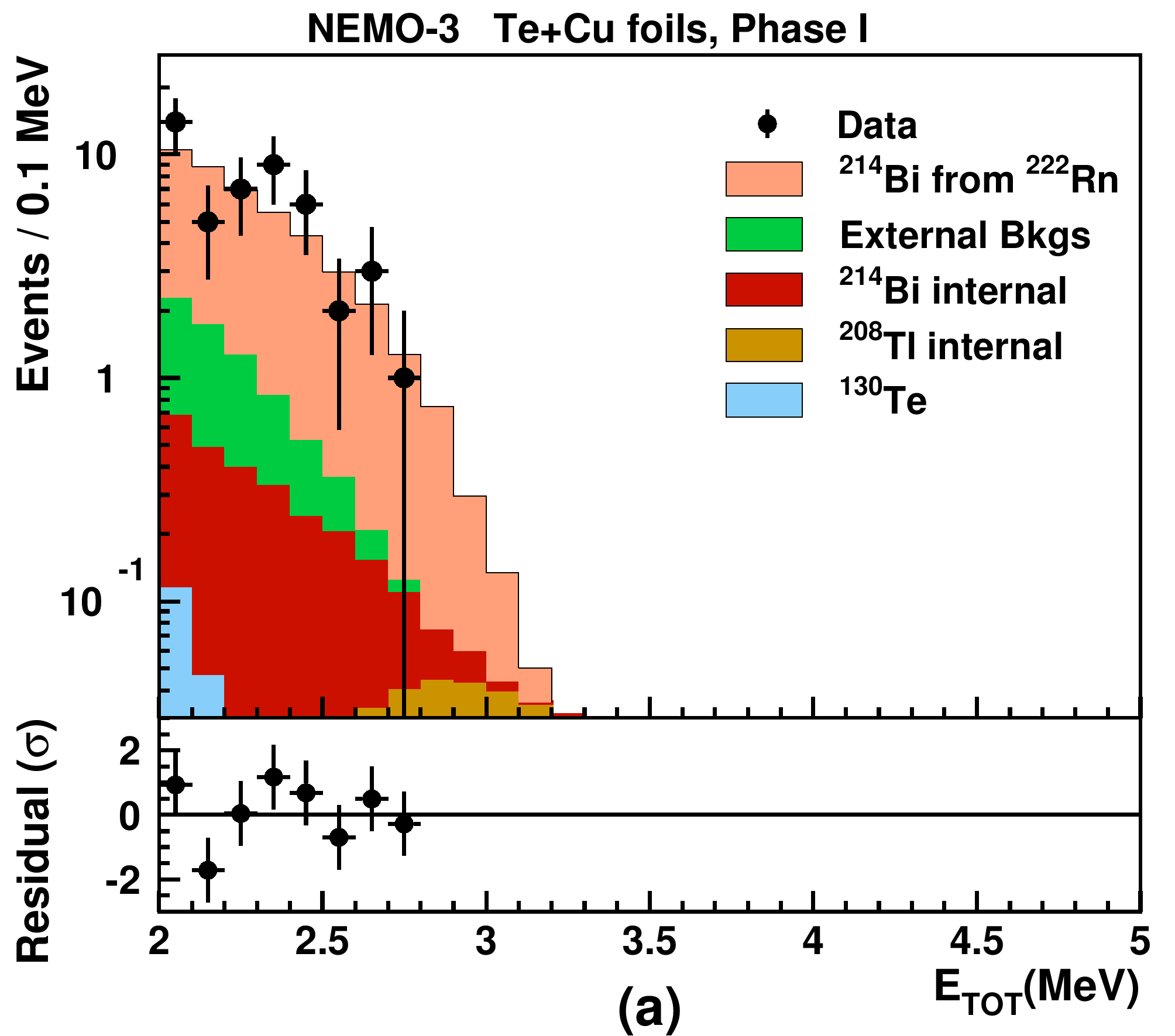}
\includegraphics[width=0.38\textwidth]{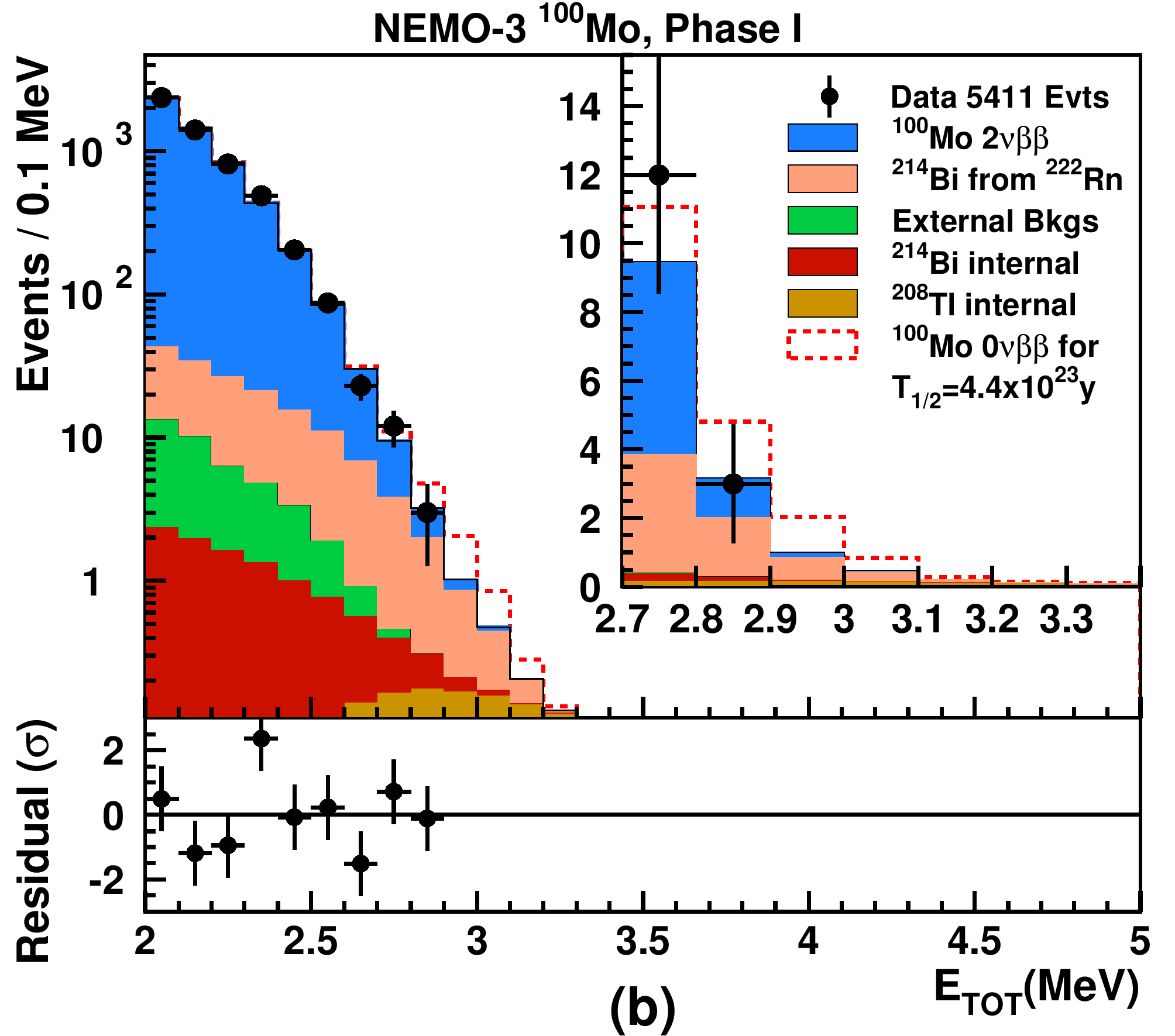}\\
\includegraphics[width=0.38\textwidth]{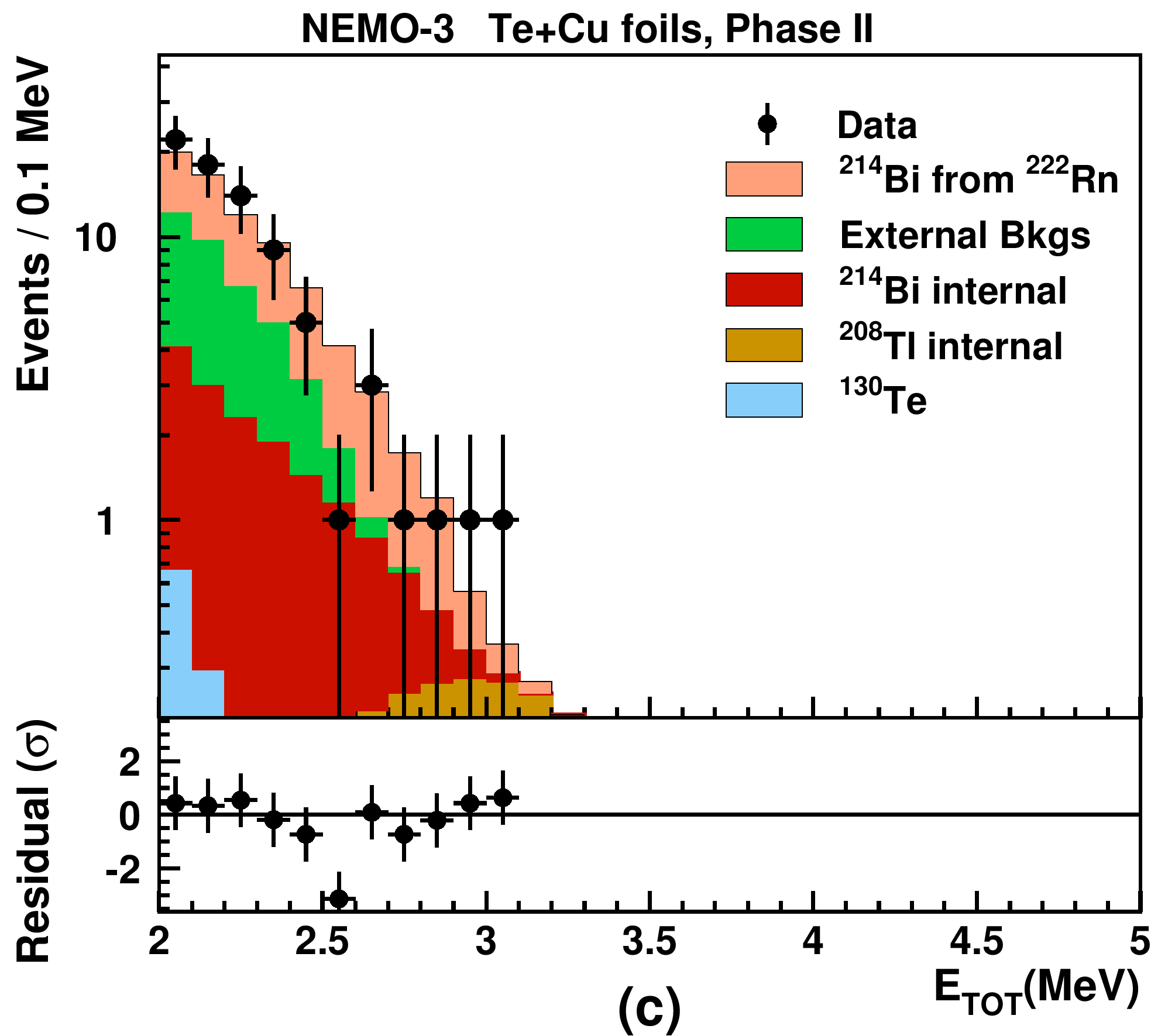}
\includegraphics[width=0.38\textwidth]{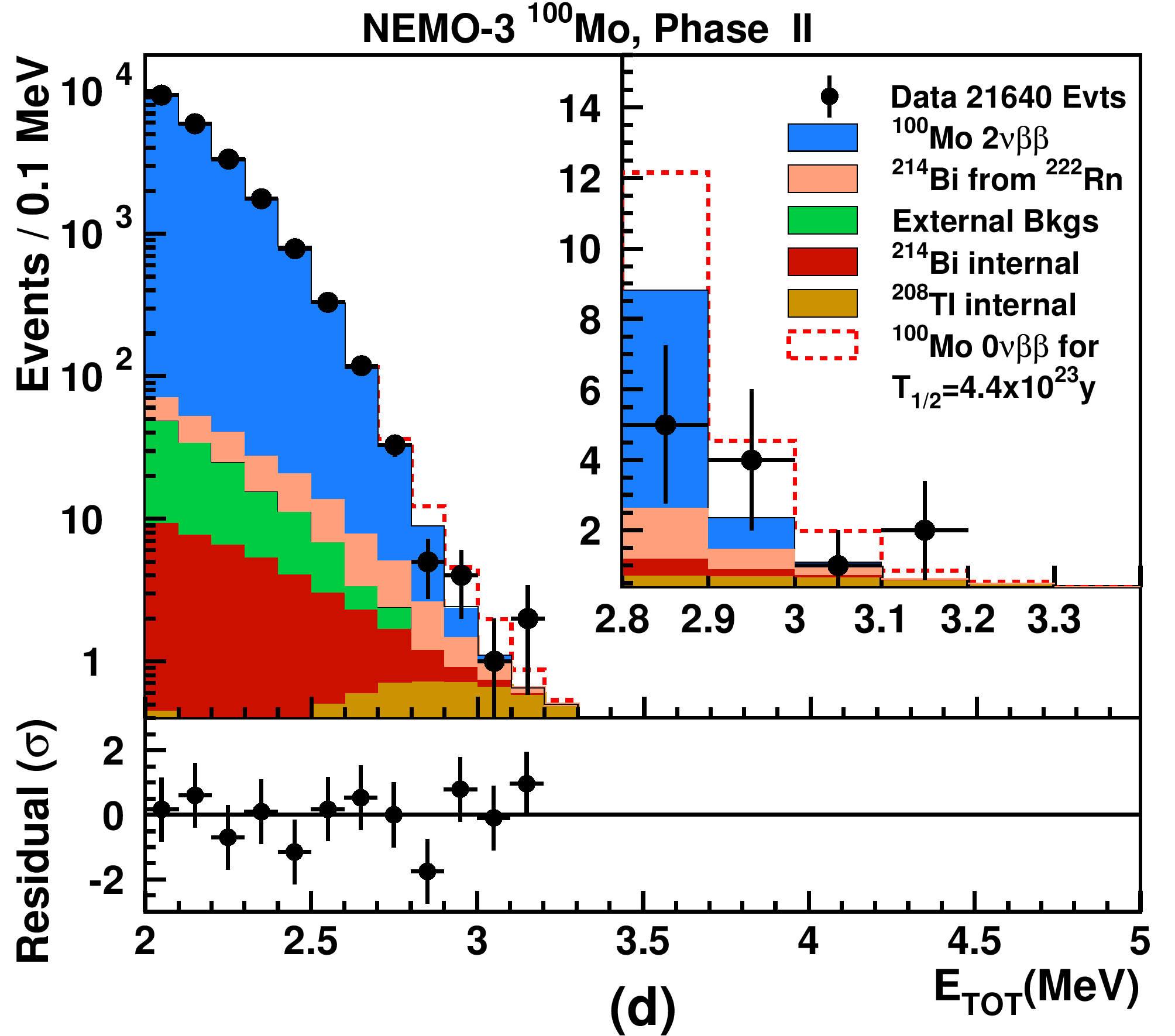}\\
\includegraphics[width=0.38\textwidth]{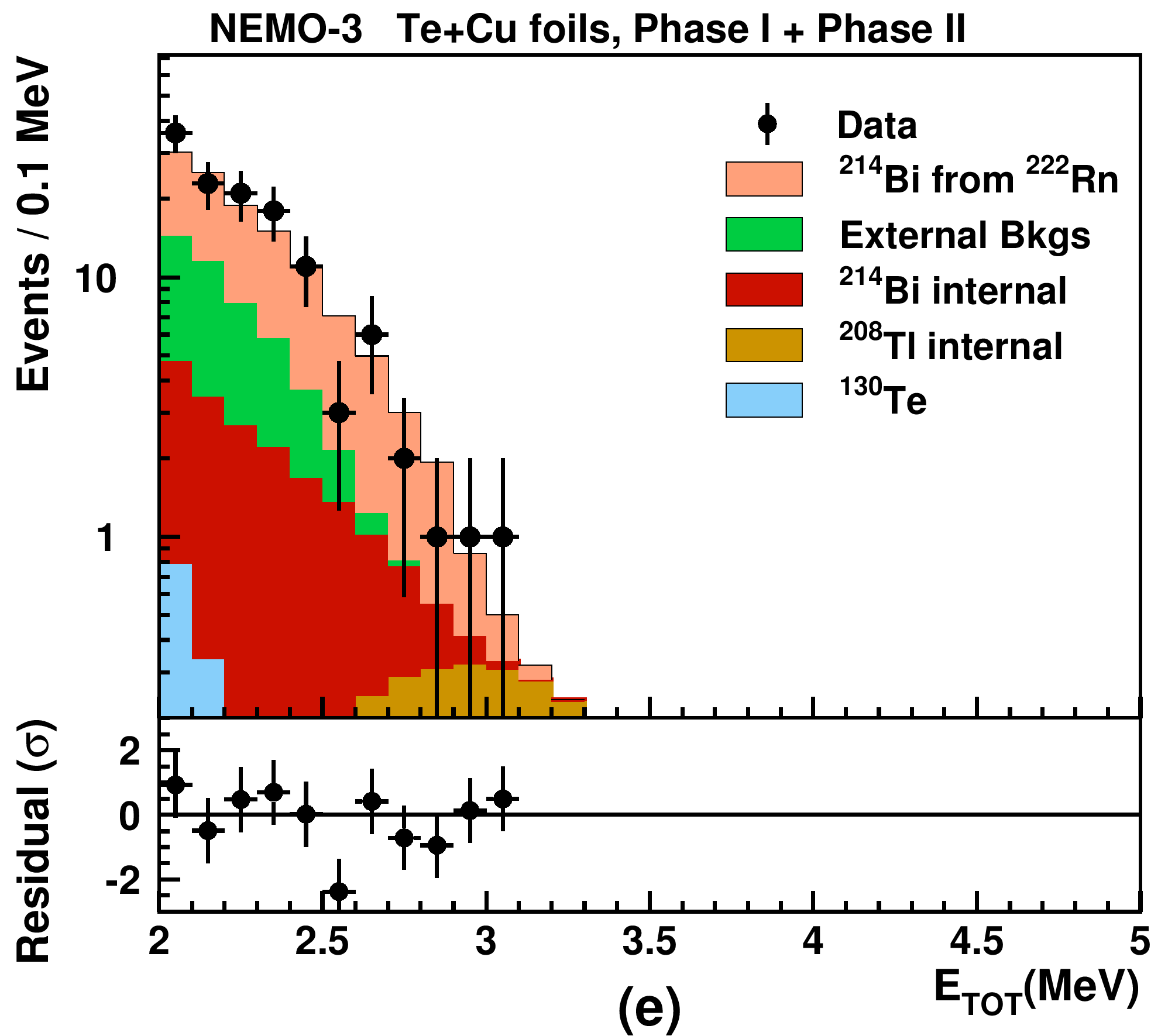}
\includegraphics[width=0.38\textwidth]{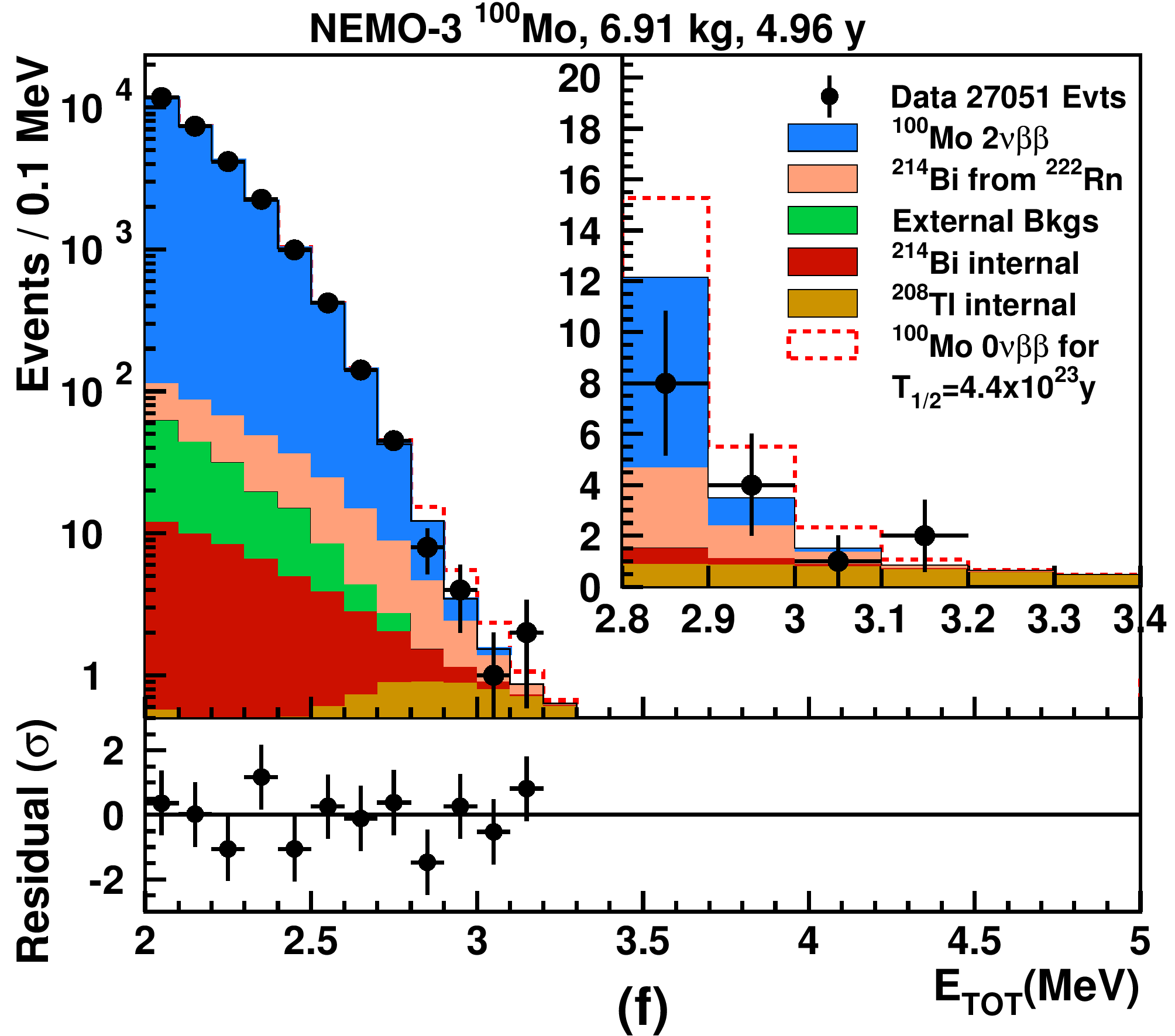}\\
\end{center}
\caption{Distribution of $E_{\rm tot}$ for two-electron events with $E_{\rm tot}> 2$~MeV for the copper, $^{130}$Te, and natural tellurium foils (a,c,e), and for $^{100}$Mo foils (b,d,f), for Phase~I (a,b) and Phase~II (c,d), and combined (e,f). The combined data correspond to an exposure of $13.5$~kg$\cdot$yr for the copper, $^{130}$Te, and natural tellurium foils, and $34.3$~kg$\cdot$yr for the $^{100}$Mo foils. The data are compared to the sum of the expected background from $2\nu\beta\beta$ decays of $^{100}$Mo, radon, external backgrounds, and from internal $^{214}$Bi and $^{208}$Tl contaminations inside the foils. Only the $2\nu\beta\beta$ background contribution is fitted to the data, while the other background components are set to the measured values given in Section~\ref{sec:bkg}.  Lower panels show residuals between data and expected background, normalized to the Poisson error, ignoring bins with 0 events.}
\label{fig:result}
\end{figure*}

The complete background model is validated by selecting two-electron events emitted from the copper, natural tellurium, and $^{130}$Te foils ($Q_{\beta\beta}= 2527.518 \pm 013$~keV) using the criteria described in Section~\ref{sec:bb-selection}. 
The data correspond to an exposure of $13.5$~kg$\cdot$yr.
The internal contaminations of these foils in $^{208}$Tl and $^{214}$Bi are measured using the same methods as those used for the Mo foils (see sections\ref{sec:tl_int}and \ref{sec:bi_int}). 
Results of the internal contaminations measurements are given in Tables~\ref{table:tl208} and \ref{table:bi214}. 
Figure~\ref{fig:result} shows the distributions of the sum of the  energies of the two electrons for $E_{\rm tot}>2$~MeV, and
Table~\ref{tab:background-cu-te} gives the number of events with  $E_{\rm tot}>2$~MeV. 
The observed numbers of two-electron events agree with the expectation from the MC simulation calculated using the background model, which is
dominated by radon background. 
The number of $2\nu\beta\beta$ decays of $^{130}$Te in 
this energy region is expected to be negligible~\cite{articleTe130}. 
In the full data set, only $3$ events with two electrons from the sectors containing copper, $^{130}$Te, and natural tellurium foils remain in the energy region 
$E_{\rm tot}=[2.8-3.2]$~MeV, compared to a MC expectation of $3.6 \pm 0.2$ events.  

\begin{table}[htbp]
\begin{center}
\begin{tabular}{ccccc}
\hline
\hline
Data Set               & Phase I & Phase II & Combined \\
\hline
External Background      &  $4.77 \pm 0.48$   &  $24.94 \pm 2.49~$   &   $29.71 \pm 2.97$  \\
$^{214}$Bi from Radon     & $36.1 \pm 3.6$  &  $34.0 \pm 3.4$   &   $70.0 \pm 7.0$  \\
$^{214}$Bi Internal       &  $2.34 \pm 0.23$  &  $13.83 \pm 1.38~$   &   $16.17 \pm1.62$  \\
$^{208}$Tl Internal       &  $0.49 \pm 0.05$  &   $2.93 \pm 0.29$   &    $3.42 \pm 0.34$  \\
$^{130}$Te                &  $0.12 \pm 0.02$  &   $0.75 \pm 0.15$   &    $0.87 \pm 0.17$  \\
Total Expected           & $43.8 \pm 3.7$  &   $76.4 \pm 4.5$  & $120.2 \pm 8.1$   \\
\hline
Data                     & 47             &    76              & 123  \\
\hline
\hline
\end{tabular}
\caption{Numbers of expected background and observed two-electron events 
with $E_{\rm tot}>2.0$~MeV in Phases~I and~II, and for the combined data set,
in the copper, natural tellurium, and $^{130}$Te foils. The combined data 
correspond to an exposure of $13.5$~kg$\cdot$yr. The contribution from
$2\nu\beta\beta$ decays of $^{130}$Te is negligible.}
\label{tab:background-cu-te}
\end{center}
\end{table}

\section{Search for neutrinoless double $\beta$ decay}
\label{sec:bb0nu}

The search for $0\nu\beta\beta$ decays is performed by first selecting two-electron events using the criteria described in Section~\ref{sec:bb-selection}, where we require two electrons emitted from a common 
vertex in one of the $^{100}$Mo foils with a combined energy $E_{\rm tot}>2$~MeV.  We then search for an excess in data above the background expectation in the $E_{\rm tot}$ distribution for energies close to the 
value of $Q_{\beta\beta}$. 
The contributions of the background from external sources, from
radon, and from the internal $^{214}$Bi and $^{208}$Tl foil contaminations are fixed to the measured values given 
in Section~\ref{sec:bkg}. 

We obtain the $2\nu\beta\beta$ background contribution 
 by fitting the $E_{\rm tot}$ distribution in the range $E_{\rm tot}>2$~MeV using the shape of the spectrum predicted by the Single State Dominance model for the  $2\nu\beta\beta$ decay of $^{100}$Mo~\cite{ssd}.
The other background components are also taken into account in the fit. 
Figure~\ref{fig:result} shows that the fitted $E_{\rm tot}$ distributions for Phase~I, Phase~II, and for the combined data set agree with the data.
The fitted number of $2\nu\beta\beta$ events for $E_{\rm tot}>2$~MeV corresponds to a $^{100}$Mo half-life of 
\begin{equation}
T_{1/2}(2\nu\beta\beta) = [ 6.93 \pm  0.04\ (\mbox{stat}) ] \times 10^{18}~{\rm yr} , 
\end{equation} 
after correcting for the signal efficiency,
which is in agreement with the previously published result for Phase~I~\cite{article0nuphase1} and with the world average~\cite{PDG}. 

The $E_{\rm tot}$ distribution in the region $2.8\le E_{\rm tot} \le 3.2$~MeV is shown in Figure~\ref{fig:result}, and
the different components of background in this energy window, and the number of observed  two-electron events are given in Table \ref{tab:background-mo}. 
In Phase~II, the observed background rate for $2.8\le E_{\rm tot} \le 3.2$~MeV
 is $0.44 \pm 0.13$~counts/yr/kg, with about $55\%$
 originating from $2\nu \beta \beta$ decays of $^{100}$Mo, about $20\%$ 
 from the radon gas contamination inside the tracking chamber, 
 and about $20\%$ from internal $^{208}$Tl contamination in the $^{100}$Mo foils.
We estimate the internal $^{214}$Bi contamination in the composite $^{100}$Mo foils to be $5\%$, while this background is negligible
for metallic foils. 
The contributions from external backgrounds are also negligible. 

\begin{table}
\begin{center}
\begin{tabular}{ccccc}
\hline\hline
Data Set & Phase~I & Phase~II & Combined \\
\hline
External Background                & $<$ 0.04 & $<$ 0.16 &      $<$ 0.2         \\
$^{214}$Bi from Radon                   & 2.8 $\pm$ 0.3 & 2.5 $\pm$ 0.2 & 5.2 $\pm$ 0.5      \\
$^{214}$Bi Internal      & 0.20 $\pm$ 0.02  & 0.80 $\pm$ 0.08 & 1.0 $\pm$ 0.1   \\
$^{208}$Tl Internal                  & 0.65 $\pm$ 0.05 & 2.7 $\pm$ 0.2 & 3.3 $\pm$ 0.3  \\
$2 \nu \beta \beta$ Decays                 & 1.28 $\pm$ 0.02 & 7.16 $\pm$
0.05 & 8.45 $\pm$ 0.05  \\
Total Expected           & 4.9 $\pm$ 0.3  & 13.1 $\pm$ 0.3 & 18.0 $\pm$ 0.6    \\
\hline
Data & 3 & 12 & 15 \\
\hline\hline
\end{tabular}
\caption{Numbers of expected background and observed two-electron events in Phases~I and~II 
in the $^{100}$Mo foil for an exposure of $34.3$~kg$\cdot$yr in the 
range $E_{\rm tot} = [2.8-3.2]$~MeV. The $0 \nu \beta\beta$ signal detection efficiency is $4.7\%$ in this energy range.}
\label{tab:background-mo}
\end{center}
\end{table}

Since we observe no significant excess in data above the background expectation, a limit on the $0\nu\beta\beta$ decay of $^{100}$Mo is derived. 
The uncertainties on the efficiency to
detect $0\nu\beta\beta$ events and on the estimated background contributions are the two main components of the systematic
uncertainty.
As discussed in Section~\ref{sec:bb-selection}, the systematic uncertainty on the $0\nu\beta\beta$ detection efficiency is $5\%$. 
The systematic uncertainties on the estimated background contributions are due to the activities of $2\nu\beta\beta$ decays, 
and the $^{214}$Bi and $^{208}$Tl backgrounds. 
An uncertainty of $0.7\%$ on the $2\nu\beta\beta$ activity is obtained from the fit to two-electron events in the energy range $E_{\rm tot}>2$~MeV. 
As discussed in Section~\ref{sec:bkg}, the systematic uncertainty on the normalisations of the background contributions from radon, $^{214}$Bi, and $^{208}$Tl radioactive contaminants is $10\%$.
This systematic uncertainty is taken into account in setting the limit on the $0\nu\beta\beta$ decay of the $^{100}$Mo isotope. The contributions of the external backgrounds and from thoron are negligible.

The limit on the $0\nu\beta \beta$ half-life is set using a modified frequentist analysis that employs a log-likelihood ratio test statistics~\cite{CLs}. 
The method uses the full information of the binned energy sum distribution in the $E_{\rm tot} = [2.0-3.2]$~MeV energy range for signal and background (see Figure~\ref{fig:result}), as well as the statistical and systematic uncertainties and their correlations, and is described in more detail in~\cite{CLs,Nd-PRC}.  All limits are given at the $90\%$ C.L. 
The data are described well by the background-only hypothesis with a $p$ value of $p=1-CL_{b} = 0.647$.
Taking into account the $0 \nu \beta\beta$ detection efficiency of $11.3 \%$ for the combined data set and the total exposure of $34.3$~kg$\cdot$yr, we obtain a limit of $T_{1/2}(0\nu\beta\beta)>1.1 \times 10^{24}$~yr for the $0\nu\beta\beta$ decays of $^{100}$Mo with decay kinematics similar to that for the light Majorana neutrino exchange.

The result agrees with the median expected sensitivity of the experiment of $T_{1/2}(0\nu\beta\beta)=1.0  \times 10^{24}$~yr within the $\pm 1$ standard deviation (SD) range of $[0.7,1.4] \times 10^{24}$~yr.
This result is a factor of two more stringent than the previous best limit for this isotope~\cite{article0nuphase1}. 
The corresponding upper limit on the effective Majorana neutrino mass is $\langle m_{\nu} \rangle < 0.33$--$0.62$~eV, where the range is determined by existing uncertainties on the calculations of the NMEs~\cite{Suhonen2015,Simkovic2013,Iachello2015,Rath2010,Rodriguez2010} and phase space factors~\cite{Iachello2012,Stoica2013}. The upper value 0.62~eV is lower than the upper value previously reported in our rapid communication~\cite{nemo3-mo100-pr-shortcomm}, because of the use of the new NME calculation from~\cite{Suhonen2015}, which is an update of the previous calculation~\cite{Suhonen2012}.
%This new calculation, based on the proton-neutron quasiparticle random-phase approximation, is an update of the previous one~\cite{Suhonen2012}, and has been done by using realistic two-body interactions and single-particle bases, and including up-to-date nucleon-nucleon short-range correlations, nucleon form factors and induced weak currents of nucleons. This new calculation uses also a new method to improve on the isospin properties of the neutrinoless double Fermi nuclear matrix elements. 

We also derive constraints on other lepton-number violating models: the supersymmetric models, the right-left symmetric models, and Majoron emission.

In supersymmetric models, the $0\nu\beta\beta$ process can be mediated by the exchange of a gluino or neutralino. 
Using the obtained limit of $T_{1/2}(0\nu\beta\beta)>1.1 \times 10^{24}$~yr and the NME from~\cite{MatrixElementsSupersymmetry} an upper bound is obtained on the trilinear $R$-parity violating supersymmetric coupling of $\lambda^{'}_{111} < (4.4-6.0) \times 10^{-2} f$,  where 
\begin{equation}
f = \left(\frac{M_{\tilde{q}}}{1~{\rm TeV}} \right)^2\left(\frac{M_{\tilde{g}}}{1~{\rm TeV}} \right)^{1/2} \; ,
\end{equation} and $M_{\tilde{q}}$ and $M_{\tilde{g}}$ represent the squark and gluino masses. 

Right-left symmetric models include right-handed currents in the electroweak Lagrangian that predict different angular and energy distributions of the final state electrons from the $0\nu\beta\beta$ decays. 
The NEMO-3 experiment, with the topological information for the two final-state electrons, can discriminate between the topologies from different mechanisms~\cite{SNemo}. The corresponding half-life limits are given in Table~\ref{tab:limits} and translate into an upper bound on the coupling between right-handed quark and lepton currents of $\langle \lambda \rangle < (0.9-1.3) \times 10^{-6}$  and into an upper bound on the coupling between right-handed quark and left-handed lepton currents of $\langle \eta \rangle < (0.5-0.8) \times 10^{-8}$. The constraints are obtained using the NME calculations from~\cite{Suhonen2002,Tomoda1991,Muto1989}. 

The $0\nu\beta\beta$ decay could also be accompanied by a Majoron ($M$), which is a light or massless boson that weakly couples to the neutrino~\cite{majoronth}. 
In this case the energy sum of the two emitted electrons, $E_{\rm tot}$, will have a broad spectrum 
in the range $[0$--$Q_{\beta\beta}]$. The shape will depend on the spectral index $n$, which determines the phase space dependence on the energy released in the decay, $G^{0\nu} \propto (Q_{\beta\beta}-E_{\rm tot})^{n}$. The lower bound on the half-life of the $0\nu\beta\beta$ decay with the spectral index $n=1$ is given in Table~\ref{tab:limits}. 
%--------------------------
% Added to referee
%{\color{red} 
The limit is set using the same method as the one used to extract the limit on the $0\nu\beta\beta$ half-life with the energy sum of the two emitted electrons, $E_{\rm tot}$. 
%}
%-------------------------
This limit is almost a factor of two more stringent than the previous best limit for this isotope~\cite{majoron}. 
Taking into account the phase space factors given in~\cite{Doi1988} and the NME calculated in~\cite{Suhonen2015,Simkovic2013,Iachello2015,Rath2010,Rodriguez2010}, an upper bound on the Majoron-neutrino coupling constant is obtained, $\langle g_{ee} \rangle < (1.6-3.0) \times 10^{-5}$. 

\begin{figure}[htbp]
\begin{center}
\includegraphics[width=0.5\textwidth]{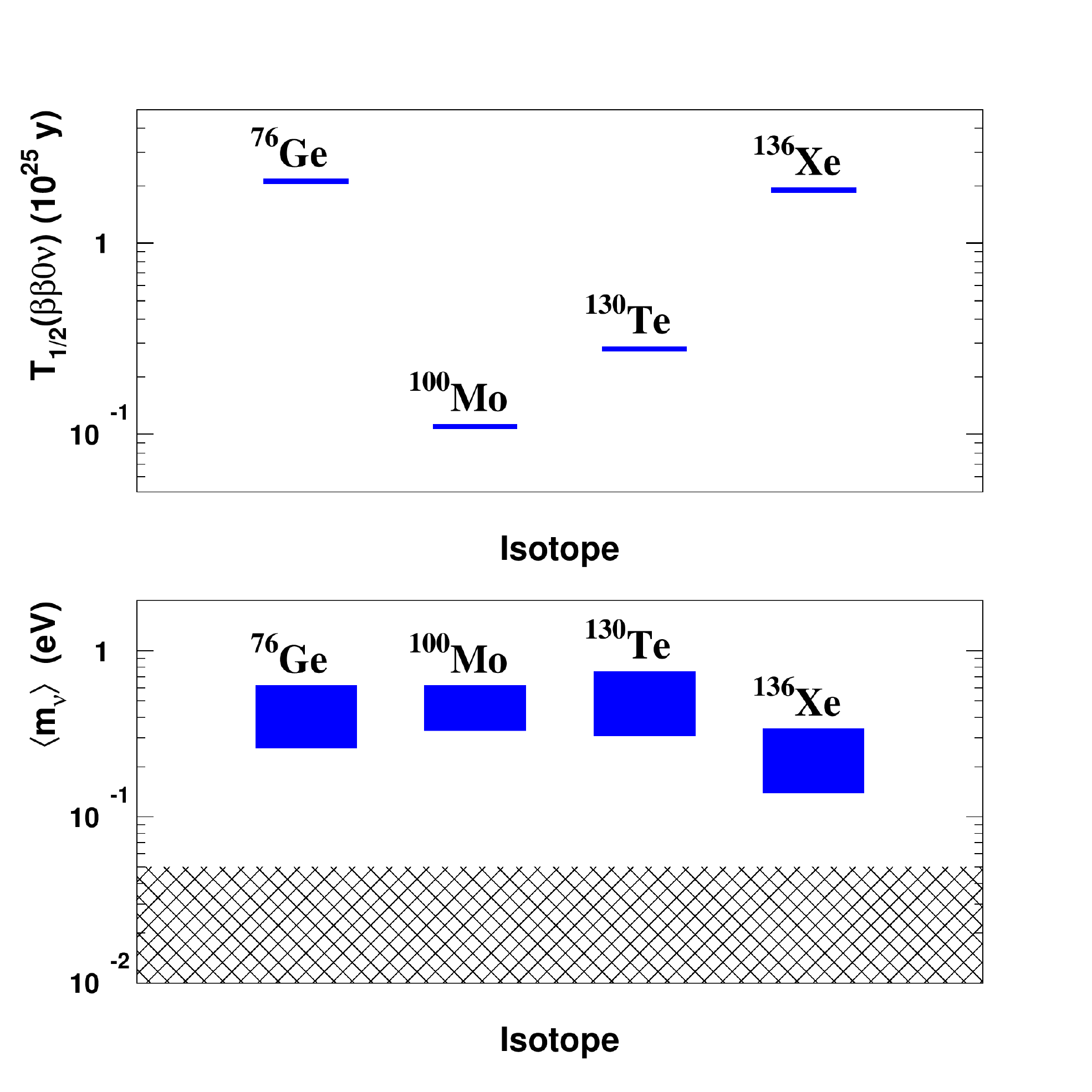}
\caption{The 90\% C.L. lower limits on $T_{1/2}(0\nu\beta\beta)$ for the light Majorana neutrino mass mechanism and upper limits on the effective Majorana neutrino mass $\langle m_{\nu} \rangle$ 
using the same NME calculations~\protect\cite{Suhonen2015,Simkovic2013,Iachello2015,Rath2010,Rodriguez2010} and recent phase space calculations~\protect\cite{Iachello2012,Stoica2013}. The shaded regions correspond to the ranges from using different NME calculations. The hatched area corresponds to the expected range for $\langle m_{\nu} \rangle$, calculated from the neutrino oscillation parameters and assuming the inverted neutrino mass hierarchy.}
\label{fig:current-limits}
\end{center}
\end{figure}

The limits on lepton number violating parameters obtained here have comparable sensitivity to the best current results obtained with other isotopes, as shown in Table~\ref{tab:comparison} and in Figure~\ref{fig:current-limits} for the light Majorana neutrino mass mechanism.

\begin{table}[htbp]
\begin{center}
\begin{tabular}{c|c|c|cccc}
\hline\hline
                                          & Statistical  & \multicolumn{4}{c}{Including Systematics} \\
                                          &               &   & \multicolumn{3}{c}{Expected} & \\
$0\nu\beta\beta$ Mechanism  &  Obs. & Obs. &   $- 1$~SD & Median &  $+1$~SD  \\
\hline
Mass Mechanism & 1.1 & 1.1 & 0.7 & 1.0 & 1.4 \\
RH Current $\langle \lambda \rangle$ & 0.7 & 0.6 & 0.4 &  0.5  & 0.8 \\
RH Current $\langle \eta \rangle$ & 1.0 & 1.0 & 0.6& 0.9 & 1.3 \\
Majoron & 0.050 & 0.044 & 0.027& 0.039 & 0.059 \\
\hline\hline
\end{tabular}
\caption{\label{tab:limits} Observed and median expected lower limits on half-lives of lepton number violating  processes (in units of $10^{24}$~yr) at the $90\%$ C.L. using statistical and systematical uncertainties. The observed lower limits are also given using only statistical uncertainties.}
\end{center}
\end{table}

\begin{table*}[htbp]
\begin{center}
\begin{tabular}{ccccccccc}
\hline\hline
          & Half-Life & $\langle m_{\nu} \rangle$ & $\langle m_{\nu} \rangle_{\rm rec}$  & $\langle \lambda \rangle$ & $\langle \eta \rangle$   & $\mathrm{\lambda^{'}_{111}/f}$ & $\langle g_{ee} \rangle$  \\ 
        & ($10^{25}$ yr) & (eV)   &  (eV) & ($10^{-6}$)  & ($10^{-8}$)   & ($10^{-2}$)   & ($10^{-5}$)   \\
\hline
$^{100}$Mo [This Work] & 0.11 & 0.33--0.62 & 0.33--0.62 & 0.9--1.3$\mathrm{^a}$ & 0.5--0.8$\mathrm{^a}$ & 4.4--6.0 & 1.6--3.0$\mathrm{^a}$ \\
$^{130}$Te \protect\cite{cuoricino,cuoricino2} & 0.28 & 0.3--0.71 & 0.31--0.75 & 1.6--2.4$\mathrm{^b}$ & 0.9--5.3$\mathrm{^b}$ & & 17--33$\mathrm{^c}$ \\
$^{136}$Xe \protect\cite{kamlandzen,KZ_majoron} & 1.9 & 0.14--0.34 & 0.14--0.34 & & & & 0.8-1.6 \\
$^{76}$Ge \protect\cite{gerda} & 2.1 & 0.2--0.4 & 0.26--0.62 & & & & \\
$^{76}$Ge \protect\cite{heidelbergmoscow,HeidelbergMoscowRightCurrent} & 1.9 & 0.35  & 0.27--0.65 & 1.1 & 0.64 & & 8.1 \\
\hline\hline
\end{tabular}
\caption{\label{tab:comparison} Limits at the $90 \%$ C.L. on
half-lives and lepton number violating parameters. Published experimental constraints on $\langle m_{\nu} \rangle$ 
and recalculated values with NMEs from 
Refs.~\protect\cite{Suhonen2015,Simkovic2013,Iachello2015,Rath2010,Rodriguez2010,Menendez2009} 
are also given.}
\end{center}
$\mathrm{^a}$ obtained with half-lives in Table~\ref{tab:limits}, $\mathrm{^b}$ using the half-life limit of $2.1 \times 10^{23}$~yr, $\mathrm{^c}$ using the half-life limit of $2.2 \times 10^{21}$~yr.
\end{table*}

\section{Conclusions}

We have presented results based on an analysis of the full NEMO-3 data set with an exposure of 
34.3~kg$\cdot$yr of $^{100}$Mo, which corresponds to $4.96$ effective years of data collection and 6.914~kg of $^{100}$Mo. The calibration of the calorimeter, the long-term stability of 
data taking, and the determination of the backgrounds are discussed in detail. 
No evidence for $0\nu \beta \beta$ decays of $^{100}$Mo has been found,
as previously reported in our rapid communication~\cite{nemo3-mo100-pr-shortcomm}.
Taking into account statistical and systematic uncertainties, the limit on the $0\nu \beta \beta$ decay half-life with decay kinematics similar to that for light Majorana neutrino exchange is $T_{1/2}(0\nu\beta\beta)>1.1 \times 10^{24}$ yr ($90\%$ C.L.). 
The corresponding limit on the effective Majorana neutrino mass is in the range 
$\langle m_{\nu} \rangle < 0.33$--$0.62$~eV, depending on the NME calculation used in the derivation.

Studies of the backgrounds using various decay channels, radioactive sources, and HPGe measurements before the installation of the detector are used to construct and validate a detailed model of the background.
In Phase~II, the expected background rate in the $0\nu\beta\beta$ signal region $E_{\rm tot}= [2.8-3.2]$~MeV is $0.44 \pm 0.13$~counts/yr/kg. 
About half of this background is expected to be $2\nu \beta \beta$ decays of $^{100}$Mo, and the remaining background is caused in roughly equal parts by the radon gas contamination inside the tracking chamber, which is about $5$~mBq/m$^3$, and by $^{208}$Tl  contamination inside the $^{100}$Mo foils, which is
between $90$--$130 \ \mu$Bq/kg depending on the type of foil. 
No background events are observed in the region of $E_{\rm tot} = [3.2-10]$~MeV for  NEMO-3 sources containing isotopes with $Q_{\beta\beta}<3.2$~MeV ($^{100}$Mo, $^{82}$Se, $^{130}$Te, $^{116}$Cd), or in the copper foil, which is not a double $\beta$ emitter, during the entire running period corresponding to an exposure of $47$~kg$\cdot$yr. 

This low level of background demonstrates that an extremely low level of non double $\beta$ background can be achieved by the future SuperNEMO experiment, which will employ the NEMO-3 technique.
%, if high-$Q_{\beta\beta}$ isotopes such as $^{82}$Se, $^{48}$Ca, $^{96}$Zr, or $^{150}$Nd are used. 
The SuperNEMO Collaboration proposes to search for $0\nu\beta\beta$ decays using $100$~kg of double $\beta$ isotopes~\cite{SNemo}. 
The $2\nu \beta \beta$ background will be further reduced by improving the energy resolution and by measuring an isotope with a long $2\nu \beta \beta$ half-life, currently assumed to be $^{82}$Se. 
Other favorable isotopes, such as $^{150}$Nd and $^{48}$Ca, are also studied. 
%As demonstrated by the study of background contaminations for NEMO-3, the radiopurity of the source foils and the radon purity inside the tracking chamber must be maximized.
A first SuperNEMO demonstrator module, currently under construction, will contain $7$~kg of $^{82}$Se. The objective is to demonstrate that the background can be reduced by $1$--$2$ orders of magnitude compared to the NEMO-3 detector.

\section*{Acknowledgments}
The authors would like to thank the Modane Underground Laboratory staff for their technical assistance in running the experiment. 
We acknowledge support by the grants agencies of the Czech Republic, CNRS/IN2P3 in France, RFBR in Russia, STFC in U.K. and NSF in U.S.

\end{document}